\documentclass[a4paper,11pt]{article}
\pdfoutput=1 

\usepackage{jheppub} 
\usepackage[T1]{fontenc} 
\usepackage[italian,english]{babel}
\usepackage{hyperref}
\usepackage{ifpdf}
\usepackage{subfigure}
\usepackage{tikz}
\usepackage[compat=1.1.0]{tikz-feynman}
\usepackage{slashed}
\usetikzlibrary{arrows,shapes}
\usetikzlibrary{trees}
\usetikzlibrary{matrix,arrows} 
\usetikzlibrary{positioning}
\usetikzlibrary{calc,through}
\usetikzlibrary{decorations.pathreplacing}  
\usepackage{pgffor}

\usetikzlibrary{decorations.pathmorphing}
\usetikzlibrary{decorations.markings}
\tikzset{
vector/.style={decorate, decoration={snake}, draw},
fermion/.style={draw=black, postaction={decorate}}, 
scalar/.style={dashed,draw=black, postaction={decorate}}}
\tikzstyle{block} = [draw, rectangle, 
minimum height=3em, minimum width=6em]

\usepackage{amssymb}
\usepackage{amsfonts}
\usepackage{epsf}
\usepackage{rotating}
\usepackage{graphicx}
\usepackage{amsmath}
\usepackage{fancyhdr}
\usepackage{lineno}
\usepackage{babel}
\usepackage{graphics}
\usepackage{pstricks}
\usepackage{color}
\usepackage{multirow}
\usepackage{float}
\usepackage{lineno}
\usepackage{mathtools}
\usepackage{stackengine}
\usepackage{comment}

\newcommand{\lsim}{\mathrel{\mathop{\kern 0pt \rlap
{\raise.2ex\hbox{$<$}}}
\lower.9ex\hbox{\kern-.190em $\sim$}}}
\newcommand{\gsim}{\mathrel{\mathop{\kern 0pt \rlap
{\raise.2ex\hbox{$>$}}}
\lower.9ex\hbox{\kern-.190em $\sim$}}}

\newcommand{\be}{\begin{equation}}
\newcommand{\ee}{\end{equation}}
\newcommand{\bea}{\begin{eqnarray}}
\newcommand{\eea}{\end{eqnarray}}

\def\gev{\ensuremath{\mathrm{\,Ge\kern -0.1em V\,}}}
\def\tev{\ensuremath{\mathrm{\,Te\kern -0.1em V\,}}}

\newcommand{\hpp}{$H_5^{\pm \pm}~$}

\begin{document}

\title{Revisiting the decoupling limit of the Georgi-Machacek model with a scalar singlet}

\author[a]{Genevi\`{e}ve B\'{e}langer,}
\author[b]{Juhi Dutta,}
\author[c]{Rohini M. Godbole,}
\author[d]{Sabine Kraml,}
\author[e,f]{Manimala Mitra,}
\author[g]{Rojalin Padhan,}
\author[e,f]{Abhishek Roy}

\affiliation[a]{LAPTh, CNRS, USMB, 9 Chemin de Bellevue, 74940 Annecy, France}
\affiliation[b]{Homer L. Dodge Department of Physics and Astronomy, University of Oklahoma, Norman, OK 73019, USA}
\affiliation[c]{Centre for High Energy Physics, Indian Institute of Science, Bengaluru - 560012, India}
\affiliation[d]{Univ. Grenoble Alpes, CNRS, Grenoble INP, LPSC-IN2P3, 38000 Grenoble, France}
\affiliation[e]{Institute of Physics, Sachivalaya Marg, Bhubaneswar, Odisha 751005, India}
\affiliation[f]{Homi Bhabha National Institute, BARC Training School Complex, Anushakti Nagar, Mumbai 400094, India}
\affiliation[g]{Department of Physics, Chung-Ang University, Seoul 06974, Korea}

\emailAdd{belanger@lapth.cnrs.fr}
\emailAdd{juhi.dutta@ou.edu}
\emailAdd{rohini@iisc.ac.in}
\emailAdd{sabine.kraml@lpsc.in2p3.fr}
\emailAdd{manimala@iopb.res.in}
\emailAdd{rojalinpadhan2014@gmail.com}
\emailAdd{abhishek.r@iopb.res.in}


\abstract{We study the connection between collider and dark matter phenomenology in the singlet extension of the Georgi-Machacek model. 
In this framework, the singlet scalar serves as a suitable thermal dark matter (DM) candidate. Our focus lies on the region $v_{\chi}<1$ GeV, where $v_{\chi}$ is the common vacuum expectation value of the neutral components of the scalar triplets of the model.
Setting bounds on the model parameters from theoretical, electroweak precision  and LHC experimental constraints, we find that the BSM Higgs sector is highly constrained. Allowed values for the masses of the custodial fiveplets, triplets and singlet are restricted to the range $140~ {\rm GeV }< M_{H_5^0} < 350~ {\rm GeV }$, $150~ {\rm GeV }< M_{H_3^0} < 270 ~{\rm GeV }$ and 
$145~ {\rm GeV }< M_{H} < 300~ {\rm GeV }$.  
The extended  scalar sector provides new channels for DM annihilation into BSM scalars that allow to  satisfy the observed relic density constraint while being consistent with direct DM detection limits.  The  allowed region of the parameter space of the model can be explored in the  upcoming DM detection experiments, both direct and indirect.   
In particular, the possible high values of BR$(H^0_5\to\gamma\gamma)$ can lead to an indirect DM signal within the reach of CTA. The same feature also provides the possibility of  exploring the model at the High-Luminosity run of the LHC. 
In a simple cut-based analysis, we find that a signal of about $4\sigma$ significance can be achieved in final states with at least two photons for one of our benchmark points. }

\maketitle              

\section{Introduction}

There are compelling motivations for considering models of  new physics, as the Standard Model (SM) is inadequate to explain a few of the major observations in nature, such as, the observed  non-zero neutrino masses and their mixings, the measured relic abundance of the dark matter (DM), and the observed  matter-antimatter asymmetry of 
the universe. Furthermore, since the discovery of the  125 GeV SM scalar at the LHC \cite{CMS:2012qbp, ATLAS:2012yve}, it is yet unknown if  the electroweak symmetry breaking (EWSB)  is achieved  by a single or multiple scalar fields. Several extensions have been studied in the literature to address this key question, among which  the  Georgi-Machacek (GM) model \cite{Georgi:1985nv}  is one of the most interesting extensions of the SM with two $SU(2)_L$ triplet scalar multiplets. 

The scalar sector of the GM model consists of one real triplet with $Y=0$ and one complex triplet with $Y=2$ in addition to the SM Higgs doublet. 
The model preserves custodial symmetry (CS) 
 at the tree level, but hypercharge interactions induce CS violation at one loop, 
thus leading to a correction to the $\rho$ parameter~\cite{Gunion:1990dt}.  These corrections are  moderate because of the underlying tree-level CS.
The model also predicts a larger Higgs-to-vector-boson coupling than in the SM, thus favouring a larger Higgs-to-diphoton rate. 
The scalar sector consists of ten physical degrees of freedom: a fiveplet, a triplet and two CP-even singlets under the CS. The Higgs and BSM phenomenology of this model have  extensively been studied in \cite{Chanowitz:1985ug, Gunion:1990dt,Haber:1999zh,Aoki:2007ah,Logan:2010en,Chang:2012gn,Chiang:2012cn,Kanemura:2013mc,Englert:2013zpa,Chiang:2013rua,Efrati:2014uta,Hartling:2014zca,Campbell:2016zbp,Degrande:2017naf,Ghosh:2019qie,Ismail:2020zoz,Ismail:2020kqz,Bairi:2022adc,Chen:2022zsh,Chiang:2018cgb}. One simplified version of the GM potential with a $Z_2$ symmetry that excludes the two dimension-3 operators, 
has been studied in~\cite{Chanowitz:1985ug, Gunion:1990dt,Chang:2012gn,Englert:2013zpa,Efrati:2014uta,deLima:2022yvn}.
The  model also offers the possibility of generating  naturally light Majorana masses for the neutrinos through the seesaw mechanism~\cite{Chiang:2012cn}. However, this minimal version of the model is inadequate to  provide a viable stable DM candidate. This drawback of the model can be overcome by incorporating an additional isospin singlet scalar field, which transforms as $S\to -S$ under a $Z_2$ symmetry  \cite{Campbell:2016zbp}. 

Amongst the various extensions of the SM that can incorporate DM, models containing  a SM gauge singlet scalar field are the simplest extensions. The singlet scalar extension can easily accommodate a viable WIMP DM candidate~\cite{Silveira:1985rk,McDonald:1993ex,Burgess:2000yq}, owing to  its quartic interaction with the SM Higgs field. The minimal model, containing only a singlet, is however severely constrained by the precise determination of the DM relic density and the non-observation of DM through direct detection, since both depend directly on the coupling of DM to the SM-like Higgs boson~\cite{GAMBIT:2017gge}.  
Adding new particles (scalars, fermions or vector bosons) allows one to alleviate these constraints by disconnecting the processes that are responsible for elastic scattering on nuclei through Higgs exchange from  the ones that contribute to  DM  annihilation in the early Universe.  
In particular, a more elaborate scalar sector, such as the BSM scalars present in the GM model makes it easier to satisfy current constraints by providing new final states for DM annihilation.

In this article, we consider the most general scalar potential of the GM model~\cite{Aoki:2007ah,Chiang:2012cn,Chiang:2013rua,Hartling:2014zca,Campbell:2016zbp} extended 
by a real singlet scalar~\cite{Campbell:2016zbp}, which will be our DM candidate. We refer to this model as GM-S model henceforth. We focus on the decoupling limit~\cite{Hartling:2014zca,Chiang:2012cn} of the GM-S model, which can be realised for a very small triplet vev. In this limit, the coupling strengths of the observed Higgs boson resemble their SM values.
We revisit the validity of the model with respect to theoretical constraints, precision measurements of the oblique parameters,  measurements of the SM Higgs to diphoton rate and the recent results from searches for neutral and doubly charged scalars at the LHC. We show that theoretical bounds, in particular the perturbative unitarity condition for this decoupling limit force the masses of the  BSM scalars to be of the order of the weak scale. The measured values of electroweak precision observables (EWPO) further constrain large mass splitting between the fiveplet and triplet states.
The diphoton rate of the SM-like Higgs boson restricts the splitting between the fiveplet mass and one of the dimension-full parameters of the potential, while searches for a doubly charged Higgs decaying into a pair of $W^\pm$ bosons require the mass splitting between the fiveplet and the triplet to be larger than roughly 30~GeV. Finally, 
searches for a BSM resonance decaying into  diphoton states impose a lower limit on the triplet vev for the mass scales under consideration. 

Taking into account the updated constraints on the parameter space of the GM-S 
model in the decoupling limit, we determine  the parameter space that satisfies DM constraints from relic density and  direct detection. 
As expected, we find that annihilation channels into any of the BSM Higgs play an important role in DM formation once we impose the direct detection constraints. 
Including the contributions of all final states leading to photons, we show that indirect detection limits from  FermiLAT do not constrain the model. The CTA experiment, however, has the potential to probe part of the parameter space.
Finally, we show that the same large decay rate into diphotons, that can lead to a signal at CTA, may also be exploited in searches at the High-Luminosity run of the LHC~(HL-LHC). Searches in  multi-lepton channels, on the other hand, are more challenging.

The structure of this paper is as follows: section~\ref{sec:model} gives a brief review of the GM-S model. Theoretical and experimental constraints on the parameter space are discussed in section~\ref{sec:bound}. The dark matter phenomenology is investigated in section~\ref{sec:DMpheno}, and the prospects for discovering the BSM scalars at the HL-LHC in section~\ref{sec:colliderpheno}. A summary and conclusions are given in section~\ref{sec:summary}. The appendix gives expressions  for the partial decay widths of all the BSM scalars.

\section{The Model}\label{sec:model}
The  GM-S model we consider here is the original Georgi-Machacek (GM) model ~\cite{Georgi:1985nv} extended by a real scalar field $S$~\cite{Campbell:2016zbp}. The original GM model 
consists of an extended scalar sector which, in addition to  the  SM scalar doublet $\Phi$, contains two scalar triplets $\xi = (\xi^{+}, \xi^{0}, \xi^{-})^{T}$ and $\chi = (\chi^{++}, \chi^{+}, \chi^{0})^{T}$. The triplet $\xi$ has hypercharge $Y = 0$ while $\chi$ has hypercharge $Y = 2$. The neutral fields of both the doublet and the triplets 
acquire vevs $\langle\phi^0\rangle = v_{\phi}/\sqrt{2}$, $\langle\xi^0 \rangle =v_{\xi}$ and  $\langle\chi^0\rangle = v_{\chi}$.
The vev $\langle\phi^0\rangle$  breaks the electroweak symmetry  spontaneously, along with contributions from $\xi^0$ and $\chi^0$. 

The scalar potential is symmetric under a global $SU(2)_L\times SU(2)_R$ transformation which, after electroweak symmetry breaking, breaks down to a custodial symmetry $SU(2)_C$.
At the tree level CS is guaranteed by demanding
\begin{equation}
    \langle\xi^0 \rangle = \langle\chi^0\rangle = v_{\chi} .
\label{eq:custodial_cond}
\end{equation} 

In order to make the global $SU(2)_L\times SU(2)_R$  symmetry explicit in the potential, we follow the convention of  \cite{Hartling:2014zca}, and  write the scalar fields in terms of a bi-doublet $\Phi$ and a bi-triplet $X$ as follows:

\begin{equation}
    \Phi = \begin{bmatrix}
        \phi^{0*} & \phi^{+} \\
        - \phi^{-} & \phi^{0}
    \end{bmatrix}, \quad
    X = \begin{bmatrix}
        \chi^{0*} & \xi^{+} & \chi^{++} \\
        - \chi^{-} & \xi^{0} & \chi^{+} \\
        \chi^{--} & - \xi^{-} & \chi^{0}
        \end{bmatrix} .
    \label{eq:matrix}
    \end{equation}

As mentioned before in the GM-S model, in addition to the triplet fields $\xi$ and $\chi$, the model is further extended by a real scalar field $S$~\cite{Campbell:2016zbp}, which serves as the DM candidate. 
The stability of the DM is ensured by imposing a $Z_2$ symmetry, under which the DM is odd and all the other particles are even. Note that, the neutral scalar states from the $X$ field, $\xi^0$ and $\chi^0$, cannot be considered as DM candidates due to the lack of a discrete symmetry which can stabilize it.
The  general gauge-invariant tree-level scalar potential with $\Phi$, $X$ and $S$ obeying CS has the form \cite{Chanowitz:1985ug,Hartling:2014zca}:  
\begin{align}
    V\left(\Phi, X, S\right)   = & \; \frac{\mu^{2}_{2}}{2}\text{Tr}\left[\Phi^{\dagger}\Phi\right] + \frac{\mu^{2}_{3}}{2}\text{Tr}\left[X^{\dagger}X\right] + \lambda_{1}\left(\text{Tr}\left[\Phi^{\dagger}\Phi\right] \right)^{2} \nonumber \\ 
    &  +  \lambda_{2}\text{Tr}\left[\Phi^{\dagger}\Phi\right]\text{Tr}\left[X^{\dagger}X\right] + \lambda_{3}\text{Tr}\left[X^{\dagger}XX^{\dagger}X\right]  \label{eq:scalarpot}
 \\
    &  + \lambda_{4}\left(\text{Tr}\left[X^{\dagger}X\right]\right)^{2} - \lambda_{5}\text{Tr}\left[\Phi^{\dagger}\tau^{a}\Phi\tau^{b}\right]\text{Tr}\left[X^{\dagger}t^{a}Xt^{b}\right] \nonumber \\
    &  - M_{1}\text{Tr}\left[\Phi^{\dagger}\tau^{a}\Phi\tau^{b}\right]\left(UXU^{\dagger}\right)_{ab} - M_{2}\text{Tr}\left[X^{\dagger}t^{a}Xt^{b}\right]\left(UXU^{\dagger}\right)_{ab} + V(S) \,, \nonumber 
\end{align} 
\noindent
where $\tau^a=\sigma^a/2$ with $\sigma^a$ being the Pauli matrices; $t^a$ are generators of the adjoint representation and the matrix 
\begin{equation}
U = 
\begin{bmatrix}
- \frac{1}{\sqrt{2}} & 0 & \frac{1}{\sqrt{2}} \\
- \frac{i}{\sqrt{2}} & 0 & - \frac{i}{\sqrt{2}} \\
0 & 1 & 0
\end{bmatrix}
\end{equation}
is used to rotate the bi-triplet into a cartesian basis~\cite{Aoki:2007ah}. In Eq.~\eqref{eq:scalarpot}, $V(S)$ contains the terms
involving the field $S$ and has the form  
\begin{equation}
V(S) = \frac{\mu^{2}_{S}}{2}S^{2} + \lambda_{a}\,S^{2}\,\text{Tr}\left(\Phi^{\dagger}\Phi\right) + \lambda_{b}\,S^{2}\,\text{Tr}\left[X^{\dagger}X\right]+\lambda_{S}S^{4}\,.
\label{eq:dm_potential}
\end{equation}
In addition to the  couplings $\lambda_i$ in Eq.~\eqref{eq:scalarpot} that represent the quartic interaction of the $\Phi$, $X$ fields and the interaction between $\Phi$ and $X$, the  interaction terms with  couplings $M_1$ and $M_2$ govern the tri-linear interactions between $\Phi$ and $X$ and the self-interactions of the $X$ fields, respectively. 

The field $\chi$, being a complex triplet with hypercharge $Y=2$, also interacts with the lepton doublets $L$ through the Yukawa Lagrangian 
\begin{equation}
\mathcal{L}_\nu = Y_{\nu} \bar{L}^c i\tau_2 \chi L+ \rm{h.c.}.
\label{eq:neumass}
\end{equation}
Note that the above term together with the tri-linear  $M_2$ term in Eq.~\eqref{eq:scalarpot} violates lepton number and hence can generate the Majorana masses of the light neutrinos.  
The light neutrino mass matrix can be expressed in terms of the Yukawa coupling $Y_{\nu}$ and the vev $v_{\chi}$ as follows 
\begin{equation}
M_{\nu}=Y_{\nu} v_{\chi}=U_{\rm PMNS}^* \, M^d_{\nu}\, U^{\dagger}_{\rm PMNS} \,,
\label{eq:neumass2}
\end{equation}
where $U_{\rm PMNS}$ is the diagonalization matrix for the light neutrinos, and $M^d_{\nu}=\textrm{diag}(m_1, m_2,m_3)$ are the neutrino masses.

\subsection{The scalar spectrum}
To obtain the scalar spectrum in the physical mass basis, one needs to impose the minimisation conditions 
\begin{equation}
\frac{\partial V}{\partial v_{\phi}} = v_{\phi}\left[\mu^{2}_{2} + 4\lambda_{1}v^{2}_{\phi} + 3\left(2\lambda_{2} - \lambda_{5}\right)v^{2}_{\chi} - \frac{3}{2} M_{1}v_{\chi}\right] = 0 \,,
\label{eq:minima}
\end{equation}
\begin{equation}
\frac{\partial V}{\partial v_{\chi}} = 3\mu^{2}_{3}v_{\chi} + 3\left(2\lambda_{2} - \lambda_{5}\right)v^{2}_{\phi}v_{\chi} + 12\left(\lambda_{3} + 3\lambda_{4}\right)v^{3}_{\chi} - \frac{3}{4}M_{1}v^{2}_{\phi} - 18M_{2}v^{2}_{\chi} = 0 \,.
\label{eq:minima1}
\end{equation}
Both the doublet and triplet vevs are responsible for the $W/Z$-boson mass generation,  
\begin{equation}
M^{2}_{W} =  M^{2}_{Z}  \cos^2\theta_W = \frac{g^{2}v^{2}}{4} = \frac{g^{2}}{4}\left(v^{2}_{\phi} + 8v^{2}_{\chi}\right) , 
\label{eq:Wmass}
\end{equation}
\noindent
where $v = 246$ GeV and $g$ is the coupling constant of the $SU(2)_{L}$ gauge group. Based upon the transformation properties under the custodial $SU(2)_{C}$ symmetry, the physical scalar states can be categorised into a fiveplet, a triplet and two CP-even singlets. The fiveplet scalars are expressed in the form: 
\begin{equation}
  H^{++}_{5} =  \chi^{++} \,, \quad 
  H^{+}_{5}  = \frac{1}{\sqrt{2}} \left(\chi^{+} - \xi^{+}\right) \,, \quad 
  H^{0}_{5} = \sqrt{\frac{2}{3}}\,\xi^{0,r} - \sqrt{\frac{1}{3}}\,\chi^{0,r} \,, 
  \label{eq:H5} \\
\end{equation}
where  $\xi^{0,r}$ and $\chi^{0,r}$ represent the real components of $\xi^{0}$ and $\chi^{0}$, respectively. 
Similarly, the triplet scalar states can be expressed as 
\begin{equation}
   H^{+}_{3} =  - s_{H}\,\phi^{+} +  \frac{1}{\sqrt{2}} c_{H} \left(\xi^{+} + \chi^{+}\right) \,, \quad
   H^{0}_{3} = - s_{H}\,\phi^{0,i} + c_{H}\,\chi^{0,i} \,,  
   \label{eq:H3}
\end{equation} 
with $\phi^{0,i}$ and $\chi^{0,i}$ being the imaginary components of $\phi^{0}$ and $\chi^{0}$, respectively. The mixing angle corresponding to $H^{\pm}_3$ and $H^0_3$ states can be written as 
\begin{equation}
c_{H} \equiv \cos\theta_{H} = \frac{v_{\phi}}{v} \,, \quad  s_{H} \equiv \sin\theta_{H} = \frac{2\sqrt{2}v_{\chi}}{v} \,.
\label{eq:vevs_relations}
\end{equation}
\noindent 
Using the minimization conditions in Eqs.~\eqref{eq:minima}, \eqref{eq:minima1}, we obtain
\begin{equation}
   M^{2}_{H_5} = \frac{M_{1}}{4v_{\chi}}v^{2}_{\phi} + 12 M_{2}v_{\chi} + \frac{3}{2}\lambda_{5}v^{2}_{\phi} + 8\lambda_{3}v^{2}_{\chi} \,,
   \label{eq:fivepletmass}
\end{equation}
and
\begin{equation}
   M^{2}_{H_3} = \left(\frac{M_{1}}{4v_{\chi}} + \frac{\lambda_{5}}{2}\right)v^{2}.
   \label{eq:tripletmass}
\end{equation}
$M_{H_5}$ and $M_{H_3}$ are the masses of the  fiveplet and the triplet fields.  
The mass of the gauge singlet field $S$ is given by 
\begin{equation}
    M_{S}^2 = \mu_{S}^2 + 2\lambda_{a}v_{\phi}^{2} + 6\lambda_{b}v_{\chi}^{2} \,.
   \label{eq:dm_mass}
\end{equation}

The neutral components of the $SU(2)_L$ doublet and triplets mix, leading to two CP-even $SU(2)_C$ singlets  $(h,\,H)$, 
\begin{equation}
\begin{split}
h & = c_{\alpha}\phi^{0,r} - s_{\alpha}H^{0'}_{1} \,,\\
H & = s_{\alpha}\phi^{0,r} + c_{\alpha}H^{0'}_{1} \,,
\end{split}
\end{equation}
\noindent
where $H^{0'}_{1}$ is defined as 
\begin{equation}
   H^{0'}_{1} = \sqrt{\frac{1}{3}}\,\xi^{0,r} + \sqrt{\frac{2}{3}}\,\chi^{0,r} \,.
\end{equation}
The mixing angle $\alpha$ and the mass eigenvalues are obtained by diagonalisation of the $2\times2$ mass-squared matrix corresponding to these scalars. This matrix  reads
\begin{equation}
\mathcal{M}^{2} = 
\begin{bmatrix}
\mathcal{M}^{2}_{11} & \mathcal{M}^{2}_{12} \\
\mathcal{M}^{2}_{12} & \mathcal{M}^{2}_{22} 
\end{bmatrix} 
\end{equation}
with 
\begin{align}
  \mathcal{M}^{2}_{11} &= 8\lambda_{1}v_{\phi}^{2}\,,\notag\\
  \mathcal{M}^{2}_{12} &= \frac{\sqrt{3}}{2}v_{\phi}[-M_{1}+4(2\lambda_{2}-\lambda_{5})v_{\chi}]\,,\notag\\
  \mathcal{M}^{2}_{22} &= \frac{M_{1}v_{\phi}^{2}}{4v_{\chi}}-6M_{2}v_{\chi}+8(\lambda_{3}+3\lambda_{4})v_{\chi}^{2}\,.
\label{eq:Mass_Matrix_Neutral}
\end{align}

\noindent
The mixing angle $\alpha$ between $h$ and $H$, and the respective mass eigenvalues are given by  
\begin{align}
  & \tan2\alpha = \frac{2\mathcal{M}^{2}_{12}}{\mathcal{M}^{2}_{22} - \mathcal{M}^{2}_{11}}\,, \\
  & M^{2}_{H,h}=\frac{1}{2} \left[ \mathcal{M}^{2}_{11}+ \mathcal{M}^{2}_{22}\pm \sqrt{ (\mathcal{M}^{2}_{11}-\mathcal{M}^{2}_{22})^2 + 4 (\mathcal{M}^{2}_{12})^2}  \right] ,
\end{align}
where we assume $M^{2}_{H} > M^{2}_{h}$. 
In our subsequent analysis, we consider $h$ as the 125 GeV SM-like Higgs boson while $H$ is a heavier BSM scalar.

The dimensionless couplings in the scalar potential, $\lambda_{1,\ldots, 5}$, can be expressed in terms of the physical  parameters $M_{H_5}$, $M_{H_3}$, $M_{H}$, $M_{h}$ and $\alpha$ as follows: 
\begin{align}
\lambda_1 &=\frac{1}{8v^2c_{H}^2}(M_h^2c_\alpha^2+M_{H}^2s_\alpha^2)\,,\notag\\
\lambda_2&= \frac{1}{6v^2s_{H}c_{H}}\left[\frac{\sqrt{6}}{2}s_{2\alpha}(M_h^2-M_{H}^2)+3s_{H}c_{H}(2M_{H_3}^2-\widetilde{M_1^2})\right],\notag\\
\lambda_3 &= \frac{1}{v^2s_{H}^2}\left[c_{H}^2(2\widetilde{M_1^2}-3M_{H_3}^2)+M_{H_5}^2-{M}^2\right],\notag\\
\lambda_4 &=\frac{1}{6v^2s_{H}^2}
\left[2M_{H}^2c_\alpha^2+2M_h^2s_\alpha^2+3{M}^2-2M_{H_5}^2+6c_{H}^2(M_{H_3}^2-\widetilde{M_1^2})\right],\notag\\
\lambda_5&= -\frac{2}{v^2}(\widetilde{M_1^2}-M_{H_3}^2)\,. \label{eq:couplambdas}
\end{align}
where 
$s_{H}$ and $c_{H}$ are defined in Eq.~\eqref{eq:vevs_relations} and
\begin{equation}
  \widetilde{M_1^2} \equiv \frac{v}{\sqrt{2}s_{H}}M_1 \,, \quad 
   {M}^2 \equiv 3 \sqrt{2}s_{H} v M_2 \,. 
\label{eq:param1}
\end{equation}
A few observations about the behaviour of various quantities in the limit of a small $v_{\chi}$, a case we will discuss in detail in the next section, are in order. For a small $v_{\chi}$ leading to the  limit $s_H\to 0$ (see Eq.~\eqref{eq:vevs_relations}) in order to assure that $\widetilde{M_1^2}$ remains finite
 we need to assume that $M_1 \to 0$.  Moreover, in this limit, it follows from Eq.~\eqref{eq:param1} that a finite value for $M$ implies a very large value of $M_2$, eventually leading to very large couplings and violation of perturbativity. Note that, in the  $s_H \to 0$ limit, the quartic couplings $\lambda_{2,4}$ diverge unless $s_{\alpha}$ is set to 0. This then implies  complete decoupling of the scalar $H$ from the SM Higgs state $h$ in this limit.  

\subsection{The case of small $v_\chi$}
A number of studies have focused on relatively high triplet vev $v_{\chi} > 1 $ GeV \cite{Li:2017daq,Das:2018vkv,Bairi:2022adc,Song:2022jns,Chakraborti:2023mya,Englert:2013zpa, Ahriche:2022aoj,Degrande:2017naf,Chang:2017niy,Ismail:2020kqz,Chen:2022ocr}. The main objective of this paper 
is to analyse the Higgs and DM phenomenology for a low value of the triplet vev, {$v_{\chi} \lsim 1$ GeV}
For a small  $v_{\chi}$, the decoupling limit characterised by $v_{\chi} \to 0$ ($s_{H}\to 0$)  along with $s_{\alpha} \to 0$ can be realised. 
For a small $v_{\chi}$, the expression of the fiveplet mass simplifies to 
\begin{equation}
M_{H_5}^2 \simeq  \frac{3}{2}\lambda_5v^2 + \widetilde{M_1^2} + {M}^2 \,.
\label{eq:massdec1}
\end{equation}
With the additional condition $\sin \alpha=0$, required to get rid of the divergence in the $\lambda$ couplings (Eq.~\eqref{eq:couplambdas}) for $s_H\to0$, we obtain for the singlet masses
\begin{equation}
M_H^2 \simeq  \widetilde{M_1^2} -\frac{1}{2}{M}^2, \quad 
M_h^2 \simeq 8\lambda_1 v^2. 
\label{eq:massdec2}
\end{equation}

\noindent 
Substituting $\lambda_5$ from Eq.~\eqref{eq:couplambdas} and  
$M_H$ from Eq.~\eqref{eq:massdec2} into Eq.~\eqref{eq:massdec1}, we obtain 
\begin{equation}
    M_{H_5}^2 \simeq 3 M_{H_3}^2 - 2 M_H^2 \,.
\label{eq:mH53relation}
\end{equation} 
In order to allow for a departure from  $v_{\chi}=0$ ($s_H=0$), we introduce $\overline{M}$, a free parameter of mass dimension one. Then  we can write $M_H^2$ and $\widetilde M_1^2$ as 
\begin{equation}
   M_{H}^2 = \frac{1}{2}\left(3M_{H_3}^2-M_{H_5}^2+3s_{H}^2 \overline{M}^2\right) , \quad 
   \widetilde{M}_1^2=\frac{1}
   {2}\left(3M_{H_3}^2-M_{H_5}^2+M^2\right) ,
\label{eq:mass_set}
\end{equation}

All the dimensionless couplings can then be rewritten as~\cite{Chiang:2012cn}
\begin{align}
   \lambda_1 &=\frac{M_h^2}{8v^2c_{H}^2}\,,\quad 
   \lambda_2 = \frac{M_{H_3}^2+M_{H_5}^2-M^2}{4v^2}\,,\quad 
   \lambda_3 = \frac{M_{H_5}^2-M^2}{v^2} \,, \notag\\
    \lambda_4 &= \frac{M_{H_3}^2-
    M_{H_5}^2+M^2+\overline{M}^2}{2v^2}\,, \quad 
    \lambda_5 = \frac{-(M_{H_3}^2-M_{H_5}^2+M^2)}{v^2}\,.
 \label{eq:decoupling_couplings}
\end{align}

It is evident from Eqs.~\eqref{eq:decoupling_couplings} that with this parametrization all the $v_{\chi}$ dependence drops out and there are no apparent divergences in the $\lambda$'s for $v_{\chi} \rightarrow 0$. It is also important to highlight that the off-diagonal element of the CP-even neutral mixing matrix $\mathcal{M}^{2}_{12}$
is identically zero in this parameterization (see Eqs.~\eqref{eq:Mass_Matrix_Neutral}) and consequently $\sin \alpha \equiv 0$ (instead of $\sin\alpha \ll 1$). 

In the following, we fix $m_h=125$~GeV and $\sin\alpha=0$ and take   
\begin{equation}
    v_\chi,~M_{H_3},~M_{H_5},~M,~\overline{M}\, .
\end{equation}
as free parameters. The value of $M_H$ is then fixed by Eqs.~\eqref{eq:mass_set} and \eqref{eq:vevs_relations}. The singlet scalar sector is characterised by free parameters $M_S$, $\lambda_a$, $\lambda_b$, $\lambda_S$. In this work, we consider the DM to be relatively heavy, so that  invisible decays of $h$ and $H$ are kinematically forbidden. Moreover $H_3$ and $H_5$ states do not couple to DM pairs via cubic interaction as it violates the CS. Hence $M_S$, $\lambda_a$, $\lambda_b$, $\lambda_S$  do not have any impact on collider searches and are only relevant for discussions of DM, only the parameters  $v_\chi,~M_{H_3},~M_{H_5},~M,~\overline{M}$ have direct influence on the collider searches.

Before concluding this section, we note that, for small values of $v_\chi$, which is the focus of this study, the couplings of $h$ to fermions are equal to those of the SM while its couplings to gauge bosons ($V=W^\pm,Z$) receive a small correction, $C_{hVV}=c_H C_{hVV}^{SM}$~\cite{Hartling:2014zca,Hartling:2014aga}. The heavy CP-even neutral scalar $H$  does not couple to fermions as $\sin \alpha=0$ and  $H_5^0$ couples  only to $\bar{\nu}\nu$. Their couplings to gauge bosons are suppressed by a factor $s_H$ (see section~\ref{sec:bound} or appendix). The couplings of the CP-odd neutral scalar $H_3^0$ to fermions are also suppressed by a factor $s_H$. Explicit expressions for the trilinear couplings of the neutral scalars can be found in section~\ref{sec:higgs}. We refer the reader to \cite{Chiang:2012cn,Hartling:2014zca,Hartling:2014aga} for a detailed description of all the couplings of the SM and BSM scalars of this model with SM fermions, gauge bosons, as well as with other BSM scalars.

\section{Existing theoretical and experimental constraints} \label{sec:bound}

In this section, we assess the parameter space of the GM-S model that is consistent with existing theoretical constraints and experimental constraints from colliders. To this end, 
we perform a flat random scan over the free parameters $\{M_{H_3}, M_{H_5}, M, \overline{M}, v_{\chi}, \lambda_a, \lambda_b, \lambda_S\}$ within the ranges specified in Table~\ref{tab:scanranges1}. About $4\times10^6$ data points are sampled in this scan and confronted against the constraints detailed below. 

Before proceeding, a comment is in order regarding the subset of parameters considered. 
As discussed in the previous section, $\lambda_{1,\ldots,5}$ are expressed in terms of the physical scalar masses $M_h, M_{H_3}, M_{H_5}$, $M$, $\overline{M}$ and $v_{\chi}$; together with the quartic couplings $\lambda_{a,b,S}$ they are subject to the condition of perturbative unitarity and potential being bounded from below. Hence, the entire subset $\{M_{H_3}, M_{H_5}, M, \overline{M}, v_{\chi}, \lambda_a, \lambda_b, \lambda_S\}$ is needed to check these and other theoretical constraints. In contrast, the DM mass $M_S$ is relevant only for DM production and detection and therefore ignored in this section. The DM-related constraints, which will affect $M_S$, $\lambda_a$, $\lambda_b$, $\lambda_S$, will be discussed in section~\ref{sec:DMpheno}.

\begin{table}[t]
\centering
\begin{tabular}{| c | c  | }
\hline
{Parameter}& \multicolumn{0}{ c |   }{\quad Scanned range }   \\
\hline
$M_{H_{3}}$ [GeV] & \quad [$10^{2}$ , $10^{3}$]   \\                                                                                
$M_{H_{5}}$ [GeV]  & \quad  [$10^{2}$ , $10^{3}$]                 \\
$M$ [GeV]        &  \quad [$10^{1}$ , $10^{3}$]                 \\
$\overline{M}$ [GeV]        &   \quad [$10^{1}$ , $10^{6}$]          \\
$v_{\chi}$ [GeV]       &   \quad [$10^{-4}$ , $10^{0}$]                \\
\hline
$\lambda_{a}$, $\lambda_{b}$, $\lambda_{S}$        & \quad  [$10^{-4}$ , $10^{0}$]                   \\
$M_{S}$ [GeV]       & \quad   [$10^{2}$ , $10^{3}$]  \\
\hline
\end{tabular}
\caption{Ranges of the input parameters that we use  in our numerical scan together with  $\sin\alpha=0$. \label{tab:scanranges1}}
\end{table}

\subsection{Theoretical constraints and constraints from oblique
parameters}\label{sec:thconstraints}
\subsubsection*{Perturbative unitarity}
By the perturbative unitarity of the $2\to2$ scalar field scattering amplitudes, the scalar couplings in Eq.~\eqref{eq:decoupling_couplings} can be constrained. We take into account the following constraints  on the quartic couplings of this model \cite{Campbell:2016zbp}
\begin{align}
8\pi>&\left| 12\lambda_1+14\lambda_3+22\lambda_4  \pm \sqrt{(12\lambda_1-14\lambda_3-22\lambda_4)^2+144\lambda_2^2} \right| , \nonumber \\
8\pi>&\left|4\lambda_1-2\lambda_3+4\lambda_4 \pm\sqrt{(4\lambda_1+2\lambda_3-4\lambda_4)^2+4\lambda_5^2}\right| , \nonumber \\
8\pi>&|16\lambda_3+8\lambda_4| , \nonumber \\
8\pi>&|4\lambda_3+8\lambda_4| , \nonumber \\
8\pi>&|4\lambda_2-\lambda_5| , \nonumber \\
8\pi>&|4\lambda_2+2\lambda_5| , \nonumber \\
8\pi>&|4\lambda_2+4\lambda_5|, \nonumber \\
8\pi>&|4\lambda_a| , \nonumber \\
8\pi>&|4\lambda_b| , \nonumber \\
\lambda_S<& \frac{1}{6}\left(4\pi + \frac{2\lambda_a^2(7\lambda_3+11\lambda_4-\pi)+9\lambda_b^2(3\lambda_1-\pi)-18\lambda_2\lambda_a\lambda_b}{2(7\lambda_3+11\lambda_4-\pi)(3\lambda_1-\pi)-9\lambda_2^2} \right), \nonumber \\
\lambda_S>& \frac{1}{6}\left(-4\pi+\frac{2\lambda_a^2(7\lambda_3+11\lambda_4+\pi)+9\lambda_b^2(3\lambda_1+\pi)-18\lambda_2\lambda_a\lambda_b}{2(7\lambda_3+11\lambda_4+\pi)(3\lambda_1+\pi)-9\lambda_2^2} \right).
\label{eq:uniconstr}
\end{align}
The condition $|4\lambda_2-\lambda_5|<8\pi $ restricts the triplet mass to $M_{H_3}\lesssim 870$ GeV. The conditions $|4\lambda_2+2\lambda_5|< 8\pi$ and 
$|4\lambda_2+4\lambda_5|<8\pi$ do not allow arbitrary mass squared splitting between $M_{H_5}^2$ and $M^2$ as $|M_{H_5}^2-M^2|/v^2 \leq 4\pi$. $|4\lambda_3+8\lambda_4|<8\pi$ leads to $|M_{H_3}^2+\overline{M}^2|/v^2 \leq 2\pi$.
As we will present later
the conditions in Eq.~\eqref{eq:uniconstr} jointly put stronger constraints on  $M_{H_3}$.

Note that, the range of $M_{H_3}$  given in Table.~\ref{tab:scanranges1}  together with the constraint $|M_{H_3}^2+\overline{M}^2|/v^2 \leq 2\pi$ dictates that  large values of $\overline{M}> \mathcal{O}(100)$ GeV will get discarded. Since $s_H$ is also small, $s_H < \mathcal{O}(10^{-2})$, the term $3s_H^2\overline{M}^2$ in Eq.~\eqref{eq:mass_set} is negligible. The condition $M^2_H > 0$ therefore implies  $M_{H_5}\lesssim\sqrt{3} M_{H_3}$, which will be visible in the allowed points that satisfy the theory constraints. We note that out of the $4\times10^6$ data points that were sampled in the scan, about $3 \times 10^6$ points satisfy this condition.
\subsubsection*{Potential bounded from below}
In order to ensure the potential is bounded from below, we implemented the conditions computed in Ref.~\cite{Hartling:2014zca,Campbell:2016zbp} which are as follows,
\begin{align}
\lambda_1&>0, \nonumber \\
\lambda_4&>\left\lbrace \begin{matrix}
&-\frac{1}{3}\lambda_3 &\mbox{ for } \lambda_3 \geq 0,\\
&-\lambda_3 &\mbox{ for } \lambda_3 < 0,\\
\end{matrix} \right. \nonumber \\
\lambda_2&>\left\lbrace \begin{matrix}
&\frac{1}{2}\lambda_5-2\sqrt{\lambda_1\left(\frac{1}{3}\lambda_3+\lambda_4\right)} &\mbox{for } \lambda_5 \geq 0, \; \lambda_3\geq 0,\\
&\omega_+(\zeta)\lambda_5-2\sqrt{\lambda_1\left(\zeta\lambda_3+\lambda_4\right)} &  \mbox{for } \lambda_5 \geq 0, \; \lambda_3< 0,\\
&\omega_-(\zeta)\lambda_5-2\sqrt{\lambda_1\left(\zeta\lambda_3+\lambda_4\right)} &\mbox{ for } \lambda_5 <0, \lambda_3 \in R
\end{matrix}\right. \nonumber 
\end{align}
\begin{align}
\lambda_a&>-2\sqrt{\lambda_1 \lambda_S}, \nonumber \\
\lambda_b&>\left\lbrace \begin{matrix}
&-2\sqrt{\left(\frac{1}{3} \lambda_3+\lambda_4\right)\lambda_S} & \mbox{ for } \lambda_3 \geq 0, \\
&-2\sqrt{\left(\lambda_3+\lambda_4\right)\lambda_S} & \mbox{ for } \lambda_3 < 0, 
\end{matrix} \right. \nonumber \\
\lambda_S&>0.
\label{eq:bfbconstr}
\end{align}
where $\zeta$ and $\omega$ satisfies the following range,
\begin{equation}
\zeta\in \left[ \frac{1}{3},1 \right], \qquad \omega\in \left[ -\frac{1}{4},\frac{1}{2} \right].
\end{equation}
For a given value of $\zeta$, we can write $\omega\in [\omega_-,\omega_+]$, where~\cite{Hartling:2014zca}
\begin{equation}
\omega_\pm(\zeta)=\frac{1}{6}(1-B)\pm\frac{\sqrt{2}}{3}\left[(1-B)\left(\frac{1}{2}+B\right)\right]^{\frac{1}{2}},
\end{equation}
with
\begin{equation}
B\equiv\sqrt{\frac{3}{2}\left(\zeta-\frac{1}{3}\right)}\in [0,1].
\end{equation}

\subsubsection*{Absence of deeper custodial symmetry-breaking minima}
To ensure no deeper custodial symmetry-breaking minima in our chosen parameter space, we passed the allowed parameters from perturbative unitarity and bounded from the below condition through Vevacious~\cite{Camargo-Molina:2013qva} to check the global minima condition.

\subsubsection*{Mass of CP-even BSM scalar:}

As we have discussed before, the positivity condition for the mass of the CP-even BSM scalar, $M_H$,  discards the region where $\sqrt{3}M_{H_3} < M_{H_5}$.  Among the SM and BSM Higgs states $h, H$, we demand that $H$ is heavier than the SM Higgs so that it does not have any influence in the measurement of  $h \to \gamma \gamma$ rate. We choose $M_H$ value sufficiently  larger than the SM Higgs mass,  $M_H\ge130$ GeV. Since $s_H$ is almost negligible and the allowed value of $\overline{M}$ can at most be  $\mathcal{O}(10^2)$ GeV, this in turn allows for $3M_{H_3}^2-M_{H_5}^2> \mathcal{O}(10^4)~\text{GeV}^2$, see   Eqs.~\eqref{eq:mH53relation} and \eqref{eq:mass_set}. 

\begin{figure}[t]
\centering
\mbox{
\subfigure[]{\includegraphics[width=0.5\textwidth]{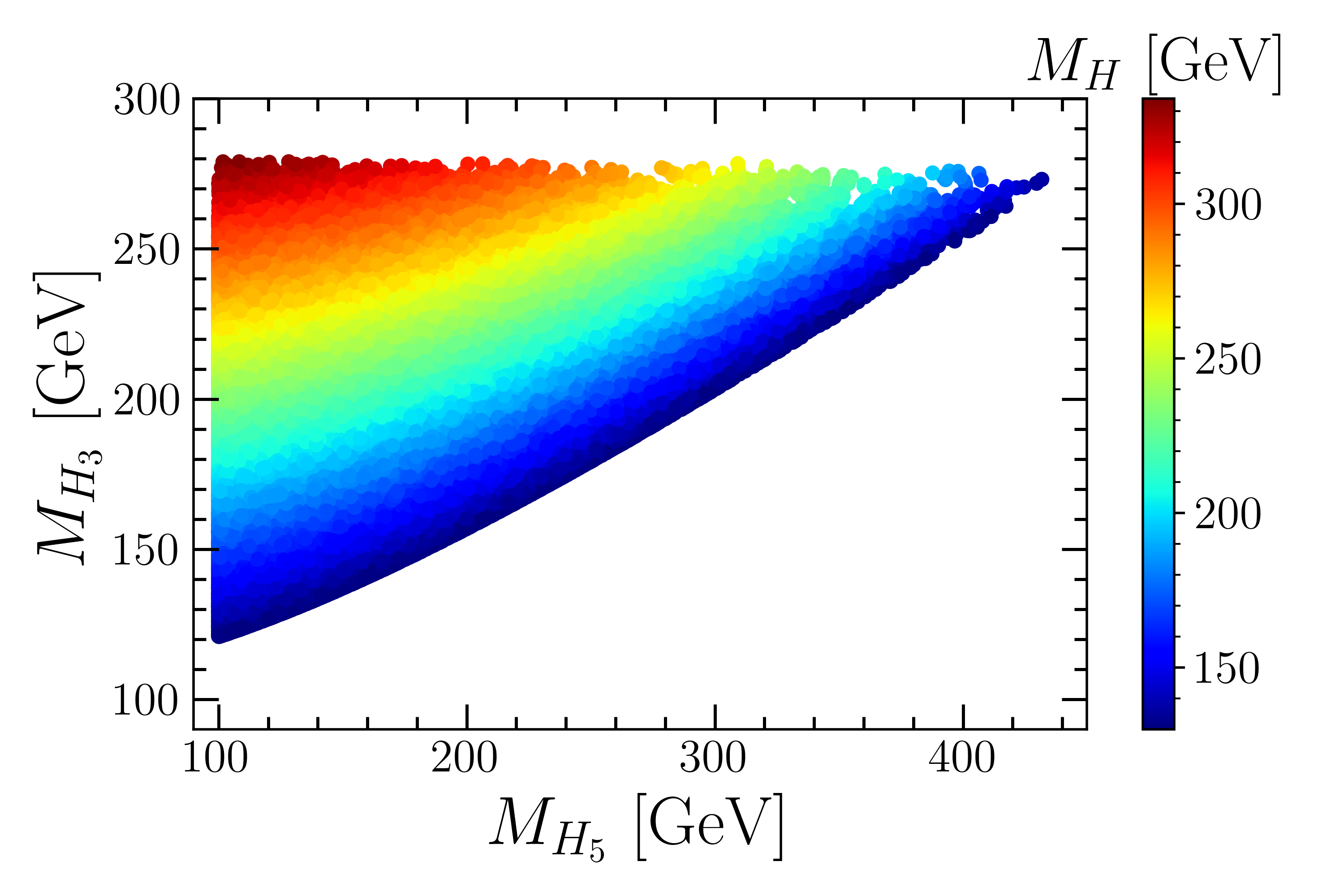}\label{Fig:TH_Constraint1}}
\subfigure[]{\includegraphics[width=0.5\textwidth]{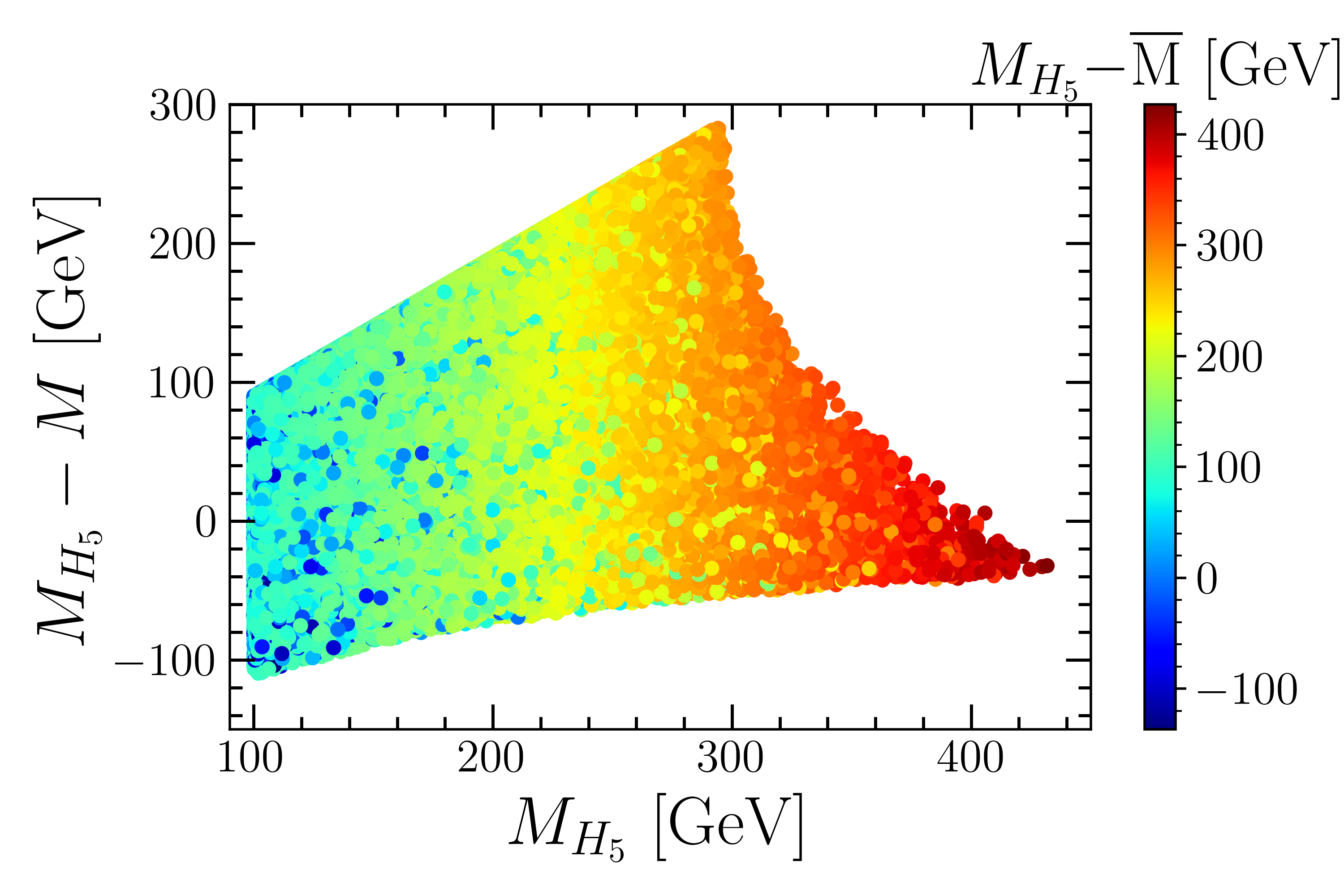}\label{Fig:TH_Constraint2}}
}
\caption{ Points that satisfy  theoretical constraints, such as perturbative unitarity, potential bounded from below and absence of deeper custodial symmetry breaking minima, mentioned in section~\ref{sec:thconstraints}: 
(a)~in the plane $M_{H_5}$ vs.\ $M_{H_3}$ with the value of $M_H$ shown in color and (b)~in the  $M_{H_5}$ vs.\ $M_{H_5}-M$ plane with $M_{H_5}-\overline{M}$ shown in color.} 
\label{Fig:th_constraint}
\end{figure}

The viable parameter space  satisfying all the theoretical constraints in different planes is shown in Fig.~\ref{Fig:th_constraint}. In Fig.~\ref{Fig:TH_Constraint1}, we show the allowed points in the $M_{H_5}$ vs. $M_{H_3}$ plane. The color code following Eq.~\eqref{eq:mass_set} indicates $M_H$. As can be seen,  for our chosen scenario with a small $v_{\chi}$ and $\sin \alpha=0$, TeV scale  fiveplet and triplet scalar states  are ruled out. 
All the perturbative unitarity conditions from Eq.~\ref{eq:uniconstr} jointly exclude  $M_{H_3}> 280$ GeV and $M_{H_5}> 435$ GeV while the mass of the singlet scalar, $H$, is restricted to $ M_H < 330$ GeV. The conditions on the potential to be bounded from below do not set additional limit on the parameters which pass the perturbative unitarity conditions.  The condition on $M_H\ge130$ GeV largely excludes points in the lower right corner of the  $M_{H_5}$ vs. $M_{H_3}$ plane as mentioned above.
In Fig.~\ref{Fig:TH_Constraint2} we show the allowed values for the differences $M_{H_5}-M$ and  $M_{H_5}-\overline{M}$.  The values of $M$ and $\overline{M}$ are constrained to $M<435$ GeV and $\overline{M}<250$ GeV. Moreover $|M_{H_5}^2-M^2|/v^2 \lesssim 4\pi $  follows for the perturbativity constraint on the $\lambda_3$ coupling. After including the theory constraint with a small $v_{\chi}<1$ GeV and $\sin \alpha=0$, it is evident that the independent parameters $M_{H_5}$, $M_{H_3}$, $M$ and $\overline{M}$ are strongly constrained.  The triplet vev $v_\chi$ and the mass of DM $M_S$, and  couplings $\lambda_{a,b,s}$ are  allowed over the full range considered.

\subsubsection*{Oblique parameters}
\begin{figure}[!t]
\centering
\mbox{
\subfigure[]{\includegraphics[width=6.5cm,height=5.5cm]{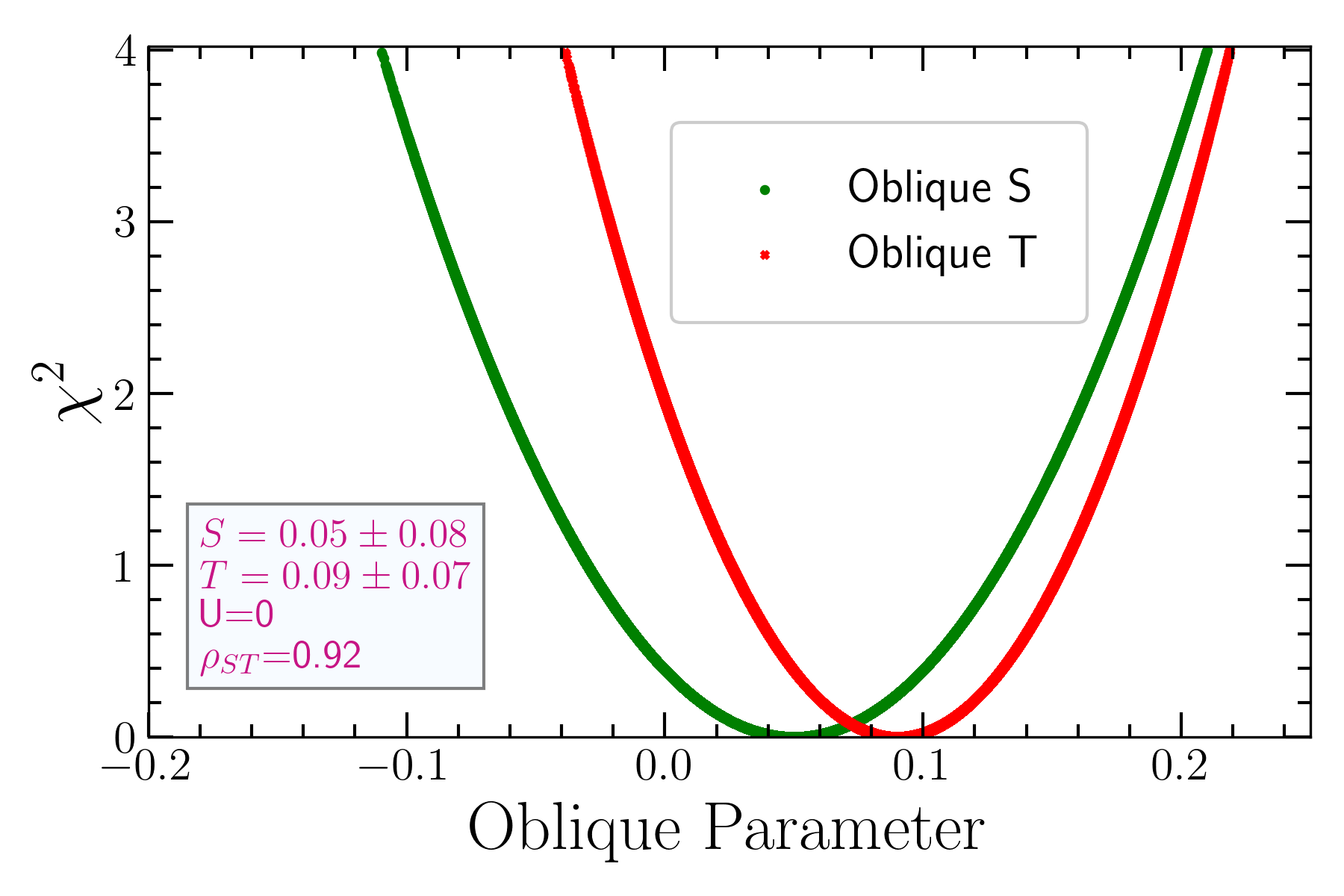}\label{Fig:Oblique1}}
\subfigure[]{\includegraphics[width=6.5cm,height=5.5cm]{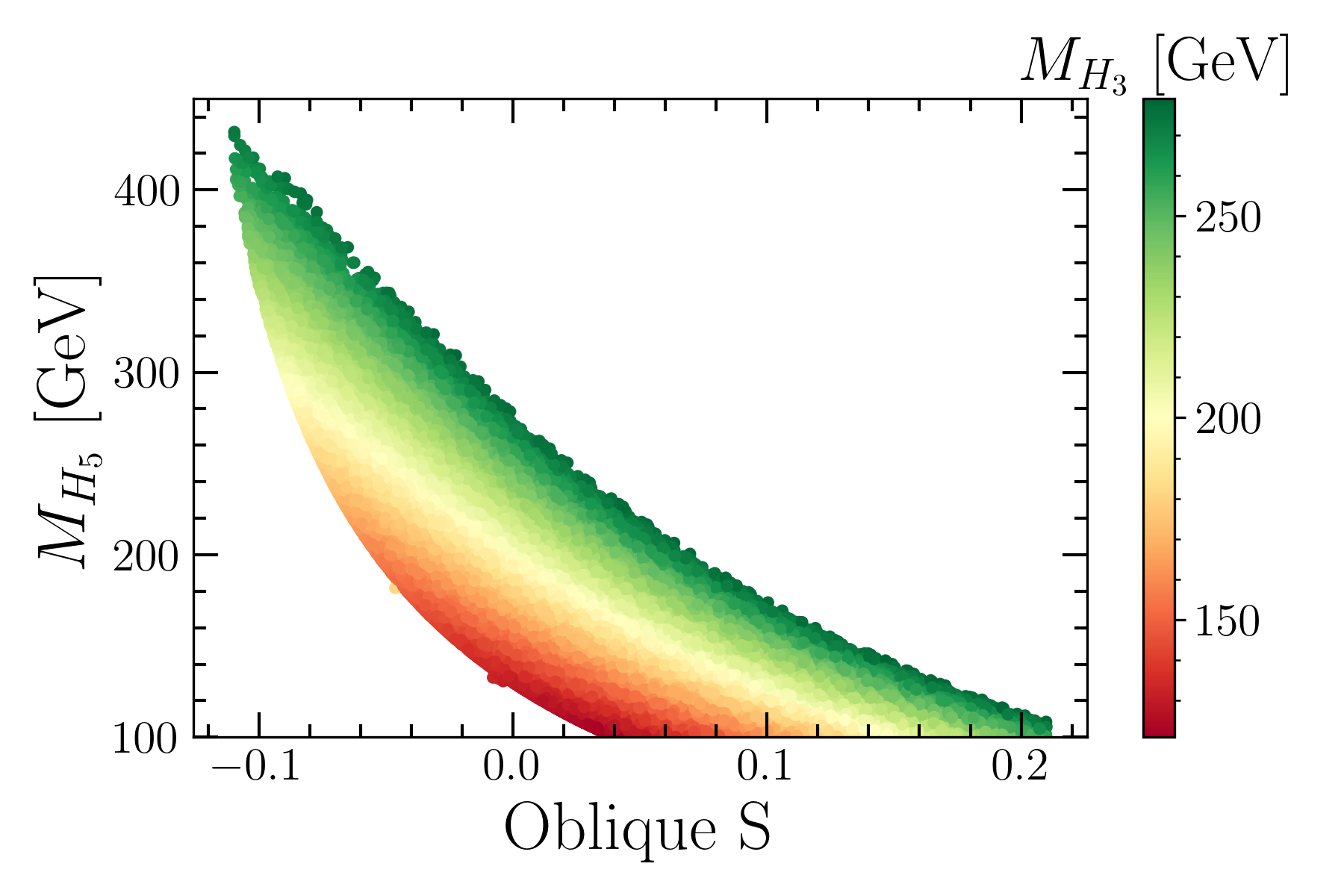}\label{Fig:Oblique2}}
}
\caption{(a)~Variation of  $\chi^2$ w.r.t. the oblique parameters, S (red line) and T (green line). (b)~Points in the $M_{H_5}$ vs.\ $S$ plane that satisfy the constraint from oblique parameters  as well as previously mentioned theoretical constraints. The color bar shows the value of  $M_{H_3}$. In deriving the constraints, we assume $U=0$. }
\label{Fig:Oblique}
\end{figure}

The $S$ parameter contributions in the GM model, relative to the SM are computed in \cite{Hartling:2014aga}.
They involve corrections to the SM diagrams contributions to the neutral bosons self-energy such as the $Zh$ loop contribution that is proportional to  $v_\chi/v$ together with 
new contributions to the $Z$ boson self-energy that involve  loops of two scalars $HH_3,H_5H_3$ and $H_5^\pm H_3^\mp$ with couplings of electroweak strength. Finally, new contributions involving two scalars, $hH_3$, or  loops of one gauge boson and a scalar $ZH,ZH_5, W^\pm H_5^\mp$ are all suppressed by a factor $v_\chi/v$.
From the global electroweak fit~\cite{Lu:2022bgw} we have the measurements for $S,T$ parameters
\begin{equation}
S_{\rm exp} = 0.05 \pm 0.08, \qquad \qquad
T_{\rm exp} = 0.09 \pm 0.07,
\end{equation}
for the SM Higgs mass $M_h = 125$~GeV, $U = 0$, and correlation $\rho_{ST} = +0.92$.  
We compute the $\chi^2$ according to
\begin{equation}
\chi^2 =  \frac{1}{\left(1-\rho_{ST}^2\right)}\left[\frac{\left(S-S_{\rm exp}\right)^2}{\left( \Delta S_{\rm exp} \right)^2}+\frac{\left(T-T_{\rm exp}\right)^2}{\left( \Delta T_{\rm exp} \right)^2}-\frac{2 \rho_{ST} \left(S-S_{\rm exp}\right)\left(T-T_{\rm exp}\right)}{\Delta S_{\rm exp} \Delta T_{\rm exp}}\right],
\end{equation}
where $\Delta S_{\rm exp}$ and $\Delta T_{\rm exp}$ are the $1\sigma$ experimental uncertainties.
As hypercharge interactions break the custodial symmetry  at one-loop level, the $T$ parameter in the GM model has divergent value~\cite{Gunion:1990dt}. Therefore following Ref.~\cite{Hartling:2014aga} we marginalise over the  $T$ parameter and set
\begin{equation}
T = T_{\rm exp} + \rho_{ST} (S-S_{\rm exp})\frac{\Delta T_{\rm exp}}{\Delta S_{\rm exp}}.
\end{equation}
In Fig.~\ref{Fig:Oblique1}, we show the $\chi^{2}$ distribution taking into account current experimental limits. In Fig.~\ref{Fig:Oblique2}, we show the allowed values of $M_{H_{3}}$ and $M_{H_{5}}$ satisfying $\chi^2<4$ and the corresponding values of $S$. We expect the oblique parameters to constrain the mass difference between two scalars and in particular $M_{H_{3}}-M_{H_{5}}$. Indeed we find that almost all points that pass theoretical bound are also consistent with the measurements of the oblique parameters. Only points where the mass difference is large are excluded, they are found in the region $M_{H_{5}}\sim100$ GeV and $M_{H_{3}}\ge260$ GeV. A summary of the different constraints that we consider in this work will be presented in  Fig.~\ref{Fig:Th_Exp_Allowed}.

\subsection{Constraints on  $\Gamma(H_i)/M_{i}$ for scalars \label{sec:widthcon}}

Before checking  the current experimental limits on BSM scalars, we impose a constraint on the total $\Gamma(H_i)/M_i$ ratio for the neutral and charged scalars, where $\Gamma(H_i)$ represent the total decay width of the BSM scalar $H_i$ and $M_i$ represents its mass. Mainly tri-linear scalar couplings lead to large $\Gamma(H_i)/M_{i}$ for the neutral scalars $H^0_5$ and $H$ in two ways: the first one is when scalars decay to two other scalars $H_i \to H_j H_j$ and the  second one is when the  tri-linear scalar couplings becomes  large leading to an  enhanced loop-induced partial with for  $H_i \to V \gamma$ with $V=\gamma, Z$ states. This can happen when for a fixed value of $M$, $v_\chi$ becomes small.

The partial decay width of a CP-even neutral scalar $H_i$ decaying into $2\gamma/Z\gamma$  is~\cite{Degrande:2017naf}
\begin{equation}
\Gamma(H_i \to V \gamma) = \frac{m_{H_i}^3}{32 \pi \eta_V} 
\left[ 1 - \frac{M_V^2}{m_{H_i}^2} \right]^3 |S|^2 ,
\label{eq:decayHi}
\end{equation}
where $S$ is the form factor and $ V = \gamma$, $Z$.  Here $\eta_V$ is a symmetry factor that accounts for identical particles in the final state, with $\eta_{\gamma}=2$ and $\eta_Z = \eta_W = 1$.
For the CP-even scalars $h$, $H$, and $H_5^0$, the form factor is\footnote{For the CP-odd scalar $H_3^0$, there is no dependency on tri-linear scalar couplings as only fermions mediate the loop decay. Therefore the BR for $H_3^0 \to V \gamma$ is negligible as compared to the one for  $H_3^0 \to ZH/W^\pm H_5^\pm$. }
\begin{equation}
S_{H_i\gamma\gamma} =  \frac{\alpha_{\rm em}}{2 \pi v}
\left[\beta_W^{H_i} F_1(\tau_W) + \sum_f N_{cf} Q_f^2 \beta_f^{H_i} F_{1/2}(\tau_f) 
+ \sum_s \beta_s^{H_i} Q_s^2 F_0(\tau_s) \right], \label{eq:hH2gam}
\end{equation}
where $N_{cf} = 3$ for quarks and 1 for leptons, $Q_j$ is the electric charge of particle $j$ in units of $e$, and the sums run over all fermions and scalars that can propagate in the loop for the parent scalar $H_i$. In this model, the charged scalars are $s=\{H_3^{\pm},H_5^{\pm},H_5^{++},H_5^{--}\}$ and we  only keep the top quark contribution to the fermion loop, $f=t$. The coupling coefficients $\beta_{j}^{H_i}$ are defined as,
\begin{equation}
\beta_W^{H_i} = \frac{C_{H_i W^+W^-} e^2 }{g M_W}, \quad\quad 
\beta_f^{H_i} = -\frac{C_{H_i f\bar{f}} v}{m_f}, \quad\quad 
\beta_s^{H_i} = \frac{C_{H_i ss} v}{2 m_s^2},
\label{eq:beta}
\end{equation}
for a propagating $W$ boson, fermion $f$, and scalar $s$, respectively.  In the case of the $W$ boson and fermion loops, these factors $\beta_{W,f}^{H_i}$ are equal to the usual ratios $\kappa_{W,f}^{H_i}$ of the scalar coupling to $WW$ or $f\bar f$ normalized to the corresponding SM Higgs coupling as described in Ref.~\cite{Hartling:2014zca,Degrande:2017naf}.  Note that $\beta_f^{H_5^0} = 0$ because the $H_5$ states are fermiophobic.

The loop factors are given in terms of the usual functions for particles of spin $0$, $1/2$ and $1$~\cite{Gunion:1989we},
\begin{eqnarray}
F_1(\tau_W) &=& 2+3\tau_W+3\tau_W(2-\tau_W) f(\tau_W), \nonumber \label{F1}\\
F_{1/2}(\tau_f) &=& -2\tau_f[1+(1-\tau_f) f(\tau_f)], \nonumber \label{eq:F12} \\
F_0(\tau_s) &=& \tau_s[1-\tau_s f(\tau_s)],\label{F0}
\end{eqnarray}
where $\tau_j = 4 m_j^2/m_{H_i}^2$ and
\begin{equation}
f(\tau) = \left\{ \begin{array}{l l}
\left[\sin^{-1} \left(\sqrt{\frac{1}{\tau}}\right) \right]^2 & \quad  {\rm if} \ \tau \geq 1, \\
-\frac{1}{4}\left[ \log \left(\frac{\eta_+}{\eta_-}\right) - i \pi \right]^2 & \quad  {\rm if} \ \tau < 1, \\
\end{array} \right.
\label{feq}
\end{equation}
with $\eta_{\pm} = 1 \pm \sqrt{1-\tau}$.
The relevant vertex factors $C_{ijk}$ that enter the calculation of $\Gamma(H,H_5^0 \to \gamma \gamma)$  are as follows: 
\begin{eqnarray}
C_{H_3^+H_3^{-}H} &=& \frac{1}{\sqrt{3} v^2} \left[ 8 (\lambda_3 + 3 \lambda_4 + \lambda_5) v_{\phi}^2 v_{\chi}
	+ 16 (6 \lambda_2 + \lambda_5) v_{\chi}^3 + 4 M_1 v_{\chi}^2 - 6 M_2 v_{\phi}^2 \right] , \nonumber \\
 & = & \frac{1}{\sqrt{3}v^2}\left[ \frac{8v_\phi^2 v_\chi}{v_2}(M_{H_5}^2 -M^2+ M_{H_3}^2 +3\overline{M}^2)+\frac{32v_\chi^3}{v^2}( M_{H_3}^2 +  M_{H_5}^2) -\frac{6v_\phi^2 M^2}{12v_\chi}  \right] , \nonumber \\
 C_{H_5^+H_5^{-}H} &=& C_{H_5^{++}H_5^{--}H} =\sqrt{3} \left[ 8 (\lambda_3 + \lambda_4) v_{\chi} + 2 M_2 \right] \nonumber \\
 &=& \frac{4\sqrt{3}v_\chi}{v^2}(M_{H_5}^2-M^2+M_{H_3}^2+\overline{M}^2)+ \frac{\sqrt{3} M^2}{6v_\chi} \,,
 \label{Eq:C_Hgaga}
\end{eqnarray}
\begin{eqnarray}
C_{H_3^+H_3^{-}H_5^0} &=& \sqrt{\frac{2}{3}} \frac{1}{v^2}\left[ 2(\lambda_3 - 2 \lambda_5) v_\phi^2 v_\chi - 8 \lambda_5 v_\chi^3+ 4 M_1 v_\chi^2 + 3 M_2 v_\phi^2 \right] \nonumber\\
&=&\sqrt{\frac{2}{3}} \frac{1}{v^2}\left[ 4M_{H_3}^2 v_\chi+ \frac{16 v_\chi^3}{v^2} (M^2-M_{H_5}^2)+\frac{M^2 (v^2-8v_\chi^2) }{4 v_\chi}\right], \nonumber\\
C_{H_5^+H_5^{-}H_5^0} &=& \sqrt{6} \left( 2 \lambda_3 v_\chi - M_2 \right)=\sqrt{6} \left(2v_\chi\frac{(M_{H_5}^2-M^2)}{v^2}-\frac{M^2}{12 v_\chi}\right),  \nonumber \\
C_{H_5^{++}H_5^{--}H_5^0} &=& -2 \sqrt{6} \left( 2 \lambda_3 v_\chi - M_2 \right)=-2 \sqrt{6} \left( 2v_\chi\frac{(M_{H_5}^2-M^2)}{v^2}-\frac{M^2}{12 v_\chi} \right).  \nonumber 
\end{eqnarray}

\begin{figure}[!t]
\centering
\mbox{
\subfigure[]
{\includegraphics[width=0.5\textwidth]{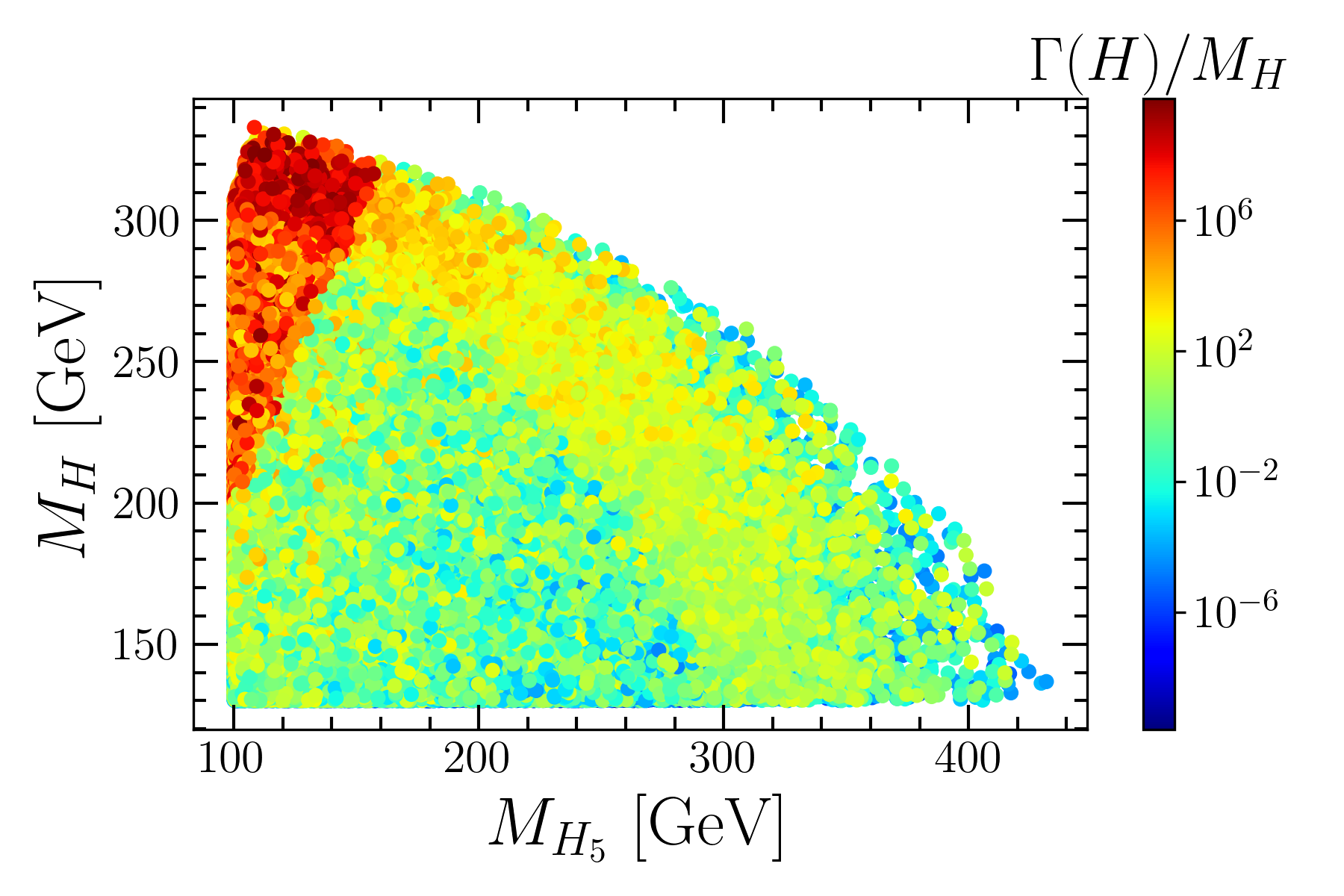}\label{Fig:width2mass1}}
\subfigure[]
{\includegraphics[width=0.5\textwidth]{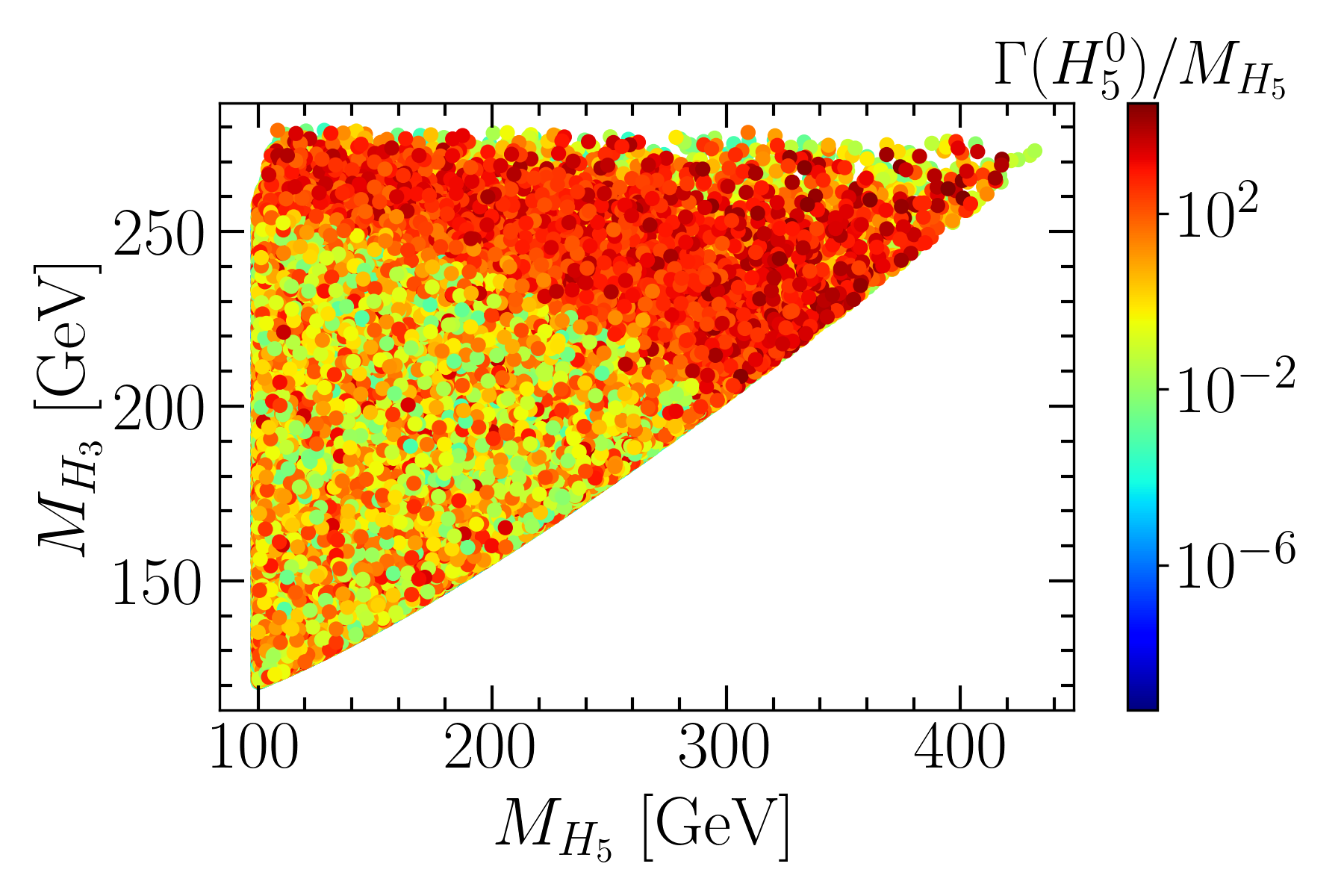}\label{Fig:width2mass2}}}
\mbox{
\subfigure[]
{\includegraphics[width=0.5\textwidth]{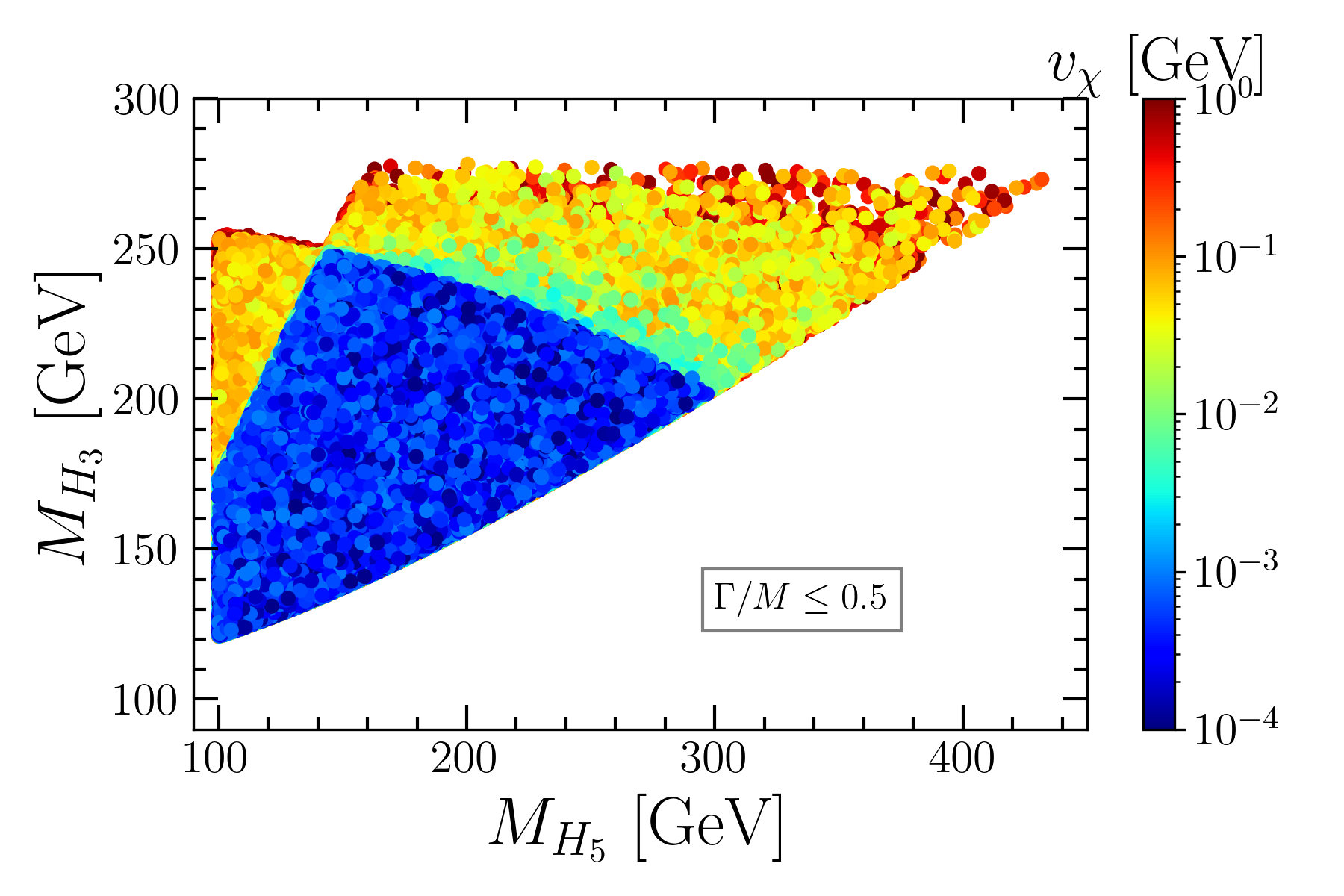}\label{Fig:width2mass4}}
}
\caption{Allowed points (a)~in the  $M_{H_5}$ vs.\ $M_{H}$ plane with $\Gamma_H/M_{H}$ in the color bar, (b)~in the  $M_{H_5}$ vs.\ $M_{H_3}$ plane with $\Gamma_{H_5}/M_{H_5}$ in the color bar, and  (c)~in the
$M_{H_5}$ vs.\ $M_{H_3}$ plane with  $v_{\chi}$ in the color bar. In the latter, we impose the constraint $\Gamma(H_i)/M_i<0.5$ for all the scalars. All points satisfy the constraints discussed in sections~\ref{sec:thconstraints} and \ref{sec:widthcon}.
}
\label{Fig:width2mass}
\end{figure}

In Fig.~\ref{Fig:width2mass1} and Fig.~\ref{Fig:width2mass2} we show the  variation of $\Gamma(H_i)/M_{i}$ for $H$ and $H_5^0$ respectively for the parameters consistent with all previously mentioned constraints. As evident from the plots, the widths of $H$ and $H_5^0$ can be huge  which is due to the involvement of the tri-linear scalar couplings which contain a term $~M^2/v_\chi$. To ensure all the BSM scalars have natural widths  we demand $\Gamma(H_i)/M_{i}<0.5$ and the resulting points are shown in Fig.~\ref{Fig:width2mass4}. Note that $\Gamma(H_i)/M_{i}$ for the charged scalars and $H^0_3$ naturally satisfy $\Gamma(H_i)/M_{i}\ll 1 $. It indicates that not all values of $v_\chi$ are allowed for all points in the plane of  $M_{H_5}$ vs. $M_{H_3}$. Note that the two-body decays $H\to H_5^{\pm \pm} H_5^{\mp \mp},H_5^{\pm}H_5^{\mp},H_5^{0}H_5^{0}$ are open for $M_H>2M_{H_5}$.
The region  $M_{H_3}>250$ GeV and $M_{H_5}= [100,150]$ GeV is not allowed for any value of $v_\chi<1$ GeV mainly because of such two body decays.
There exists a small orange triangular shape starting around $M_{H_5} \sim [100,140]$ GeV and $M_{H_3} \sim [170,250]$ GeV, where the two body decays are open but nevertheless  $\Gamma(H_i)/M_{i}<0.5$ is satisfied  for $v_\chi>0.05$ GeV. We note that, in this region $M<50$ GeV. Towards smaller $M_{H_5},M_{H_3}$ all values of $v_\chi$ satisfy $\Gamma(H_i)/M_{i}<0.5$ as a suitable values of $M$ is available in this region which can regulate the ratio $~M^2/v_\chi$. Whereas for $M_{H_5}>300$ GeV only $M>100$ GeV is favoured by the previous constraints \footnote{ The correlation between $M$ and $M_{H_5}~(M_{H_3})$ is shown in Fig.~\ref{Fig:Th_Exp_Allowed}.} (mainly perturbative unitarity bound). Therefore, a value of $v_\chi$ larger than  $0.01$ GeV is required to control the width. Further in this region $\Gamma(H_5^0 \to 2\gamma)/M_{H_5} \propto M_{H_5}^2$ as $M_{H_5}\simeq M$.
\subsection{Experimental constraints}
\label{sec:higgs}
Existing experimental constraints come from measurements of the properties of the SM-like Higgs boson,  in particular the $h\to\gamma\gamma$ rate and searches for additional scalar states in a variety of channels. Among the BSM Higgs searches, we will consider $H_5^{\pm \pm}\to W^{\pm}W^{\pm}$ as well as $H_5^0,H\to \gamma\gamma$. 
We will not include searches for the pseudoscalar $H_{3}^0\to \gamma\gamma$ since this process does not depend on the tri-linear scalar coupling and, therefore, always features a small branching ratio. The contributions coming from the $t \to H_{3}^{+}b$  and  B physics observables like $B_s^{0} \to \mu^{+} \mu^{-}$, $b \to s \gamma$ are also suppressed because of small $v_{\chi}$~\cite{Hartling:2014aga} and hence do not provide any constraints on our parameter space. For definiteness note that the experimental constraints discussed below will be applied step-by-step in a cumulative manner to determine the model's final, allowed parameter space.

\begin{figure}[h]
\centering
\mbox{
\subfigure[]{
\includegraphics[width=0.45\textwidth]{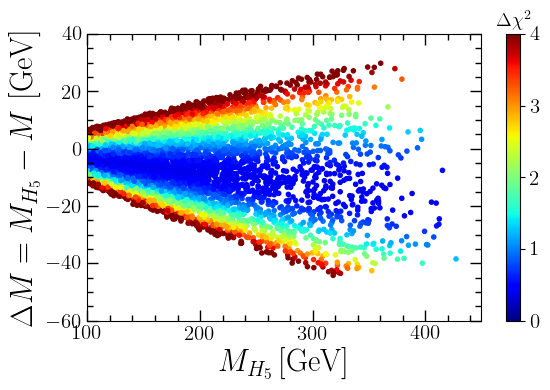}\label{Fig:hgaga2}}
\subfigure[]{
\includegraphics[width=0.48\textwidth]{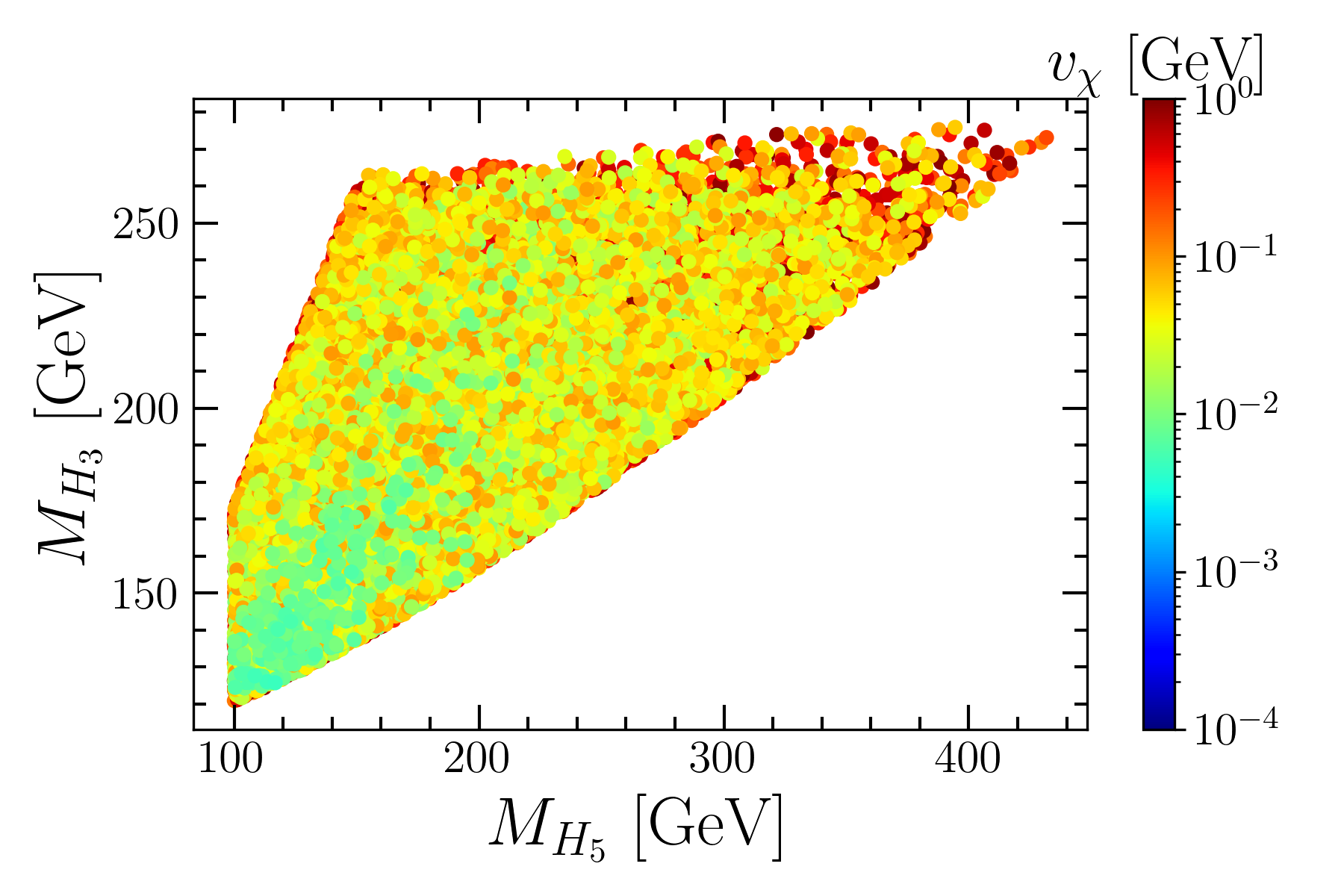}\label{Fig:hgaga4}}
}
\caption{ (a)~Allowed mass splitting between $M_{H_{5}}$ and $M$ after imposing the constraint from $h\to\gamma\gamma$ and all theoretical constraints, with $\Delta\chi^2$ shown in color.   
(b)~Allowed points in the $M_{H_{5}}$ vs.\ $M_{H_{3}}$ plane with $v_\chi$ in the color bar.}
\label{Fig:hgaga}
\end{figure}
  
\subsubsection{Higgs-to-diphoton measurements ($h \to \gamma \gamma$) }
Higgs-to-diphoton  rate is well measured by the ATLAS \cite{ATLAS:2022tnm} and CMS~\cite{CMS:2021kom} searches for $h \to \gamma \gamma$.  Charged scalars in this model give rise to additional contributions to $h \to \gamma \gamma$ decay and hence can alter the predicted rate for this channel compared to its SM-predicted value. 
The signal strength of the SM Higgs  for channel $h\to \gamma\gamma$ is given by,
\begin{equation}
\mu_{h\to \gamma\gamma}=\frac{\sigma_{h}}{\sigma_{h}^{SM}}\frac{{\rm BR}(h\to \gamma\gamma)}{{\rm BR}^{\rm SM}(h\to \gamma\gamma)} \approx \frac{S_{h\gamma\gamma}}{S_{h\gamma\gamma}^{\rm SM}},
\end{equation}
where $\sigma_{h}$ and $\sigma_{h}^{\rm SM}$ is the production cross section of $h$  through $\text{ggF}$ and $\text{VBF}$  channel in the GM and SM model respectively. For $v_{\chi}\leq 1$ GeV, the ratio $\sigma_{h}/\sigma_{h}^{\rm SM} \approx1 $. In the above $S_{h\gamma\gamma}$ is defined in Eq.~\eqref{eq:hH2gam} and 
the couplings $C_{ijk}$~\cite{Degrande:2017naf} relevant for $S_{h \gamma \gamma}$ are:
\begin{eqnarray}
C_{H_3^+H_3^{-}h} &=& \frac{c_{\alpha}}{v^2}  \left[ (4 \lambda_2 - \lambda_5) v_{\phi}^3 + 8 (8 \lambda_1 + \lambda_5) v_{\phi} v_{\chi}^2 + 4 M_1 v_{\phi} v_{\chi} \right] \nonumber \\
& \simeq & \frac{1}{v^2 (v^2-8 v_\chi^2)^{1/2}}\left[ 2 M_{H_3}^2 v^2 + 8 v_\chi^2(M_h^2 -2 M_{H_3}^2 )  \right], \nonumber \\
C_{H_5^+H_5^{-}h} = C_{H_5^{++}H_5^{--}h} &=& (4 \lambda_2 + \lambda_5) v_{\phi}=\frac{2 (v^2-8 v_\chi^2)^{1/2} }{v^2} (M_{H_5}^2-M^2),  \nonumber\\
C_{hW^+W^{-}} &=& c_W^2 C_{hZZ} = \frac{(v^2-8 v_\chi^2)^{1/2}}{2 s_W^2}. 
\end{eqnarray}
The decay of $h$ to $HH,H_{3}^{+}H_{3}^{-},H_{5}^{+}H_{5}^{-},H_{5}^{++}H_{5}^{--}$ channels is kinematically forbidden for the parameters favoured by the theoretical constraints, so the only decays of $h$ are into SM final states.

We perform a $\chi^{2}$ analysis to obtain the parameter space consistent with the recent $h\to\gamma\gamma$ signal strength measurement by ATLAS \cite{ATLAS:2022tnm} and CMS \cite{CMS:2021kom},
\begin{equation}
\mu_{\rm ATLAS} = 1.04^{+0.1}_{-0.09}, \qquad \qquad
\mu_{\rm CMS} = 1.12 \pm 0.09.
\end{equation}
The $\chi^{2}$ from multiple independent measurements can be computed as,
\begin{equation}
\chi^{2} =\sum_{i}\frac{ \mu_{i}^2 - \hat\mu^{2} }{\sigma^+_i\sigma^-_i + (\sigma^+_i -\sigma^-_i)(\mu_{i} -\hat\mu)}
\label{chisq}
\end{equation}    
where $\mu_i$ and $\sigma^{\pm}_i$ are the experimental mean values and uncertainties for CMS and ATLAS measurements, and $\hat\mu$ is the predicted value. We impose the condition $\Delta \chi^2=\chi^2-\chi^2_{min}  <4$. 

In Fig.~\ref{Fig:hgaga2}, we show the allowed points satisfying theoretical, oblique parameter and $h\to \gamma\gamma$ rate constraints in the $M_{H_{5}}$ vs.\ $\Delta M $ plane where the color bar corresponds to the different values of $\Delta\chi^2$. It is evident from the figure that mass splitting between $M_{H_{5}}$ and $M$ cannot be arbitrarily large. The vertex factors $C_{H_5^+H_5^{-}h}$ and $C_{H_5^{++}H_5^{--}h}$  depend crucially on 
$\Delta M$ and large values of $\Delta M$ will lead to an enhanced diphoton rate and  hence  in disagreement with the measured rate.

Figure~\ref{Fig:hgaga4} shows the variation of $v_\chi$ in the $M_{H_{5}}$ vs.\ $M_{H_{3}}$ plane. As can be seen from the figure, the measured diphoton rate discards the points $v_\chi < 4.8 \times 10^{-3}$ GeV.  
As discussed above $M^2/v_\chi$ is restricted due to $\Gamma/M_i<0.5$. Therefore, lower  $M$ is required for lower $v_\chi$. Although  $v_\chi\sim10^{-4}$ GeV satisfy $\Gamma/M_i<0.5$, the required $M$ for that is so small that the difference between $M_{H_{5}}$ and $M$ is large and hence the measured $\Gamma(h\to \gamma\gamma)$  rate is disturbed. The combined effect of $\Gamma/M<0.5$ and $\Gamma(h\to \gamma\gamma)$  rate set lower limit on $v_\chi$.
\subsubsection{$H^{\pm\pm}_5\to W^\pm W^\pm$ search}
\label{subsec:h5pp}
\begin{figure}[b]
	\mbox{
		\subfigure[]{\includegraphics[width=0.5\textwidth]{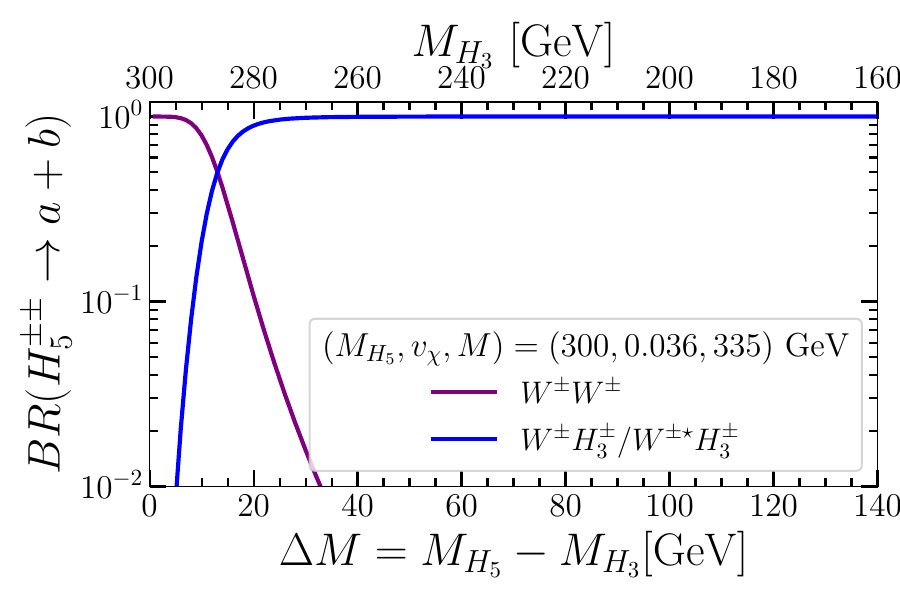} \label{fig:H5ppallowed1}}
	}
	\mbox{
		\subfigure[]{\includegraphics[width=0.5\textwidth]{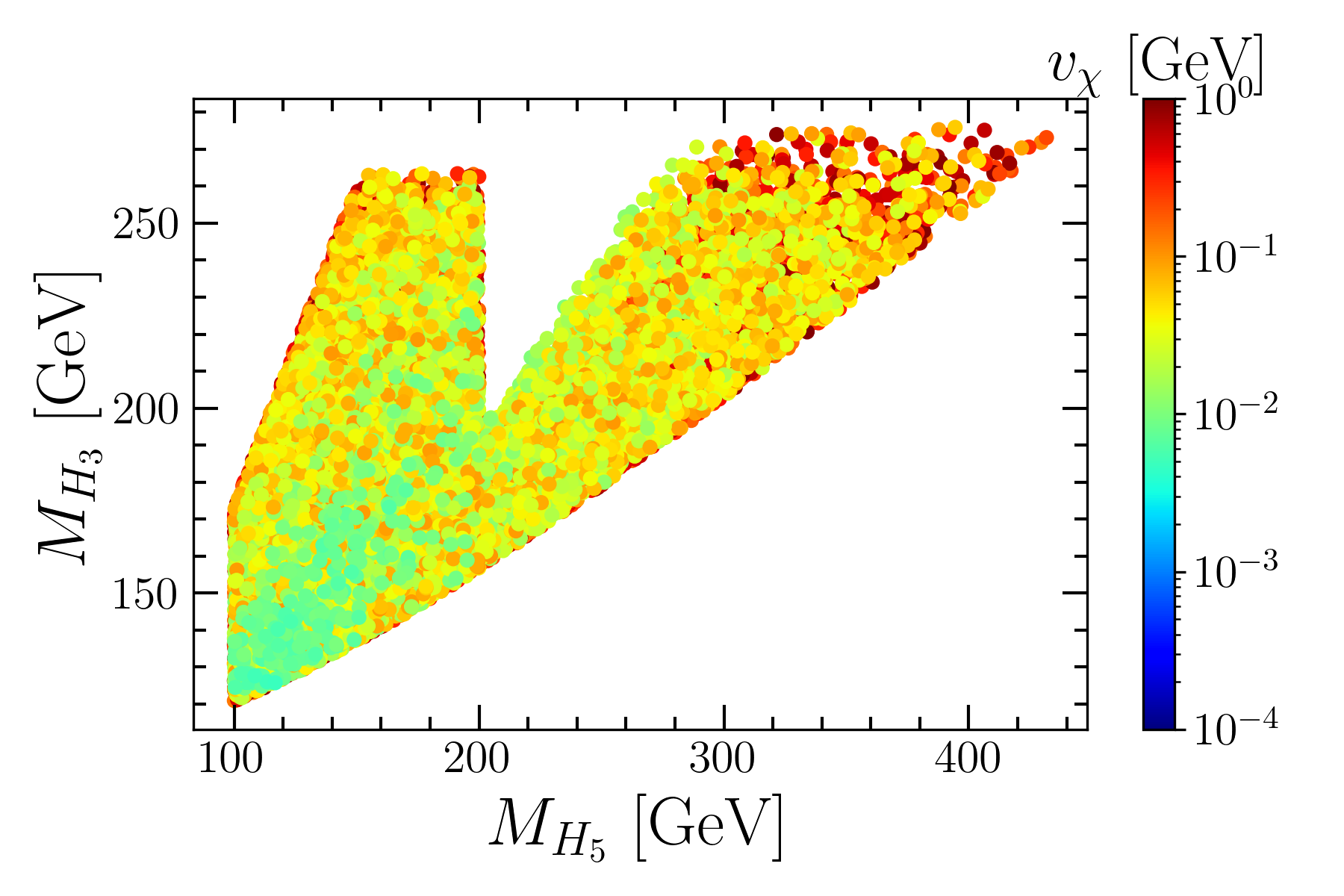} \label{fig:H5ppallowed2}}
	}
	\caption{(a)~Variation of branching ratios  of $H^{\pm \pm}_5$ decays to different channels w.r.t the mass difference between the fiveplet and triplet states for a specific benchmark point. 
 (b)~Allowed points in the $M_{H_5}$ vs.\ $M_{H_3}$ plane after applying the limit from the ATLAS search for \hpp decay to same-sign gauge bosons~\cite{ATLAS:2021jol} as well as all  previously discussed theoretical constraints, constraints from the measurement of oblique parameters, constraints on the decay width of the BSM scalars, and the measured rate $h \to \gamma \gamma$.
 } 
	\label{fig:H5ppallowed}
\end{figure}

The CMS~\cite{CMS:2017fhs} and ATLAS searches \cite{ATLAS:2018ceg,ATLAS:2021jol} investigated the presence of $H^{\pm \pm}$ decaying into same-sign gauge bosons leading to the multi-lepton final state. In the GM model, the scalar \hpp can lead to such signature and hence its production cross section  is constrained by the non-observation of such a signal at the LHC. The CMS~\cite{CMS:2017fhs} search targets the production of \hpp via vector boson fusion~(VBF) channel, however, the $H_5^{++}W^-W^-$ coupling depends on   $v_\chi$ such that the 
\hpp  production rate via VBF for  $v_\chi \le 1$ GeV is suppressed. Hence this CMS search does not set strong limit on our chosen parameter space. 

The ATLAS collaboration has  searched for pair production of \hpp via the Drell-Yan process $pp\to H^{\pm\pm}_5 H^{\mp\mp}_5$, which is sensitive to  the parameter space considered in our work. 
The current ATLAS search~\cite{ATLAS:2021jol} constrains the doubly charged Higgs mass in the range of 200 GeV to 350 GeV assuming 100$\%$ branching ratio for $H_5^{\pm \pm} \to W^\pm W^\pm$. 
To reinterpret this constraint for our scenario, where the branching ratio of $H^{\pm \pm} \to W^{\pm} W^{\pm}$ can vary, we calculate $\sigma(pp \to H_5^{\pm \pm} H_5^{\mp \mp}) \times {\rm BR}^2(H_5^{\pm \pm} \to W^\pm W^\pm)$ and compare with the observed limit. Similar multi-leptonic final state can also arise from  $pp \to H_5^{\mp \mp} H_5^{\pm \pm} (\to W^\pm H_3^\pm)$ followed by the decay $H_3^\pm \to \ell^\pm \nu$. 
However, for the range of $v_\chi$ that has survived previous constraints, i.e.\  $v_\chi>4.8\times10^{-3}$ GeV, BR$(H_3^\pm \to \ell^\pm \nu)$ is suppressed and hence this secondary contribution is negligible in our case.

As an understanding of the variation BR$(H_5^{\pm \pm} \to W^\pm W^\pm)$ is crucial to study the ATLAS limit, in  Fig.~\ref{fig:H5ppallowed1}, we present the branching ratio of the \hpp of mass $300$ GeV w.r.t.\ $\Delta M=M_{H_5}-M_{H_3}$ for $M=335$~GeV and $v_\chi=0.036$~GeV. We choose this benchmark for illustrative purpose, as it corresponds to a point that is allowed by all theoretical  and experimental constraints imposed  in the following sections. For this value of   $v_\chi$, the leptonic decay mode $H^{\pm\pm}_5\to l^{\pm}l^{\pm}$ is suppressed\footnote{From Eq.~\ref{eq:neumass2}, it can be seen that for $v_{\chi}=0.1$~GeV, we need $Y_{\nu}\approx10^{-10}$ to satisfy light neutrino mass. Consequently, the partial decay width for $H^{\pm\pm}_5\to l^{\pm}l^{\pm}$ process which is proportional to  $(Y_{\nu}/v_{\chi})^2$ is suppressed for small $Y_{\nu}$.} and hence only  decay channels containing at least one gauge boson are shown. The expressions for the partial decay widths for  $H^{\pm\pm}_5 \to l^{\pm} l^{\pm}$, $W^{\pm} W^{\pm}, H_3^\pm H_3^\pm, W^{\pm}H^{\pm}_3/{W^{\pm}}^*H^{\pm}_3$ are given in appendix .~\ref{appen1}.
The partial decay widths $\Gamma(H_5^{\pm \pm} \to W^\pm W^\pm)\sim v_\chi^2$, $\Gamma(H_5^{\pm \pm} \to \ell^\pm \ell^\pm)\sim (Y_\nu/v_\chi)^2$ and $\Gamma(H_5^{\pm \pm} \to W^\pm H_3^\pm)$ are independent of $v_\chi$. Thereby for sufficiently large $v_{\chi}$ the branching ratio $H^{\pm \pm} \to W^{\pm} W^{\pm}$ will be significantly large.
As can be seen from Fig.~\ref{fig:H5ppallowed1}, for $\Delta M=0$ GeV, the decay $H_5^{\pm \pm} \to W^\pm W^\pm$ is 100\%.  With the increase of $\Delta M$, $M_{H_3}$ is decreasing thus opening the decay modes $W^{\pm \star}H_3^\pm/W^{\pm}H_3^\pm$. This  channel  governs the decay of \hpp beyond $\Delta M \sim 30$ GeV for this benchmark point. For the shown mass range, the other decay mode  $H^{\pm\pm}_5\to$ $H_3^\pm H_3^\pm$ is not open kinematically. For $\Delta M <M_{W}$, the decays $H_5^{\pm \pm} \to W^\pm H_3^\pm$ is forbidden, however  the decay $H_5^{\pm \pm} \to {W^\pm}^* H_3^\pm$ is  still open. As we shift towards a smaller mass difference $\Delta M \sim {20}$~GeV, the branching ratio BR$(H_5^{\pm \pm} \to W^\pm W^\pm)$ increases, eventually for smaller mass differences  $H_5^{\pm \pm} \to W^\pm W^\pm$ becomes the leading decay mode, {thereby we expect a strong constraint for a small $\Delta M$.} {This is clearly evident from Fig.~\ref{fig:H5ppallowed2}, where we show all the points in the $M_{H_5}$ vs. $M_{H_3}$ plane that satisfy theoretical constraints as well as constraints from oblique parameters measurement, from the measured $h \to \gamma \gamma$ rate and  from ATLAS same-sign diboson search. The values of $v_\chi$ are shown in  the color-bar.  As expected, the parameter space around $M_{H_5} \sim M_{H_3}$ is ruled out, as in this region $H^{\pm \pm}_{5} \to W^{\pm} W^{\pm}$ has almost 100$\%$  branching ratio. Note that there is no constraint from the ATLAS search for $M_{H_5}< 200$ GeV, as this range has not been covered in \cite{ATLAS:2021jol}.} 

As stated above, BR$(H_5^{\pm \pm} \to W^\pm W^\pm)$ depends on  $M_{H_5} ,M_{H_3}$ and $v_\chi$.  
In the $M_{H_5}$ vs.\ $M_{H_3}$ plane, the ATLAS limit will be satisfied when BR$(H_5^{\pm \pm} \to W^\pm W^\pm)$ is suppressed. This happens in the region where $H_5^{\pm \pm} \to W^\pm H_3^\pm/W^{\pm\star} H_3^\pm$ is open provided $v_\chi$ is not too large. Note that for $v_\chi>4.8\times10^{-3}$~GeV, $H_5^{\pm \pm} \to \ell^\pm \ell^\pm$ is suppressed and hence  $H_5^{\pm \pm}$ dominantly decays to $ W^\pm W^\pm$ with $\sim100\%$ branching ratio. Therefore the region where  $\Delta M \to 0$ is excluded by the ATLAS search. 

\subsubsection {$H^0_5\to \gamma \gamma$ search}
\label{subsec:h50}
\begin{figure}[]
\mbox{
\subfigure[]{\includegraphics[width=0.5\textwidth]{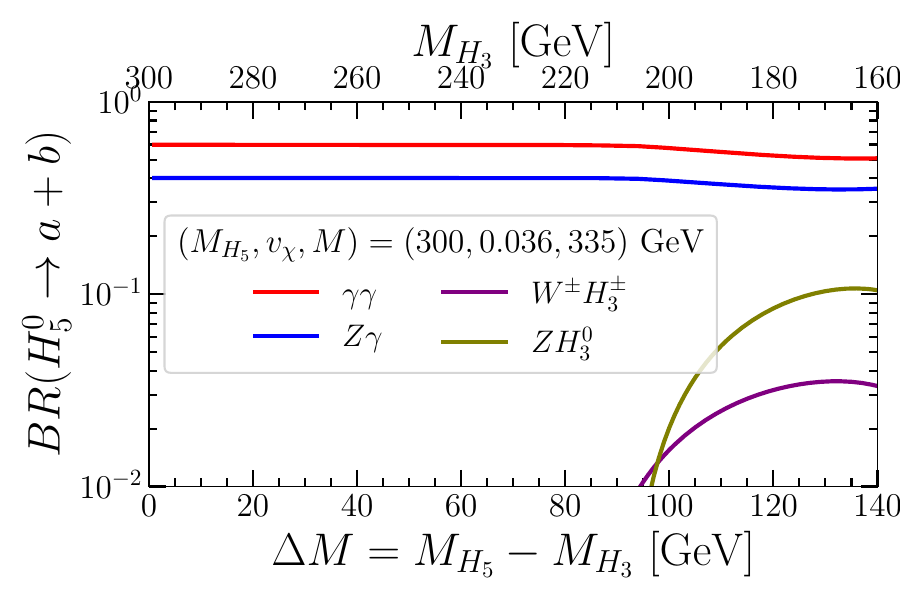}\label{fig:H50todiphoton2}}
}
	\mbox{
\subfigure[]{\includegraphics[width=0.5\textwidth]{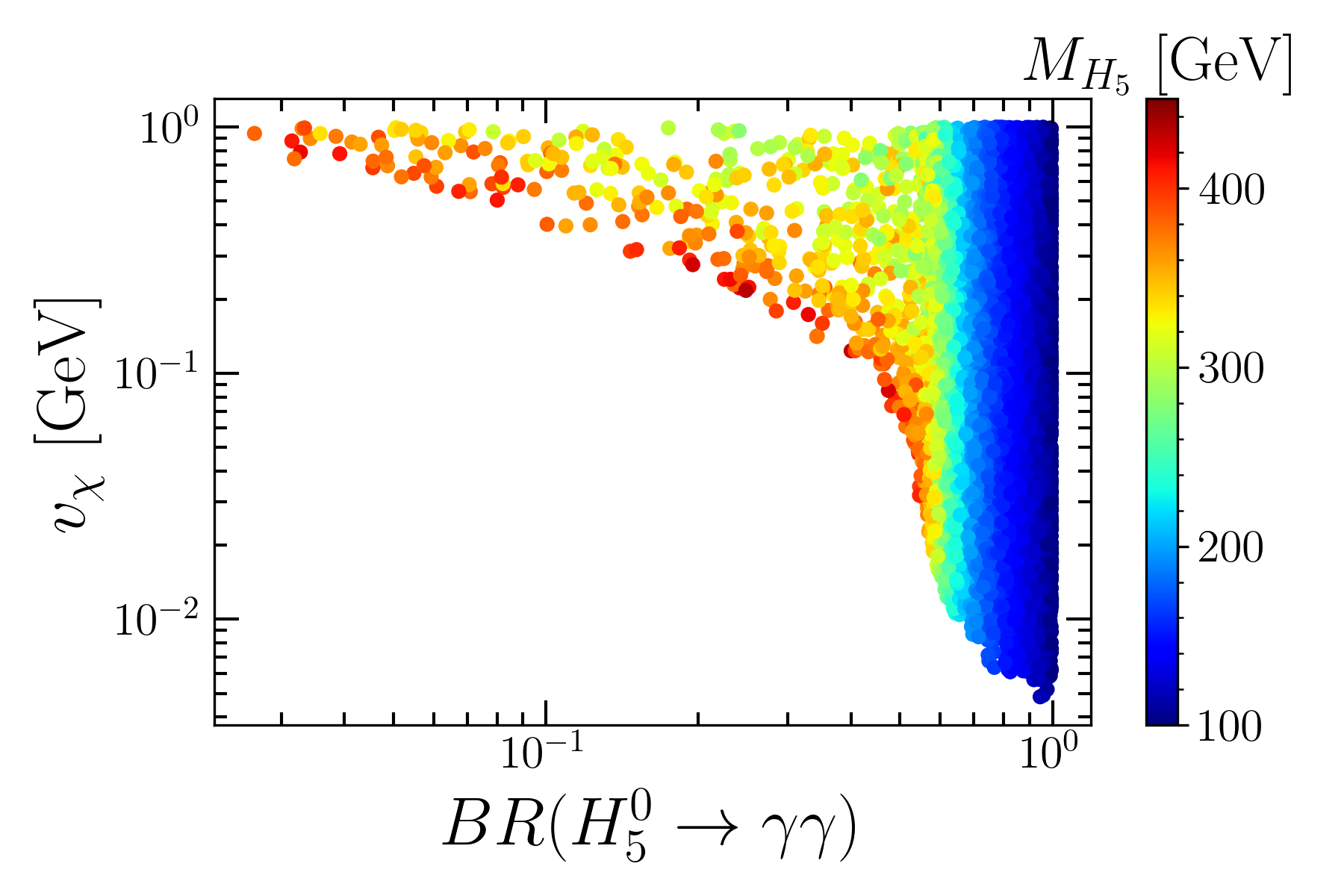}\label{fig:H50todiphoton1}}
	}
	\caption{(a) Branching ratio of $H^0_5$ to different channels for the benchmark point of Fig.~\ref{fig:H5ppallowed1}. (b) Allowed points in the 
 $BR(H_5^0\to\gamma\gamma)$ vs $v_\chi$ plane consistent with all previous theoretical and experimental constraints. 	}
	\label{fig:H50todiphoton}
\end{figure}
The ATLAS experiment has searched for spin-0 BSM resonances in the diphoton final state  using 139/fb data at $\sqrt{s}=13$~TeV~\cite{ATLAS:2021uiz}. In our model, the  neutral BSM scalars ($H_5^0,H$) decay to diphoton final states and hence this model can be constrained using this ATLAS diphoton search. In this subsection, we discuss the limit for $H^0_5\to \gamma \gamma$ and, in the following one, the limit for $H\to \gamma \gamma$.

Before turning to the implementation of the ATLAS analysis, we discuss the variation of BR$(H^0_5 \to \gamma \gamma)$ in the available parameter space.  Among the SM fermions, $H_5^0$ only interacts with neutrino via Yukawa coupling, which is suppressed unless $v_\chi\lesssim10^{-4}$ GeV. So, the preferred tree-level decays are $H^0_5 \to VV, W^{\pm} H^{\mp}_3, Z H_3^0$ or $ {W^{\pm}}^* H^{\mp}_3, W^{\pm} {H^{\mp}_3}^*$ when the two-body tree-level processes are not kinematically open, {and among the loop-level decays, $\gamma \gamma$ and $Z \gamma$ are preferred}. The partial decay width of $H^0_5 \to \gamma \gamma$ can be calculated following Eq.~\eqref{eq:decayHi} for $H_i= H^0_5$. 
The relevant vertex factors that enter  the calculation of $\Gamma(H^0_5 \to \gamma \gamma)$  are as follows:
\begin{eqnarray}
C_{H_3^+H_3^{-}H_5^0} &=& \sqrt{\frac{2}{3}} \frac{1}{v^2}\left[ 2(\lambda_3 - 2 \lambda_5) v_\phi^2 v_\chi - 8 \lambda_5 v_\chi^3+ 4 M_1 v_\chi^2 + 3 M_2 v_\phi^2 \right] \nonumber\\
&=&\sqrt{\frac{2}{3}} \frac{1}{v^2}\left[ 4M_{H_3}^2 v_\chi+ \frac{16 v_\chi^3}{v^2} (M^2-M_{H_5}^2)+\frac{M^2 (v^2-8v_\chi^2) }{4 v_\chi}\right], \nonumber\\
C_{H_5^+H_5^{-}H_5^0} &=& \sqrt{6} \left( 2 \lambda_3 v_\chi - M_2 \right)=\sqrt{6} \left(2v_\chi\frac{(M_{H_5}^2-M^2)}{v^2}-\frac{M^2}{12 v_\chi}\right),  \nonumber \\
C_{H_5^{++}H_5^{--}H_5^0} &=& -2 \sqrt{6} \left( 2 \lambda_3 v_\chi - M_2 \right)=-2 \sqrt{6} \left( 2v_\chi\frac{(M_{H_5}^2-M^2)}{v^2}-\frac{M^2}{12 v_\chi} \right),  \nonumber \\
C_{H_5^0 W^+ W^{-}} &=& \sqrt{\frac{2}{3}} \frac{1}{s_W^2} v_{\chi} \,.  \nonumber  \\
\end{eqnarray}

Note in particular that the contributions from the $W$  bosons are suppressed while the charged Higgses contributions can be strongly enhanced for small $v_\chi$ due to the presence of $M^2/{v_{\chi}}$ in the respective vertex factors. As a result, the partial decay widths of both $H^0_5 \to \gamma \gamma$ and $H^0_5 \to \gamma Z$   are inversely proportional to $v_\chi^2$. Hence, for a small $v_{\chi}$, the branching ratio for $H^0_5 \to \gamma \gamma, Z \gamma$ can be significantly enhanced.  We provide the analytic expressions for the partial decay widths for all final states  in  Appendix~\ref{appen1} and in Eq.~\eqref{eq:decayHi}.  
For the numerical analysis, we have implemented the model in  micrOMEGAs6.0~\cite{Alguero:2023zol} using Feynrules~\cite{Alloul:2013bka} and we use the code to compute all tree-level and loop-induced decays of the scalars. When there are no tree-level two-body processes we also include 3-body processes.  In Fig.~\ref{fig:H50todiphoton2}, we show the variation of the relevant branching ratios for our chosen benchmark point. As expected, for  smaller $v_{\chi}$,  $H_5^0$ dominantly decays to loop-induced processes, such as,  $\gamma\gamma$ and $\gamma Z$.  Thus the BSM resonance search decaying into  diphotons final state  at the LHC can constrain the model significantly.

 In Fig.~\ref{fig:H50todiphoton1}, we show the dependency of BR$(H^0_5 \to \gamma \gamma)$ on the vev $v_{\chi}$ and the mass of the fiveplet $M_{H_5}$, for all the points that satisfy the previously mentioned constraints. Note that since the theory constraints require $M_{H_5} < \sqrt{3} M_{H_3}$,  the decay mode $H^0_5 \to H^{\pm}_3 H^{\mp}_3$ is kinematically forbidden. The figure indicates that for a smaller $v_{\chi} < \mathcal{O}(0.1)$ GeV and for the specified variation of the parameter $M$, the branching ratio of $H^0_5 \to \gamma \gamma$ is always larger than 40$\%$ irrespective of the $M_{H_5}$ value.
\begin{figure}[]
\centering
\mbox{
	\subfigure[]{\includegraphics[width=0.5\textwidth]{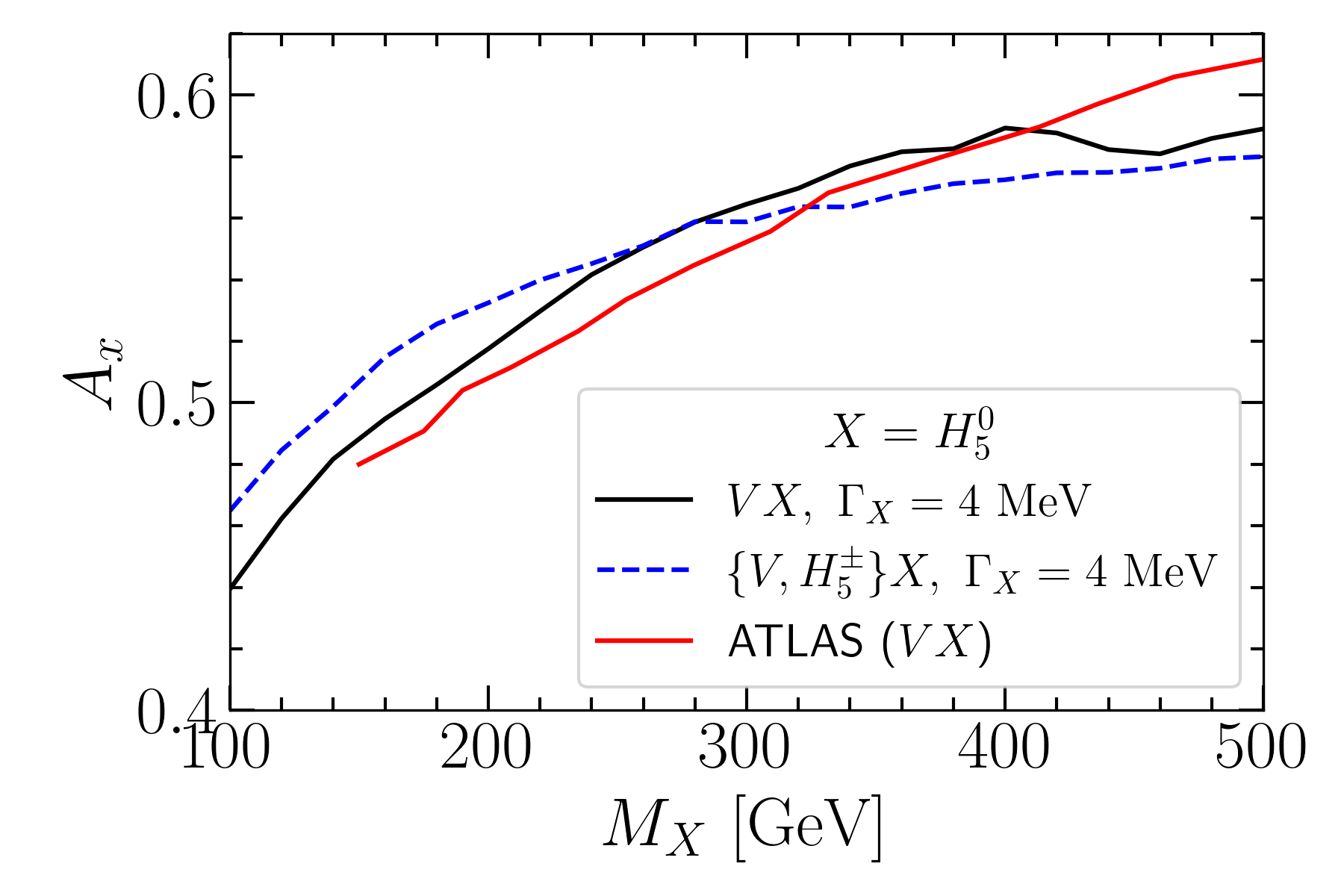}\label{fig:eff}}
  \subfigure[]{\includegraphics[width=0.5\textwidth]{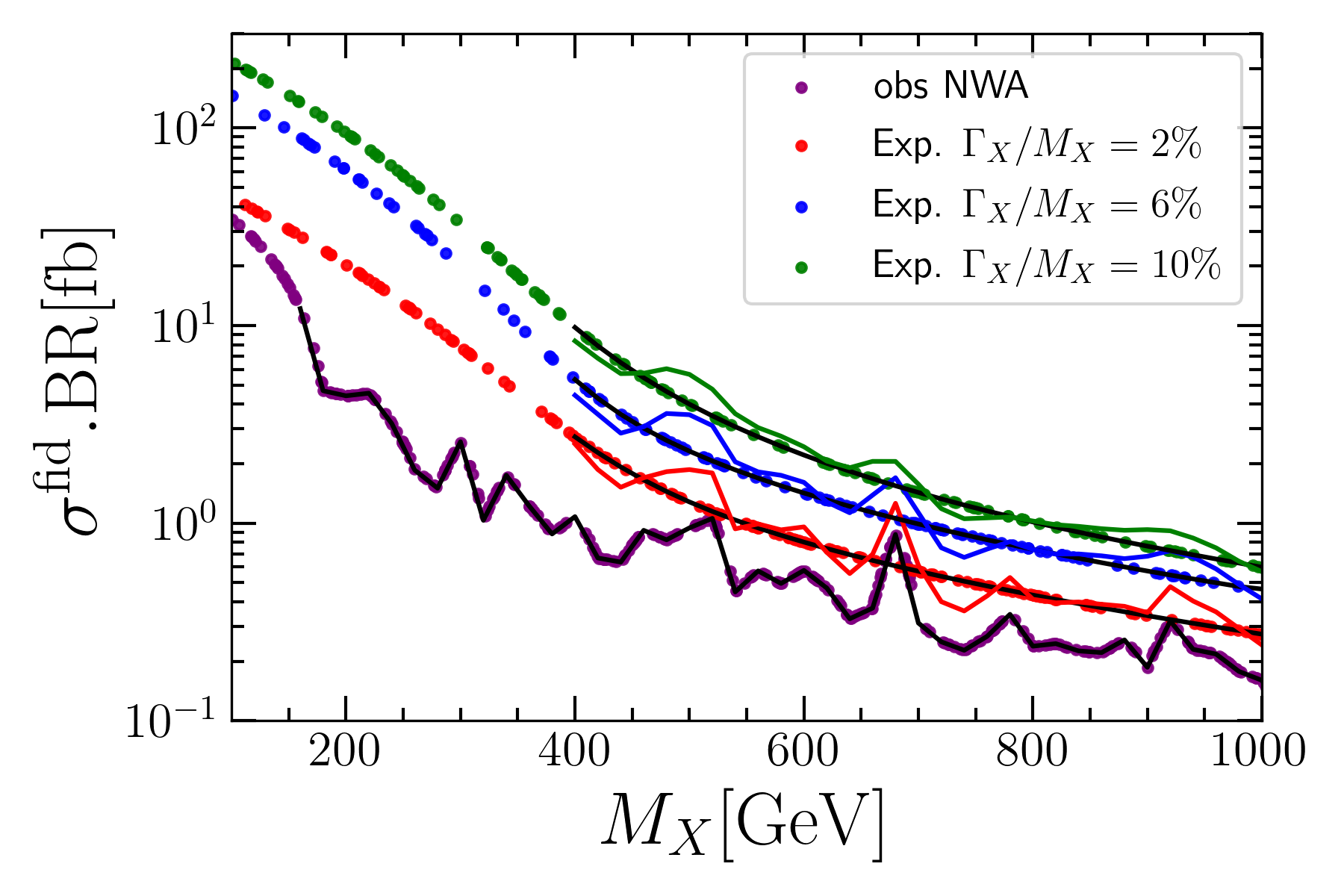}\label{fig:obsxs}}
	}
\caption{(a)~Efficiency for diphoton selection cuts as a function of the mass $M_X$ of the diphoton resonance, following the search $X \to \gamma \gamma$~\cite{ATLAS:2021uiz}. (b)~Observed and expected limit on the diphoton cross section.}
\label{fig:H502diphotonvalidation}
\end{figure}
Since over a large range of parameters, BSM Higgs decaying into diphoton has a significantly large branching ratio, hence we expect to receive tight constraint from ATLAS diphoton resonance search  $X \to \gamma \gamma$~\cite{ATLAS:2021uiz}. Following this ATLAS search, we implement the following set of cuts on our signal sample. In the ATLAS search, for the associate production mode, the experimental collaboration considered  $p p \to V X$ process, where $V$ is $W^\pm/Z$. In our case, our main production mode is somewhat different $p p \to H^0_5 H^{\pm}_5$, which will lead to a different cut-efficiency. Hence, we first validate our selection cuts with the same sample that have been considered in the ATLAS search, and later calculate the final cut-efficiency taking into account the relevant  channels.

\textbf{\large Acceptance cuts:  }
\begin{itemize}
    \item $c_1$:~ Number of photons $N^\gamma\ge 2$,
    \item $c_2:~c_1+ \{ p_T^{\gamma_{1,2}} \ge 25~\text{GeV},~ |\eta^{\gamma_{1,2}}|<2.37$ , isolation \footnote{
    For isolation we demand scalar sum of $p_T$ of all the stable particles (except neutrinos) found within a $\Delta R= 0.4$ cone around the photon direction, is required to be less than $0.05~p_T + 6$ GeV.} for $ \gamma_{1,2}$ \}
    \item $c_3:~c_2+ \{ \text{ Reject event with } 1.37<|\eta^{\gamma_{1,2}}|<1.52$ \}
    \item $c_4:~c_3+ \{ p_T^{\gamma_{1}}/M(\gamma_1 \gamma_2) > 0.3,~p_T^{\gamma_{2}}/M(\gamma_1 \gamma_2) > 0.25$ \}, {where $M(\gamma_1 \gamma_2)$ is the invariant mass of the two leading photons.}
\end{itemize}

In Fig.~\ref{fig:eff}, we show the efficiency ($A_x$) for diphoton selection cuts w.r.t.\ the mass of the diphoton resonance $M_X$ ($X=H_5^0$ ). The red line is extracted from the ATLAS analysis for associated production with a vector boson $pp \to V X$ and for a total width 
$\Gamma = 4$~MeV.  The black line corresponds to $pp \to V H_5^0(\to 2 \gamma)$ for $\Gamma(H_5^0) = 4$ MeV, which we compute. The efficiency for the mode $ \{V, H^\pm_5 \} X$ is shown by the blue-dashed line assuming $\Gamma(H_5^0) = 4$ MeV. The efficiency  for this mode is different  than that for $VX$ for two reasons. First, the photon $p_T$ for $ \{V, H^\pm_5 \} X$  is larger than that for $VX$. This occurs because larger momentum of $H_5^0$ for $ \{V, H^\pm_5 \} X$ than $VX$.  Another reason for the difference in $A_x$ between the two processes is coming from the isolation requirement of two photons. For the process $VX$, from $W^\pm/Z$ there are one lepton/jet pair, which can fall within a $\Delta R= 0.4$ cone around the photon direction. For $ \{V, H^\pm_5 \} X$, $H^\pm_5 \to W^\pm Z$ with subsequent decay of the gauge bosons to  leptons/jet pairs, there are more number of particles in the final state. Hence demanding the same isolation criterion somewhat reduces the cut-efficiency. We note that $H_5^0$ can also be produced in association with $H_3^\pm, H_3^0$ via  $pp \to H_5^0 \{H_3^\pm, H_3^0 \} $. For this channel $A_x$ depends on $M_{H_3}$ as well. However, contribution of this channel to the total production rate of $H_5^0$ is less than $10\%$. Hence, there is no notable variation of efficiency with $M_{H_3}$. Further, $H_5^0$ can be pair produced via off shell Higgs,  $pp\to h\to H_5^0 H_5^0$ but this cross section is suppressed compared to other channels. We do not include these sub-dominant channels while calculating the cut efficiency but for the production cross section we consider contribution of all possible channels.

In Fig.~\ref{fig:obsxs}, we show the  expected  and observed upper limits at 95\% CL on the fiducial cross-section times branching ratio to two photons for a spin-0 resonance as a function of its mass $M_X$ for different values of $\Gamma_X/M_X=(10,6,2)\% $~\cite{ATLAS:2021uiz}. The expected limits are shown by black solid lines and the observed limits by the coloured solid lines. The observed limit varies w.r.t.\ the width of the diphoton resonance, being much weaker for higher values of  $\Gamma_X/M_X$. For $ \Gamma_X/M_X =(10,6,2)\%$, the observed limit for a mass  lighter than 400 GeV is not given in the ATLAS paper~\cite{ATLAS:2021uiz}, thus we  extrapolate the corresponding expected limit and consider that as observed limit. The various coloured circle upto $M_X=400$~GeV  indicates extrapolated data of the expected limits for $\Gamma_X/M_X=(10,6,2)\%$. The purple points indicate the interpolated data of the observed upper limits at 95\%~CL for the narrow width approximation (NWA) and the black line under purple points is the corresponding ATLAS data. If $\Gamma_X/M_X \le 1 \%$ we consider the  NWA limit, else we follow Table.~\ref{tab:limit}. 
 
\begin{table}[]
\centering
\begin{tabular}{|l|l|l|}
\hline
 & $M_X\ge400$ GeV  &  $M_X < 400$ GeV \\ \hline
 $6\% < \Gamma_X/M_X   $& $10\%$ obs. limit &  $10\%$ extrapolated exp. limit  \\ \hline
 $2\% < \Gamma_X/M_X \le 6 \% $&   $6\%$ obs. limit & $6\%$ extrapolated exp. limit \\ \hline
$1\% < \Gamma_X/M_X \le 2\% $ &   $2\%$ obs. limit&  $2\%$ extrapolated exp. limit \\ \hline
$ \Gamma_X/M_X \le 1 \% $ &   NWA obs. limit & NWA obs. limit  \\ \hline
\end{tabular}
\caption{Selection of the ATLAS limit depending on the mass and decay width of the diphoton resonance. 
}
\label{tab:limit}
\end{table}
\begin{figure}[]
\centering
\mbox{
	\subfigure[]{\includegraphics[width=0.5\textwidth]{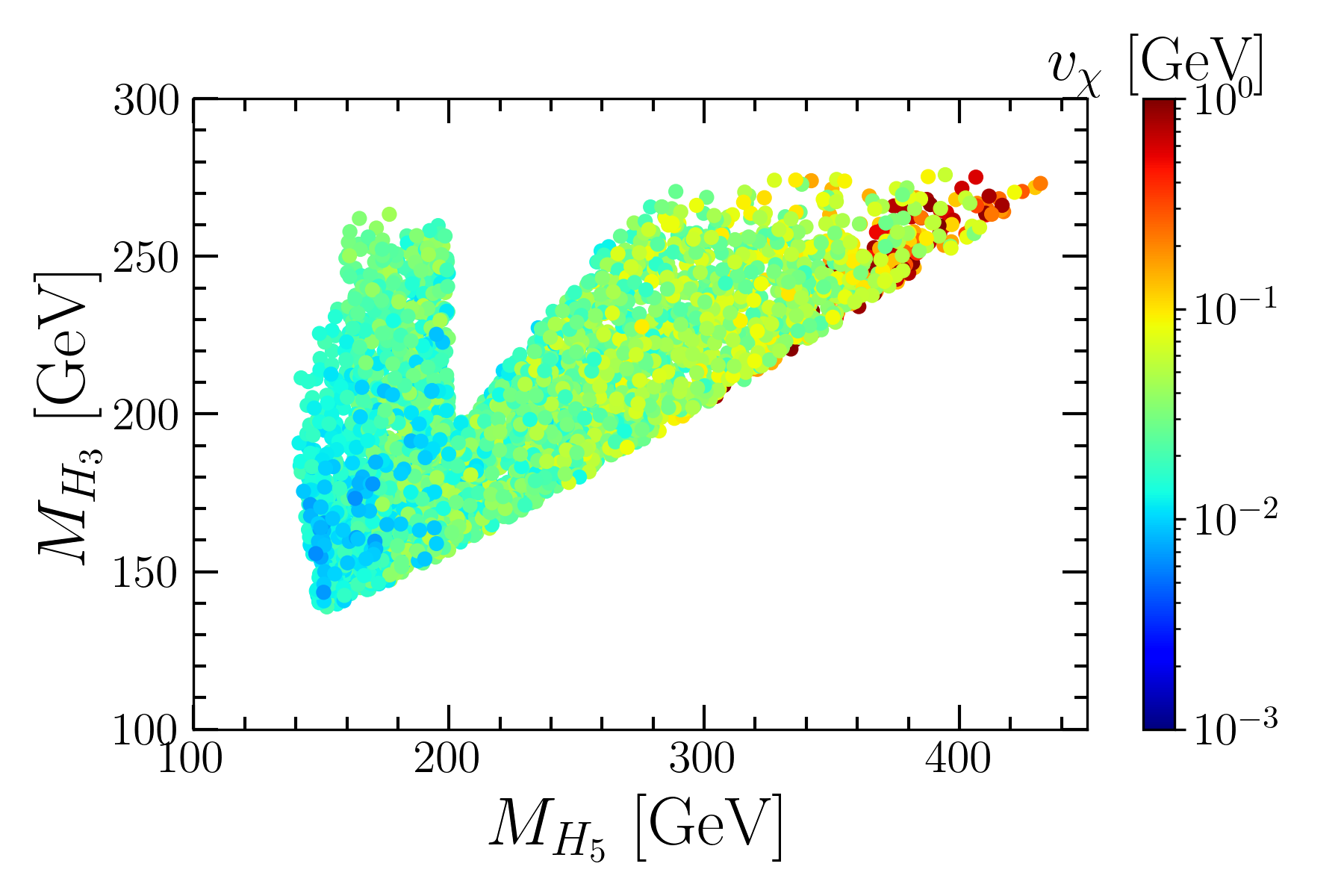}\label{fig:H50}}
  \subfigure[]{\includegraphics[width=0.5\textwidth]{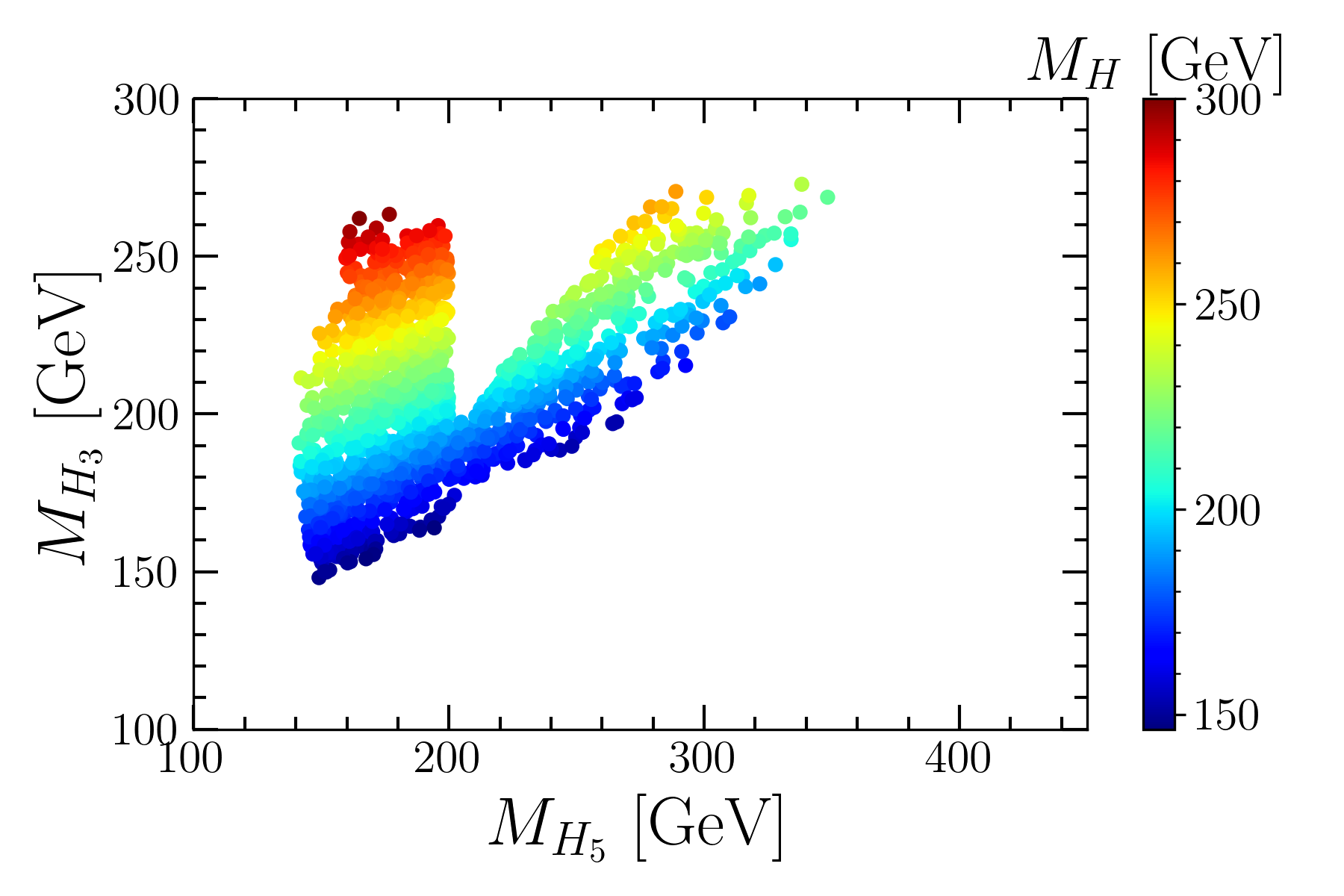}\label{fig:H}}
	}
\caption{(a)~Allowed points after imposing the  limit on $H_5^0 \to \gamma\gamma$ with $v_\chi$ in the color bar. (b)~Allowed points after imposing in addition the limit on $H \to \gamma\gamma$ with $M_H$ in the color bar.}
\label{fig:h50limit}
\end{figure}

Figure~\ref{fig:H50} shows the parameter space in the $M_{H_5}$ vs.\ $M_{H_3}$ plane that passes the diphoton limit for $H_5^0 \to 2\gamma$. We calculate the fiducial production cross-section times branching ratio to diphoton mode for $H_5^0$ and compare with the observed $95\%$ CL limit by the ATLAS search. 
The leading production channel is $pp\to H_5^0 H_5^\pm$ which depends mostly on $M_{H_5}$ while  the BR$(H_5^0 \to 2\gamma)$ depends on $M_{H_5}$, $M_{H_3}$, and $M^2/v_\chi$.  We note that, $p p \to H^0_5 H^0_5$ mode has a very small cross-section and therefore does not have any large impact.  If $H_5^0 \to VV/V H_3^{0/\pm}$ is either kinematically closed, or suppressed due to a small $v_{\chi}$, the leading decay modes are $H_5^0 \to 2\gamma/Z\gamma$ which depends on $M^2/v_\chi$. Note that $M$ is already constrained by the SM Higgs diphoton rate and the allowed value of $M$ is $\mathcal{O}(100)$~GeV. For the values of $M^2/v_\chi$ favoured by the previous bounds, $H_5^0 \to VV$ mode is suppressed. The total width is determined by the $H_5^0 \to 2\gamma/Z\gamma$ modes which varies as $\sim 1/v_\chi^2$, {as also has been emphasized before}. As the observed limit on  the cross section is stronger for a narrow resonance (see Fig.~\ref{fig:obsxs}), larger $v_\chi$ for which $H_5^0$ has smaller width is excluded. Contrary to that,  lower $v_\chi$ passes the ATLAS limit for which the width of $H_5^0$ is larger and has a weaker limit. We find that $v_{\chi} \sim \mathcal{O}(1)$ GeV is significantly constrained over a large mass range, except  $300$ GeV $< M_{H_5} < 400 $ GeV, represented by the red points.

\subsubsection {$H\to \gamma \gamma$ search}
\label{subsec:h50}
\begin{figure}[]
\centering
\includegraphics[width=1\textwidth]{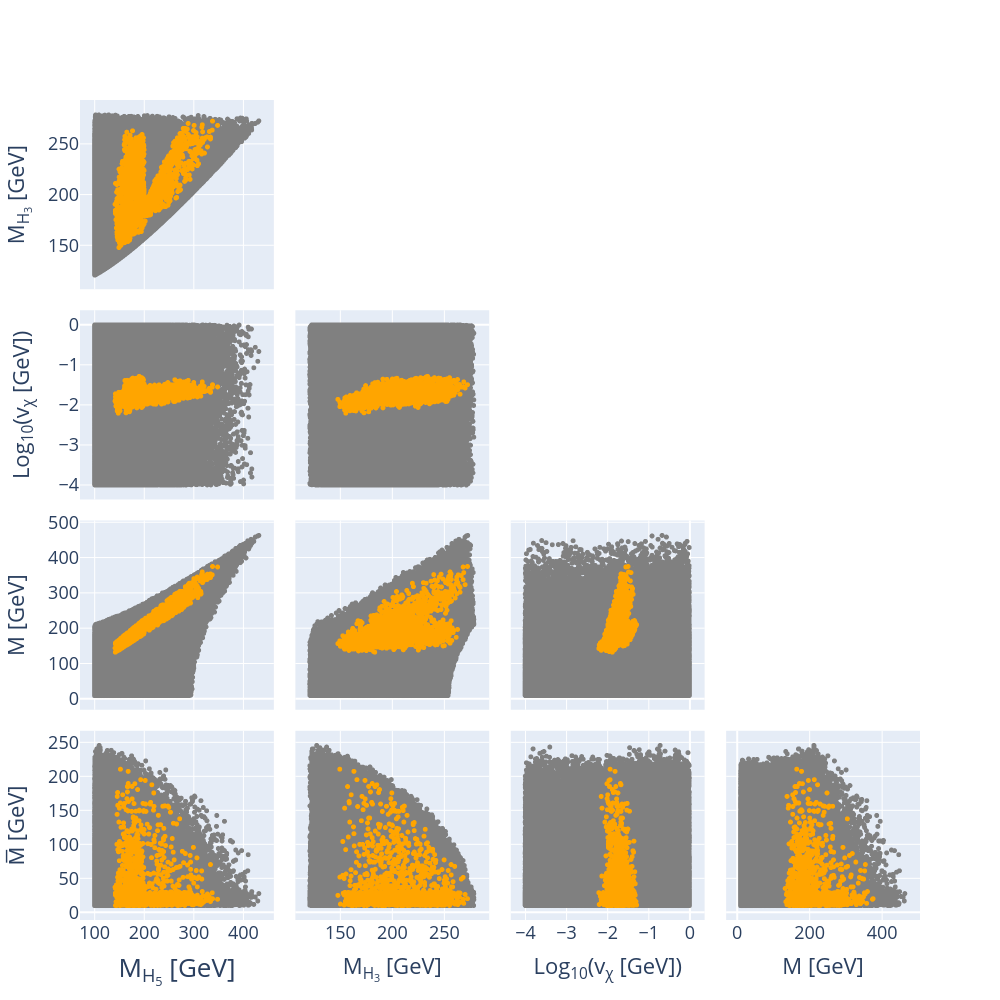}
\caption{Allowed parameter space after theoretical constraints (grey points) and experimental constraints from colliders (orange points) have been imposed.}
\label{Fig:Th_Exp_Allowed}
\end{figure}

We perform a similar analysis for the singlet scalar in the channel $H\to 2\gamma$. The  dominant production channels for $H$ is $pp \to H H_3^{\pm,0}$ and the main decays are  to $2\gamma/Z\gamma$ via loop. The corresponding vertex factor is proportional to to $M^2/v_\chi$.  The decay $H\to W^\pm W^\mp$ is suppressed   by kinematics and/or  because the coupling depends on  $v_{\chi}$, which is $\le1$ GeV, see Eq.~\eqref{eq:HVVvertex}. 
We obtain the fiducial cross section after calculating the cut efficiency ($A_x$) with the same set of cuts as mentioned for the case of $H_5^0$. Note that $A_x$ for this channel varies between 0.5 to 0.59, depending  on $M_H$ and $M_{H_3}$. To check the diphoton limit, we compare the fiducial cross section for this model with the observed limit in a similar way as for $H_5^0$.
Figure~\ref{fig:H} shows allowed values of $M_{H_5}, M_{H_3}$ and $M_H$   that pass the diphoton limit for $H\to 2\gamma$.
In Fig.~\ref{Fig:Th_Exp_Allowed}, we show the allowed parameter space by orange colour, which is consistent with $H\to 2\gamma$ limit as well as all previously discussed constraints. We consider all possible planes such as $M_{H_5}$ vs.  ${M}$, $M_{H_5}$ vs. $v_{\chi}$. From Fig.~\ref{Fig:Th_Exp_Allowed}, it is evident that the points with $v_\chi >0.05$ GeV are mostly excluded  because the width of $H$ becomes very narrow,  $\Gamma/M_H\lesssim 2\%$,  which corresponds to the stronger limit on the observed cross-section as can be seen from Fig.~\ref{fig:obsxs}.  For $v_\chi \le 0.05$ GeV, the width $\Gamma/M_H$ becomes large $\Gamma/M_H \sim \mathcal{O}(6-50)\%$ and the experimental constraint weakens significantly. 
Therefore only points where $v_{\chi} \le 0.05$ GeV are allowed. Finally when $v_\chi$ is even smaller then $\Gamma/M_H$ becomes very large, we have excluded those points by the condition on the total width. We conclude this section with the observation that after taking into account diphoton limit for $H$, the five-plet mass $M_{H_5}$ becomes  more constrained $M_{H_5}< 350$ GeV, while the allowed value of $M_H$ is in between $145$ GeV and $300$ GeV; $M_{H_3}$ can vary in between $150$ GeV and $270$ GeV, and the vev $v_{\chi} \le 0.05$ GeV. Improvement in the di-photon signal sensitivity by $3~\rm{to}~4$ order of magnitude by future LHC search can probe the complete range of  $v_\chi=[10^{-4},1]$ GeV. Furthermore, as the $H_5^{\pm\pm}\to W^\pm W^\pm$ search sets an upper bound on  $v_\chi$, tighter limits on same-sign gauge boson production can also shrink the range of $v_\chi$.  To improve the constraint on $v_\chi$ for $M_{H_5}<200$ GeV, however, one would need to  extend the $H_5^{\pm\pm}\to W^\pm W^\pm$ search to the low mass region where $H_5^{\pm\pm}\to W^\pm W^{\pm \star}$ is open.

\begin{table}[]
\centering
\begin{tabular}{|l|l|l|}
\hline
   & \text{Total number of points sampled}                                    & $4\times 10^{6}$          \\ \hline
   & \textbf{Constraints}                                                       & \textbf{Number of points survived} \\ \hline
c1: & Theory                                                            & 418410                    \\ \hline
c2: & c1 + Oblique parameters                                           & 418239                    \\ \hline
c3: & c2 + $\Gamma/M <0.5$                                              & 307117                    \\ \hline
c4: & c3 + $h\to \gamma \gamma$ measurements                            & 18476                     \\ \hline
c5: & \multicolumn{1}{l|}{c4 + $H_{5}^{++/--}\to W^{\pm}W^{\pm}$ measurements} & 16491                     \\ \hline
c6: & c5 + $H_{5}^{0}\to \gamma \gamma$ measurements                    & 3447                      \\ \hline
c7: & \multicolumn{1}{l|}{c6 + $H\to \gamma \gamma$ measurements}   & 1446                      \\ \hline
\end{tabular}
\caption{Total number of points that satisfy different constraints.}
\label{tab:points_constraint_table}
\end{table}

\section{Dark matter observables} \label{sec:DMpheno}
\subsection{The scalar singlet as cold dark matter}
As mentioned before, in our discussions of the GM-S model the singlet scalar $S$ serves as the WIMP DM candidate due to the $Z_2$ symmetry which makes it stable.
The observed relic density of DM is obtained through the pair annihilation of $S$ to bath particles in the early universe. 
The evolution of the number density of DM,  $n_S$, in the early Universe is governed by the Boltzmann equation, which reads~\cite{Gondolo:1990dk},
\bea
\frac{dn_{S}}{dt} + 3{\rm H}n_{S} = -\langle\sigma v\rangle (n_{S}^2-n_{S}^{{\rm eq}^2})\,\, ,
\label{beqn1}
\eea
where  $n_{S}^{\rm eq}$ is the  equilibrium number density of $S$,  $\langle\sigma v\rangle$ is the thermal average cross section  for the annihilating DM particles and $H$ is the Hubble parameter.  The Boltzmann equation can be written  in terms of the co-moving number density,  $Y_{S}=n_{S}/s$, where $s$ is the entropy number density of the Universe,
\begin{equation}
\frac{dY_{S}}{dx} = -\frac{s(T)}{H(T)x}\langle \sigma v \rangle (Y_S^2-{Y_S}^{{eq}^2})\,\,\, ,
\label{beqn_comov}
\end{equation}
where $x=m_S/T$ and   the Hubble parameter and entropy density~\cite{Gondolo:1990dk,kolb:1990vq},
\be
H(T) = \frac{\pi}{\sqrt{90}}\frac{\sqrt{g_{\rm eff}(T)}}{M_{pl}}T^2 \,, \quad
 s(T) = h_{\rm eff}(T)\frac{2\pi^2}{45}T^3 \,.
\label{cosomo_param}
\ee
Here, $g_{\rm eff}$, $h_{\rm eff}$ are the effective degrees of freedom for the energy and entropy densities of the Universe at temperature $T$, $Y_{S}^{\rm eq}$ is the value of $Y_S$ when $n_S = n_{S}^{\rm eq}$, ~\cite{Gondolo:1990dk} and is given by:
\begin{equation}
Y_S^{eq} = \frac{45g}{4\pi^4}\frac{x^2K_2(x)}{h_{\rm eff}(T)} \,.
\label{eq_no_density}
\end{equation}
In the above, the internal degree of freedom $g$ takes value unity for the case of $S$ which is a real scalar field. Solving Eq.~\eqref{beqn_comov}, the DM relic density can be obtained using the relation~\cite{Alguero:2022inz} 
\bea
\Omega_{S} h^2 &=& 2.755\times10^8 \left(\frac{M_S}{\rm GeV}\right) Y_S(\rm T_{now})\,\,,
\label{obs_relic}
\eea
where $h$ is the reduced Hubble constant.
\begin{figure}[]
	\centering
	\includegraphics[width=0.75\textwidth]{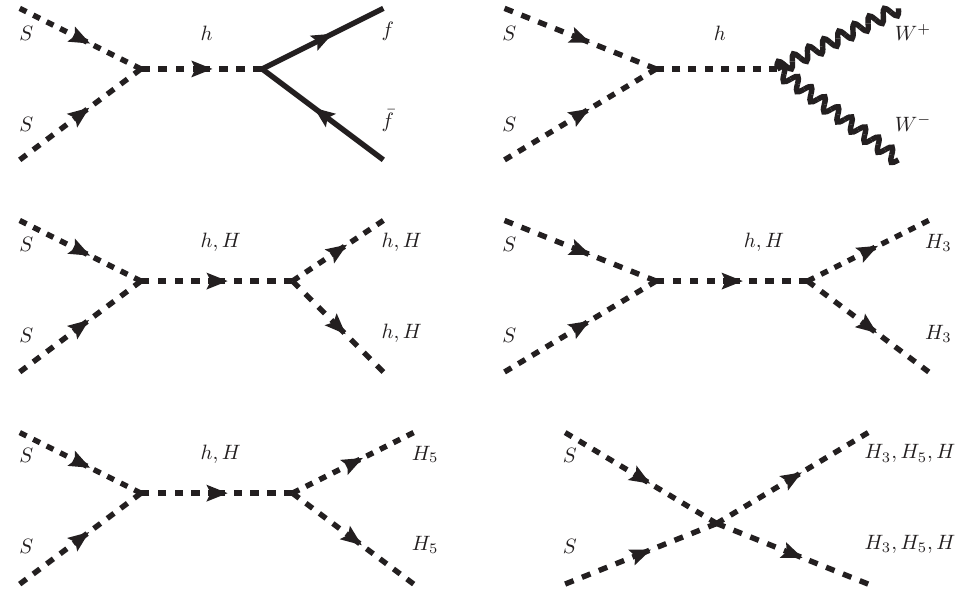}
	\caption{Tree level  Feynman diagrams for DM pair annihilation into  SM/BSM particles where we use notation $H_{3}=H_{3}^{0},H_{3}^{+}$ and $H_{5}=H_{5}^{0},H_{5}^{+},H_{5}^{++}$.}
	\label{fig:Fdiag_DM}
\end{figure}

\begin{table}[h]
\begin{center}
\vskip 0.05cm
\begin{tabular} {|c|c|}
\hline
Vertex & Vertex Factor\\
\hline
$SSh$ & $\,-4 \lambda_{a}v_{\phi}$ \\
\hline
$SSH$ & $\,-4 \lambda_{b} v_{\chi}$ \\
\hline
$SShh$ & $\,-4 \lambda_{a}$  \\
\hline
$SSHH$ & $\,-4 \lambda_{b}$ \\
\hline
$SSH_{3}H_{3}$ & $\,-4 (\lambda_{b}c_{H}^{2}+\lambda_{a}s_{H}^{2})$ \\
\hline
$SSH_{5}H_{5}$ & $\,-4 \lambda_{b}$ \\
\hline
\end{tabular}
\end{center}
\caption{Couplings of DM with SM and BSM scalars where we use the notation $H_{3}=H_{3}^{0},H_{3}^{+}$ and $H_{5}=H_{5}^{0},H_{5}^{+},H_{5}^{++}$.}
\label{vertex_table_Higgs}
\end{table}
\noindent
The precise determination of the observed relic density of DM has been achieved through PLANCK measurements of the cosmic microwave background (CMB)~\cite{Planck:2018vyg},
\begin{equation}
    \Omega h^2=0.120\pm 0.0012
\end{equation}

All  possible $S$ pair annihilation channels that can contribute to $\langle \sigma v\rangle$ in Eq.~\eqref{beqn_comov} are shown in Fig.~\ref{fig:Fdiag_DM}. The $SS\to {\rm SM}+{\rm SM}$ annihilation channels depend only on $\lambda_a$,   the coupling of  the singlet to the Higgs, see Table~\ref{vertex_table_Higgs}. These channels are the same ones that contribute to DM annihilation in the  real singlet scalar model ~\cite{Guo:2010hq,Armand:2022sjf,GAMBIT:2017gge}. In the GM-S model additional channels, such as $SS\to H_{3}H_{3},H_{5}H_{5},HH$  contribute to DM annihilation.  All the couplings involving the BSM scalars depend on $\lambda_{b}$, see Table~\ref{vertex_table_Higgs}. The two couplings $\lambda_{a}$ and $\lambda_{b}$  are therefore the only parameters that govern the DM phenomenology in addition to the mass parameters. Other than the quartic interaction governing $SS \to H_3 H_3, H_5 H_5, H H$, $s$-channel mediated process $S S \to H \to H_5 H_5, H_3 H_3$ can also give large contribution to the annihilation cross-section, if kinematically accessible. This is a result of the enhancement of couplings $H H_5H_5$ and $H H_3 H_3$, for small values of $v_\chi$, which can be seen from Eq.~\eqref{Eq:C_Hgaga}. 

\begin{figure}[!t]
\mbox{
\subfigure[]{\includegraphics[width=0.5\textwidth]{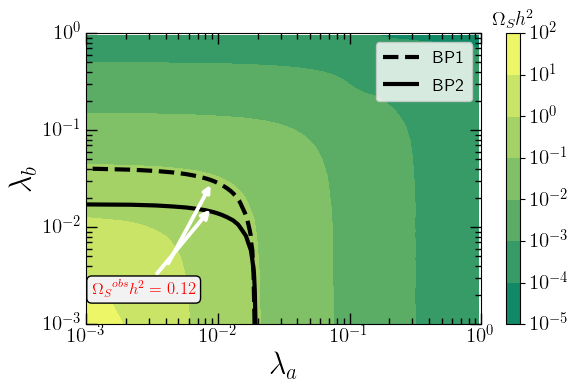}\label{fig:DM300GEV_relic1}}
}
\mbox{
\subfigure[]{
\includegraphics[width=0.5\textwidth]{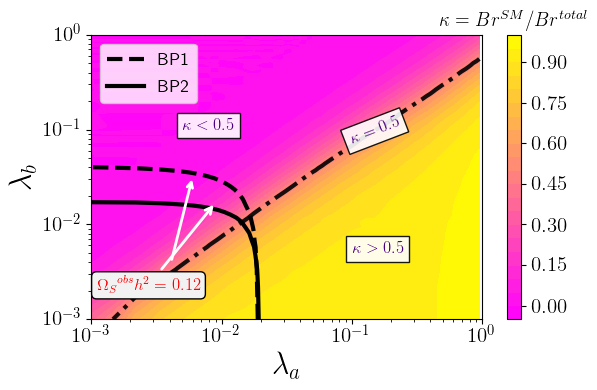}\label{fig:DM300GEV_relic2}}
}
\caption{(a)~Contours of $\Omega_{S}h^{2}$ in the plane of $\lambda_{b}$ vs.\ $\lambda_{a}$ for BP1. The black dashed and solid lines superimposed on these  indicate the contour corresponding to $\Omega_S h^2 = 0.12$ for BP1 (BP2). 
(b)~Contours of $\kappa$, the ratio of  DM annihilation to SM vs.\ (SM+BSM) particles, in the  $\lambda_{b}$ vs.\ $\lambda_{a}$ plane again for BP1.  The dot-dashed line corresponds to scenarios where DM annihilates equally to SM and BSM particles. The black dashed and solid lines superimposed on these indicate the contour corresponding to $\Omega_S h^2 = 0.12$ for BP1 (BP2). 
}
\label{fig:DM300GEV_relic}
\end{figure}

We will first consider only two benchmark points to illustrate the impact of various DM constraints on the 
parameters relevant for the discussion of DM, viz. $\lambda_a,\lambda_b$ and $M_S$, leaving a complete exploration of the entire allowed parameter space of the model to section~\ref{sec:scan}. Recall from the discussions in section~\ref{sec:scan} and section~\ref{sec:bound},
that the  current theoretical and experimental constraints restrict the mass of the new scalars in a relatively narrow range, namely  $140~{\rm GeV }< M_{H_5} < 350~{\rm GeV }$, $150~{\rm GeV }< M_{H_3} < 270~{\rm GeV }$ and 
$145~{\rm GeV }< M_{H} < 300~{\rm GeV}$. For our two benchmark points, we fix the value of $M_S$ at 280~GeV and choose the other masses so as to cover the range of allowed masses for $M_{H_{5}}, M_{H_{3}}$ and $M_H$ mentioned above. The two benchmark points  are thus defined as

\begin{itemize}
    \item BP1: $M_{H_{5}}=300\,\text{GeV}$, $M_{H_{3}}=254\,\text{GeV}$, $M_{H}=227\,\text{GeV}$, $v_{\chi}=0.036\,\text{GeV}$, \\ $M=335\,\text{GeV}$, $\overline{M}= 43\,\text{GeV}$ and  $M_{S}=280\,\text{GeV}$
    \item BP2: $M_{H_{5}}=190\,\text{GeV}$, $M_{H_{3}}=234\,\text{GeV}$, $M_{H}=253\,\text{GeV}$, $v_{\chi}=0.05\,\text{GeV}$, \\ $M=210\,\text{GeV}$, $\overline{M}= 10\,\text{GeV}$ and  $M_{S}=280\,\text{GeV}$
\end{itemize}
\noindent

We use micrOMEGAs6.0~\cite{Alguero:2023zol} to obtain the DM relic density according to Eq.~\eqref{obs_relic}. To that end, we generate the necessary CalcHEP~\cite{Pukhov:2004ca,Belyaev:2012qa} files using Feynrules~\cite{Alloul:2013bka}. 
In Fig.~\ref{fig:DM300GEV_relic1}, we show contours of relic density $\Omega_{S}h^{2} = 0.12$ in the $\lambda_{a}$ vs. $\lambda_{b}$ plane for the two benchmarks BP1 and BP2. These are superimposed on the contours in shades of green of $\Omega_{S}h^{2}$ for BP1. The ones for BP2 are qualitatively similar. As expected for a typical WIMP candidate, $\Omega_{S}h^{2}$ decreases when the values of the coupling of the DM to SM  ($\lambda_{a}$)  and/or to new scalars ($\lambda_{b}$) increase. The overabundant region corresponds to small values of both $\lambda_a$ and $\lambda_b$.  When $\lambda_{b}$ is below $\approx 10^{-2}$, annihilation into SM final states dominates and the relic density is the same for the two benchmarks since they both feature the same DM mass. When $\lambda_{a}$ drops below $\approx 10^{-2}$, annihilation into custodial multiplets dominates. The value of $\lambda_b$ corresponding to  $\Omega_{S}h^{2}=0.12$  is larger for BP1 than for BP2 because for BP1 the final state $H_5H_5$ is not kinematically accessible.  The value of $\kappa$, representing the ratio of DM annihilation to SM vs. (SM+BSM) particles is illustrated in Fig.~\ref{fig:DM300GEV_relic2}, again for BP1. Superimposed on it are the contours of $\Omega_S h^2 = 0.12$  for BP1 and BP2, indicated by the dashed and solid black curve. For $\lambda_{a}\approx\lambda_{b}$, DM annihilates equally to SM and custodial multiplets as shown by the dashed line corresponding to $\kappa=0.5$.  The behaviour of $\kappa$ for BP2 is also qualitatively similar to that for BP1. It is essential to highlight that $\sin\alpha$ is identically zero in this analysis, for $\lambda_{a}>\lambda_{b}$, the GM-S model mimics the behaviour of the real singlet scalar DM extension of SM.

\begin{figure}[!t]
\centering
\includegraphics[width=0.35\textwidth]{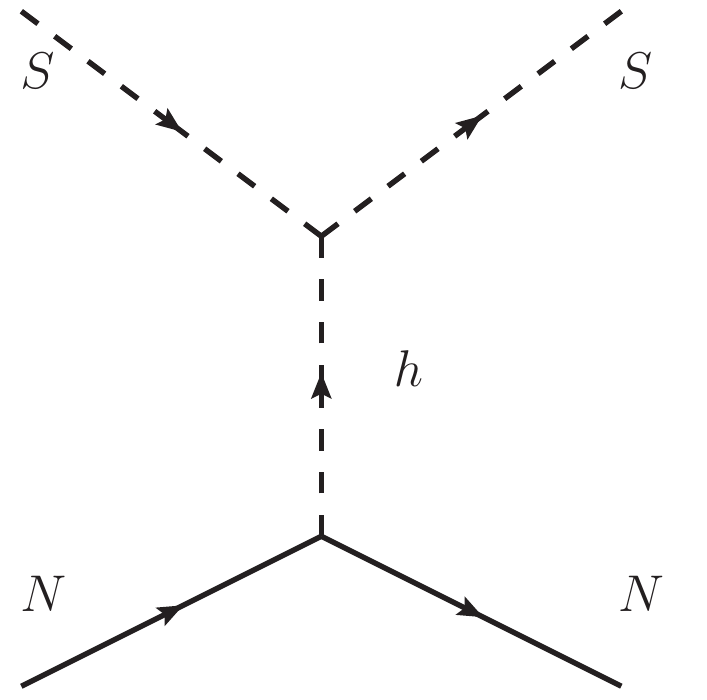}
\caption{Feynman diagram for the elastic scattering  of DM $S$ on nucleons $N$ via Higgs exchange.}
\label{fig:Fdiag_DD}
\end{figure}

\subsection{Direct and Indirect detection}

Direct searches for DM through its  scattering on nuclei pose severe constraints on WIMP DM models. 
In the  GM-S model, the elastic scattering of the DM $S$ on nucleons occurs via $t$-channel Higgs exchange as shown in Fig.~\ref{fig:Fdiag_DD}. In the decoupling limit where  $\sin\alpha=0$, there is no tree-level contribution from the  $t$-channel  exchange of  the BSM Higgs bosons $H$. The spin independent cross section, $\sigma_{SI}$, for the process $S N\to S N$ where $N$ is a nucleon is given by 
\begin{align}
\sigma_{SI}=\frac{4}{\pi}\mu_{r}|A^{SI}|^{2} \,,
\label{eq:DD}
\end{align}
where $\mu_{r}=\frac{M_{S}M_{N}}{M_{S}+M_{N}}$ with   $M_{N}$ the nucleon mass and $A^{SI}$ is the nucleon amplitude which depends on the DM interaction with nucleons and the nucleon form factors. To evaluate the spin independent cross section  we use micrOMEGAs6.0 \cite{Alguero:2023zol}. 
For the two benchmark points,  we compare the predicted cross section with the current experimental data from LUX-Zeplin(LZ)~\cite{LZ:2022lsv}. We also examine the future reach of XENONnT~\cite{XENON:2015gkh} and DARWIN~\cite{DARWIN:2016hyl}.

\begin{figure}[t!]
\mbox{
\subfigure[]{\includegraphics[width=0.48\textwidth]{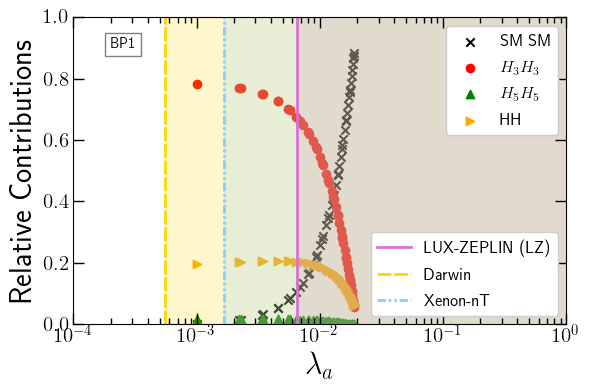}\label{fig:DM_scan_res1}}}
\mbox{
\subfigure[]{\includegraphics[width=0.48\textwidth]{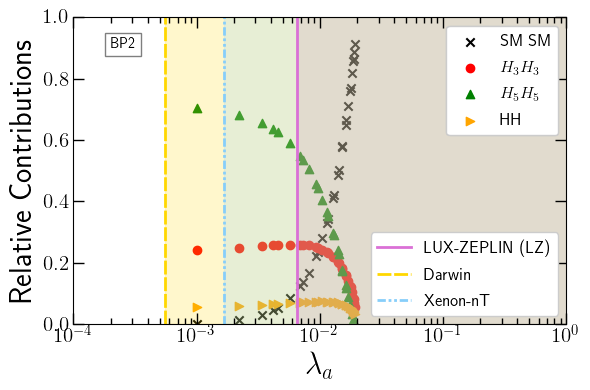}\label{fig:DM_scan_res2}}
}
\caption{Relative contribution of DM pair annihilation into various final states vs $\lambda_{a}$ for benchmark points  (a) BP1  and (b) BP2. The direct detection limits on $\lambda_{a}$ from  Lux-Zeplin (LZ)~\cite{LZ:2022lsv} as well as future projections from DARWIN~\cite{DARWIN:2016hyl} and XENONnT~\cite{XENON:2015gkh} are also displayed.  Note that all the points satisfy the  observed DM relic density. }
\label{fig:DM_scan_res}
\end{figure}

In Figs.~\ref{fig:DM_scan_res1} and \ref{fig:DM_scan_res2} we show how the relative contributions to the DM pair annihilation into the channels ${\rm SM\,SM}, H_3 H_3, H_5 H_5$ and $HH$ vary with  $\lambda_a$ for BP1 and BP2 respectively. For each benchmark point, the value of  $\lambda_b$ is fixed to the value that leads to  the observed relic density.  In both cases,  
we observe that for $\lambda_{a}>10^{-2}$, the dominant contributions  arise from  DM annihilation into the SM states. However, this region  is stringently constrained from current direct detection data from PANDAX-4T, Xenon1T and particularly LZ. 
The current limit from LZ sets an upper limit on $\lambda_a \sim 0.007$ $(0.005)$  for  BP1(BP2). 
In the future, this upper limit can be lowered to $\lambda_a \sim 1.5\times 10^{-3}$ with XENONnT and to  $\lambda_a \sim 5.5\times 10^{-4}$ with DARWIN~\cite{DARWIN:2016hyl}.
For  values of $\lambda_a$ below the current LZ limit,  the non-SM modes dominate over the SM, channels with $H_3 H_3 $ being the dominant mode for BP1 and  $H_5 H_5 $ for BP2. 

Dark matter can also be detected indirectly by observing gamma rays originating from DM annihilation in galaxies. In particular, DM annihilates to SM and BSM final states
via s-channel process mediated by the Higgses and via quartic interactions as shown in Fig.~\ref{fig:Fdiag_DM}.
The photon spectra originating from all final states are computed with micrOMEGAs and 
we use~\cite{Eckner} to determine the indirect constraints extracted   from Dwarf Spheroidal Galaxies from the Fermi-LAT and MAGIC
telescopes \cite{MAGIC:2016xys,Reinert:2017aga}. We use~\cite{CTA:2020qlo}  as  implemented in micrOMEGAs\,6.1 to compare the photon spectra with the expected sensitivity of CTA. 

\begin{figure}[ht!]
\mbox{
\subfigure[]{\includegraphics[width=0.48\textwidth]{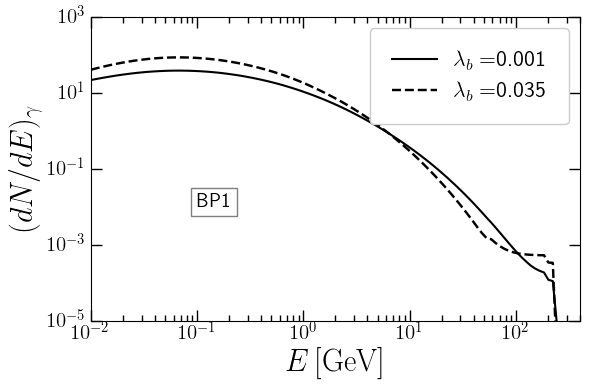}\label{fig:DM_spectrum1}}
\subfigure[]{\includegraphics[width=0.48\textwidth]{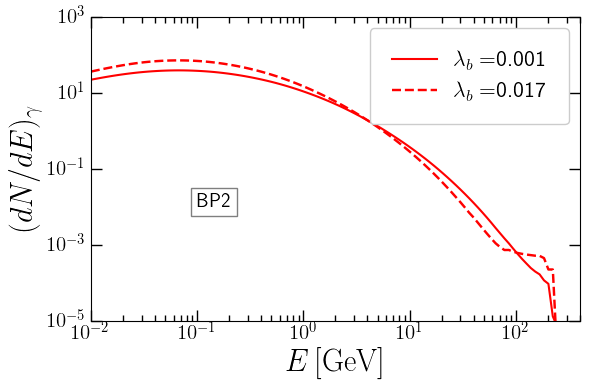}\label{fig:DM_spectrum2}}
}
\mbox{
\subfigure[]{\includegraphics[width=0.48\textwidth]{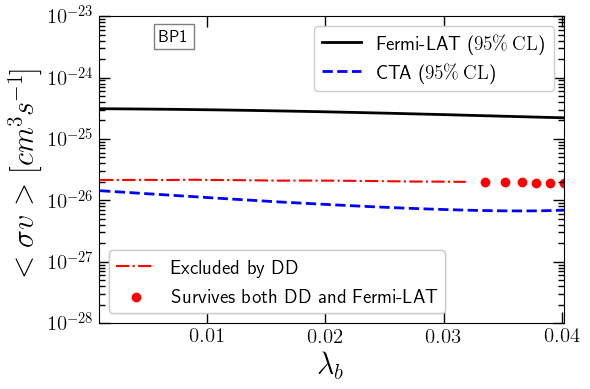}\label{fig:ID_DM_spectrum3}}
\subfigure[]{\includegraphics[width=0.48\textwidth]{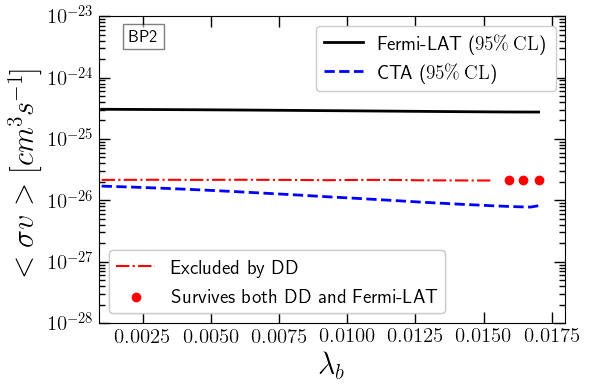}\label{fig:ID_DM_spectrum4}}
}
\caption{Upper row: Photon spectrum for two different choices of $\lambda_b$ for (a)  BP1 and (b) BP2. Lower row: Theoretical prediction for  $\langle \sigma v \rangle$ vs.\ $\lambda_{b}$ for (c) BP1 and (d) BP2. The dash-dot line shows the region excluded by LZ. The current limit from FermiLAT (full) and the sensitivity of CTA (dash) are also shown. Note that all the points satisfy the observed DM relic density.
}
\label{fig:DM_BP_ID}
\end{figure}

\begin{figure}[ht!]
\mbox{
\subfigure[]{\includegraphics[width=0.48\textwidth]{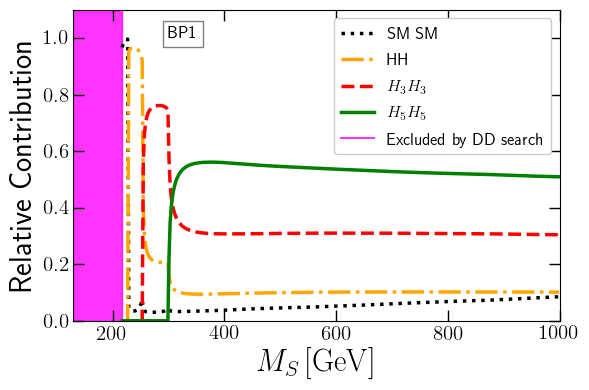}\label{fig:BP1_CTA_Scan1}}
\subfigure[]{\includegraphics[width=0.48\textwidth]{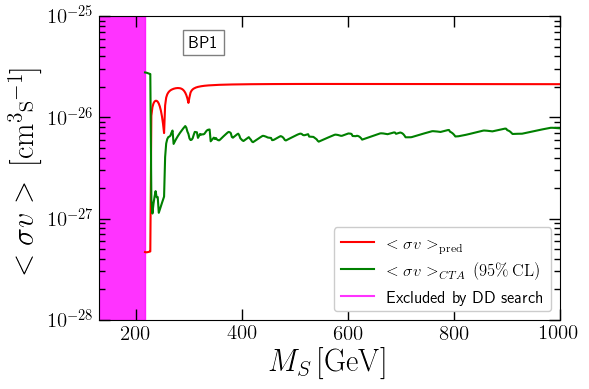}\label{fig:BP1_CTA_Scan2}}
}
\mbox{
\subfigure[]{\includegraphics[width=0.48\textwidth]{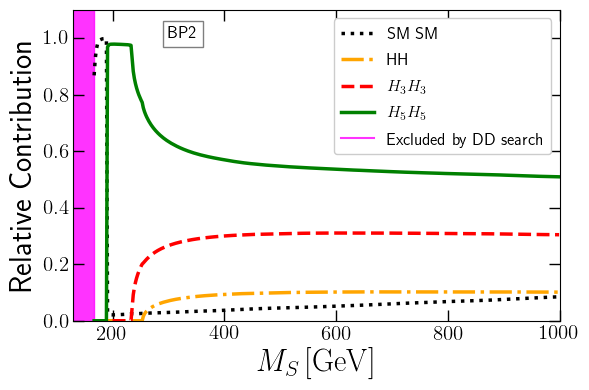}\label{fig:BP2_CTA_Scan1}}
\subfigure[]{\includegraphics[width=0.48\textwidth]{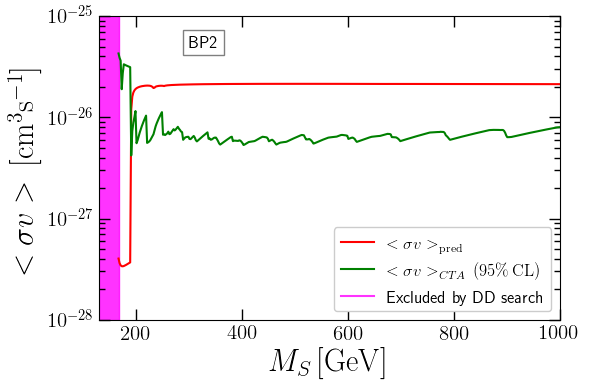}\label{fig:BP2_CTA_Scan2}}
}
\caption{(a) Relative contribution of the DM pair annihilation and (b) CTA sensitivity for different DM masses for benchmark BP1; 
(c) Relative contribution of the DM pair annihilation and (d) CTA sensitivity for different DM masses for benchmark BP2.
 Note that all the points along the lines satisfy the observed DM relic density.
}
\label{fig:BP_CTA_SCAN}
\end{figure}

Figure~\ref{fig:DM_BP_ID} shows the associated photon energy spectra  for different values of $\lambda_b$  for BP1 and BP2 satisfying the relic density constraint. For both benchmarks, there is a kinematic edge at $E\sim 227$~GeV which follows from the fact that the dominant channel which leads to photon final states is $HH$ and $H_5H_5$ for BP1 and BP2 respectively as can be seen in Fig.~\ref{fig:BP_CTA_SCAN}.  For the larger value of  $\lambda_b$,  annihilation into BSM states dominates over SM annihilation channels. This then leads to more high-energy photons above 100 GeV.  The gamma-ray flux of high-energy photons from $100$~GeV to $100$~TeV originating from the DM pair-annihilation results in better sensitivity in CTA~\cite{Gueta:2021vrf,Miener:2021ixs}. Thus, for both BP1 and BP2, CTA sensitivity increases due to the hard photon spectrum, resulting in probing the $\langle \sigma v \rangle$ well below the typical WIMP DM annihilation $\langle \sigma v \rangle \sim 3\times 10^{-26}\,\text{cm}^3\text{sec}^{-1}$.
 Figures~\ref{fig:ID_DM_spectrum3} and \ref{fig:ID_DM_spectrum4}  show variation of  $\langle\sigma v\rangle$ varies with $\lambda_b$. 
 The  limits from indirect detection searches from Fermi-LAT and the projected reach of CTA  at 95$\%$ CL (following ref.~\cite{CTA:2020qlo}) are shown as well. As can be seen, the direct detection (DD) experiments severely constrain the lower values of $\lambda_b$. The red points in Figs.~\ref{fig:ID_DM_spectrum3} and \ref{fig:ID_DM_spectrum4} represent the points allowed both by DD and Fermi-LAT, although we notice that Fermi-LAT constraints do not impose any stringent restrictions for our parameter space. In blue dashed line, we show the future projection for  indirect detection searches with CTA. As can be seen, CTA  will have the sensitivity to probe both benchmark points for any value of $\lambda_b$.  

 Figures~\ref{fig:BP1_CTA_Scan1} and \ref{fig:BP2_CTA_Scan1} show the relative contributions of DM annihilation channels versus $M_S$. Dark matter masses below 200 GeV are severely constrained from direct detection searches.
 For BP1, the dominant channel always corresponds to the heavier state that is kinematically accessible: viz.  $H_3 H_3$ for $M_S \sim 280$ GeV and  $H_5 H_5$ for $M_S \sim 300$~GeV.  For BP2, $H_5 H_5$ is the  dominant channel followed by subdominant contributions from $H_3 H_3$, $HH$ and $SM SM$ for  the entire mass range for $M_S>200$~GeV. 
 The  annihilation cross-section $\langle\sigma v\rangle$ versus the dark matter mass are shown in Fig.~\ref{fig:BP1_CTA_Scan2} and ~\ref{fig:BP2_CTA_Scan2} together with the CTA sensitivity.  
For both BP1 and BP2,   the predicted indirect detection cross-section from the GM-S model is larger than the expected reach from CTA in most of the mass range. 
The only region where the  predicted cross-section is below  the reach for CTA  corresponds to  $M_S < 227 \,(200)$ GeV for BP1 (BP2). In this region, the pair production of new scalars is not kinematically accessible at low velocities and the predicted
$\langle\sigma v\rangle$ which involves only the SM final states drop significantly at $M_S \sim 227~ (200)$~GeV as can be seen in Fig.~\ref{fig:BP1_CTA_Scan2} (\ref{fig:BP2_CTA_Scan1}) for BP1 (BP2).
Note that the subsequent dips in the indirect detection cross-section occur around $M_S \sim 250$ and 300~GeV where $H_3H_3$ and $H_5H_5$ modes open up for BP1.

\subsection{DM Global Scan}
\label{sec:scan}
\begin{figure}[t!]
\mbox{
\subfigure[]{\includegraphics[width=0.5\textwidth]{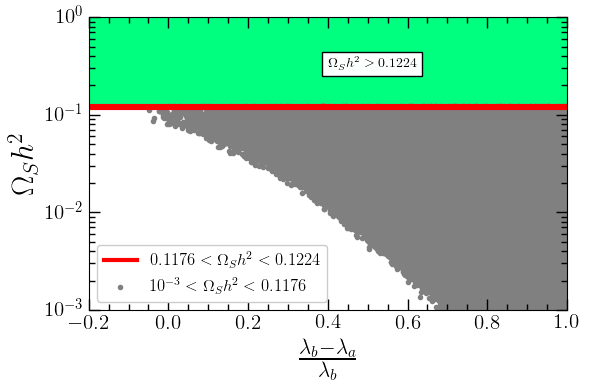}\label{fig:DM_Gscan_res1}}}
\mbox{
\subfigure[]{\includegraphics[width=0.5\textwidth]{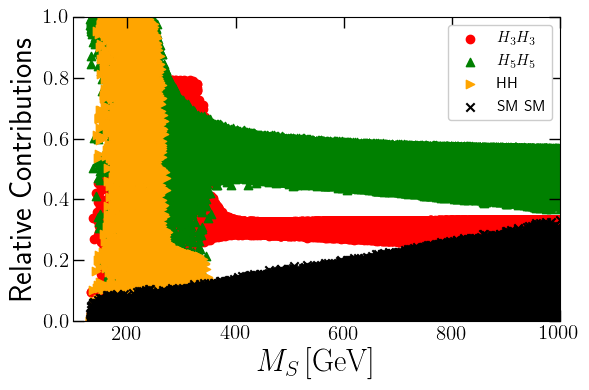}\label{fig:DM_Gscan_res2}}
}
\caption{(a) Variation of DM relic abundance $\Omega_{S}h^{2}$ vs $\frac{\lambda_{b}-\lambda_{a}}{\lambda_{b}}$ after imposing the constraint from LZ, (b)  relative contribution of different annihilation channels vs.\ $M_{S}$.}
\label{fig:DM_GScan_res}
\end{figure}

\begin{figure}[!t]
\centering
\includegraphics[width=0.6\textwidth]{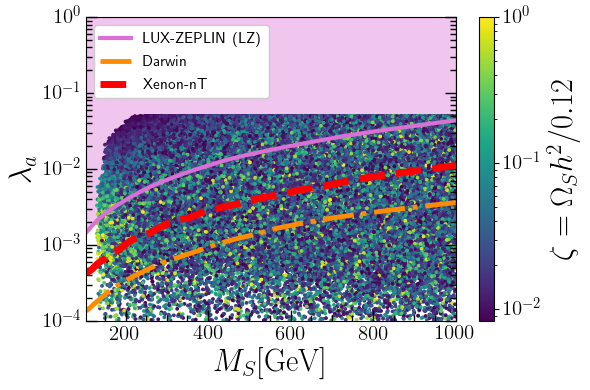}\label{fig:DM_Gscan_DD}
\caption{Impact of the constraint from DD searches from LZ 
(pink solid line) and projected reach of XENONnT~\cite{XENON:2015gkh} and  DARWIN~\cite{DARWIN:2016hyl}  shown by the red dashed and orange dash-dotted lines, respectively.  The colour pallet corresponds to the ratio of predicted dark matter relic abundance and observed dark matter relic abundance.}
\label{fig:DM_GScan_DD1}
\end{figure}

For all the points that satisfy the theoretical and experimental constraints discussed in previous sections which is about 1.4$\times 10^3$ allowed points, we have performed a global scan over the four  parameters $\{\lambda_a, \lambda_b, \lambda_S, M_S\}$ relevant for DM production and detection,  and computed the DM observables of the GM-S model. The total number of points sampled is about $3 \times 10^5$. We  ensure that the points satisfy the theoretical constraints discussed in section~\ref{sec:thconstraints}. Moreover for these points $H \to S S$ is kinematically forbidden.
Figure~\ref{fig:DM_Gscan_res1} shows the variation of the relic density versus $\frac{\lambda_b - \lambda_a}{\lambda_b}$ after imposing the direct detection constraint from LZ. The red points denote the points satisfying thermal relic within 2$\sigma$ of the PLANCK measurement. As discussed above, 
 as $\frac{\lambda_a}{\lambda_b}$ decreases, the relic density decreases leading to a large region of the parameter space being underabundant due to the proliferation of the relative contribution of the different BSM modes as shown in Fig.~\ref{fig:DM_Gscan_res2}. In the following, we keep only points for which  $\Omega_{S} h^2 > 0.001$ such that the singlet constitutes  roughly, at least 1\% of DM.  As discussed above, see also Fig.~\ref{fig:DM_GScan_DD1}, the DD constraint forces $\lambda_a$ to be small and we find that in general  $\lambda_a/\lambda_b < 1$. There are only a few points with $\lambda_a>\lambda_b$ which are associated with TeV scale DM where the DD constraint is weaker.  

 This implies that there is an important contribution from DM annihilation channels to BSM states over the entire allowed parameter space. Figure~\ref{fig:DM_Gscan_res2} shows that the relative contribution of DM annihilation to SM particles is always less than $30\%$  for the entire DM mass range, and is less than $10\%$ for $M_S< 200$~GeV corresponding to the region where the DD constraint is strongest. It also shows that  $H_5H_5$ channels are the dominant BSM modes for $M_S > 370$ GeV while any of the three modes $HH$, $H_3H_3$ and $H_5H_5$ can be  dominant  when  $M_S < 340$ GeV.  

In Fig.~\ref{fig:DM_GScan_DD1}, we show the impact of the spin-independent direct detection cross-section on  $\lambda_a  $   versus the DM mass $M_S$ with $\zeta =\Omega_S h^2/0.12$ in the coloured palette. We can infer that values of $\lambda_a>10^{-3}$ are excluded by LZ  \cite{LZ:2022lsv} for lower DM masses while the lower limit $\lambda_a>5\times 10^{-2}$ applies at $M_S=1$~TeV. For lower values of $\lambda_a$, both DD and relic density constraints are satisfied. 
Moreover some of the lower values of $\lambda_a$ remain well within  the ambit of future direct detection experiments such as XENONnT and DARWIN while a fraction of the parameter space remains out of reach of future experiments. Note that as  $M_S$ approaches the TeV scale,  we find a large number of points where the relic density is within the range measured by PLANCK ($\zeta\approx 1$).
Since we work in the decoupling limit such that the BSM Higgs  does not couple to SM fermions,  $\lambda_b$ remains unconstrained from direct detection searches.

\begin{figure}[!t]
\mbox{
\subfigure[]{\includegraphics[width=0.48\textwidth]{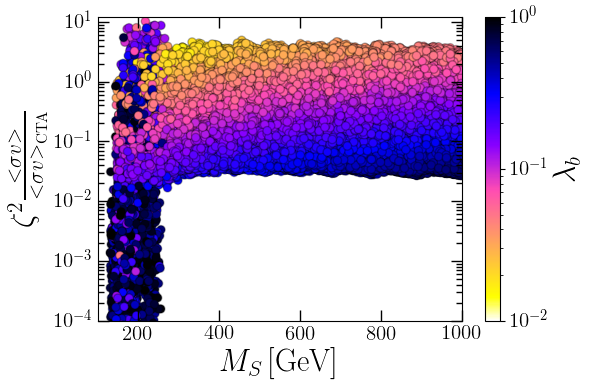}\label{fig:ID_DM_GS_res1}}
\subfigure[]{\includegraphics[width=0.48\textwidth]{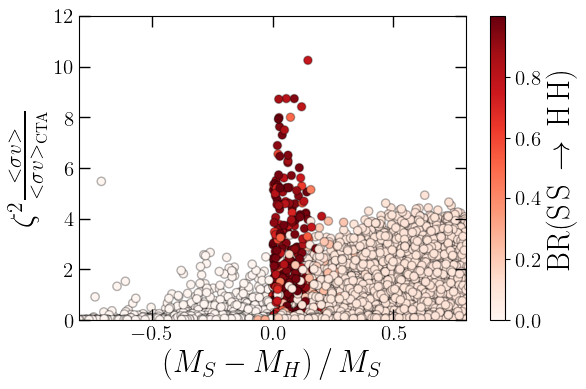}\label{fig:ID_DM_GS_res2}}
}
\caption{ Ratio of the rescaled predicted cross-section to the 95\%  reach of CTA  as a function of (a)~$M_S$ with $\lambda_b$ in the color palette and (b)~$(M_S-M_H)/M_S$ with the fraction of DM pair annihilation into $HH$ in the color palette.  }
\label{fig:DM_GS_ID}
\end{figure}

Following the procedure described in the previous section, we also explored the sensitivity of indirect detection experiments. As expected, we found that the current Fermi-LAT data does not constrain the GM-S model since the typical annihilation cross-section ($\langle\sigma v\rangle \approx 3 \times 10^{-26} {\rm cm}^3/{\rm s}$) is much below  the Fermi-LAT limit for DM masses above 100~GeV. On the other hand, CTA will have the potential to probe a fraction of the parameter space of the GM-S model. 
Figure~\ref{fig:DM_GS_ID} shows the sensitivity of the DM pair annihilation cross-section at CTA  versus the dark matter mass $M_S$. The predicted cross-section is scaled by the square of the DM  fraction ($\zeta^2$) and compared to  the cross-section that can be reached by CTA including all DM annihilation channels into photons.  The values of  $\lambda_b$ are shown in the color bar. The sensitivity ($\zeta^2 \langle \sigma v\rangle / \langle \sigma v\rangle_{\rm CTA}$) can exceed 1 for the whole range of DM masses. Many points are however beyond the reach of CTA, typically they correspond to underabundant DM ($\zeta<< 1$) found for the larger values of $\lambda_b$. In fact we found that in general, points where $\Omega_S h^2 \approx 0.12$ were well within reach of CTA. The only exception is found in the region $M_S \approx 200 {\rm GeV}$ when the DM mass is close to either $M_H$, $M_{H_3}$ or $M_{H_5}$ due to threshold effects. In these cases  the annihilation channels into any pair of BSM scalars can contribute significantly to the relic density, but they have a negligible contribution  at the lower velocities  relevant for indirect detection.  Therefore only SM final states are accessible and  $\langle \sigma v\rangle$ for Indirect Detection (ID) is much suppressed.  Note however that there is also a peak in the sensitivity around $M_S \sim 200 $ GeV in Fig.~\ref{fig:ID_DM_GS_res1}, this corresponds to DM annihilating dominantly into  $HH$ as shown by the red points in Fig.~\ref{fig:ID_DM_GS_res2}. For higher values of the dark matter mass, the $H_3H_3$ and $H_5H_5$ channels open up and again there can be a significant  increase in sensitivity although it  never exceeds 4.
We conclude that, while $\lambda_b$ is unconstrained from direct detection data, future observations may constrain $\lambda_b$ substantially.

\section{Future prospects at colliders} \label{sec:colliderpheno}
In this section, we briefly discuss the discovery prospect of the BSM Higgs $H^0_5$  at the high-luminosity run of the LHC, focusing on the diphoton channel. As discussed in previous sections, current theoretical and experimental constraints restrict the mass of the new scalars in a relatively narrow range, namely  $140~{\rm GeV }< M_{H_5} < 350~{\rm GeV }$, $150~{\rm GeV }< M_{H_3} < 270~{\rm GeV }$ and 
$145~{\rm GeV }< M_{H} < 300~{\rm GeV }$. These are well within the range of the LHC, hence we investigate in this section the prospects of searches for these new scalars at the HL-LHC.  In section~\ref{sec:bound}, we have shown that current  searches for $H^{\pm \pm} \to W^{\pm} W^{\pm}$ rule out the parameter space where this decay channel is dominant. Hence, for the doubly charged Higgs, one would need to investigate $H_5^{\pm \pm} \to H_3^{\pm} W^{\pm}$ and its cascade decays which lead to multi-leptons final states. However, the leptons being soft, this channel is challenging.  A potentially  more promising channel is the one with $H_i \to \gamma\gamma$ leading to a clean high $p_T$ diphoton signature. Although both $H_5^0$ and $H$ can decay into diphotons, as an example in this work we will consider only $H_5^0 \to \gamma \gamma$.

\begin{figure}[!t]
\centering
\includegraphics[width=0.7\textwidth]{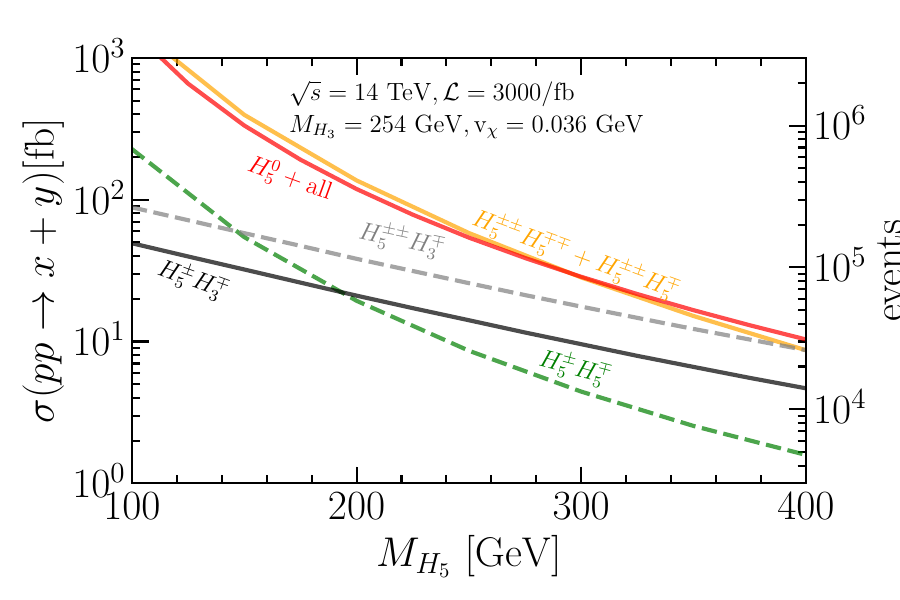}
\caption{Production  cross section (pair and associated) of BSM Higgs states as a function of $M_{H_5}$ for $M_{H_3}=254$ GeV and $v_\chi=0.036$ GeV. Here, $pp \to H_5^0 + all$, represents $pp\to H_5^0+\{V,H_5^\pm,H_3^\pm,H_3^0,H_5^0\}$. The right y-axis represents the number of events obtained for an integrated luminosity $\mathcal{L}=3000/$fb.  }
\label{fig:mvsxs}
\end{figure}

Before proceeding further with a complete analysis of the diphoton projection for HL-LHC, we show in Fig.~\ref{fig:mvsxs} the production cross sections (and the number of events along the right $y$-axis) as a function of $M_{H_5}$ for a number of processes involving at least one member of the fiveplet. Here we fix $M_{H_3}=254$~GeV and $v_\chi=0.036$~GeV corresponding to one of the benchmarks (BP1) considered in section~\ref{sec:DMpheno} to study DM production and detection.  Note that all the processes shown are driven by DY so independent of $v_\chi$).  The number of events correspond to an integrated luminosity $\mathcal{L}=3000~\rm{fb}^{-1}$ for HL-LHC. In the following, we analyse the diphoton signal from the process $pp \to H_5^0 + all$, which includes production of $H_5^0$ in association with $V,H_5^\pm,H_3^\pm,H_3^0,H_5^0$, with the dominant contribution coming from $pp \to H_5^0 H_5^\pm$.

\begin{table}[!b]
\centering
\begin{tabular}{|c|c|c|c|c|c|c|}\hline
&$M_{H_5}$ [GeV]& $M_{H_3}$ [GeV] & $v_\chi$ [GeV]&$M$ [GeV]& BR($ H_5^0 \to \gamma\gamma$)& $ \Gamma(H_5^0)$ [GeV]\\ \hline
BP1&300&254&0.036&335& 0.59& 45.7  \\ \hline
BP2&190&234&0.05&210&0.70& 3.8 \\ \hline
\end{tabular}
\caption{Benchmark points along with respective branching ratio of  the decay channel $H^0_5 \to \gamma \gamma$ and total decay width of $H_5^0$. }
\label{tab:BP}
\end{table}

We consider the  two benchmark points defined in Table.~\ref{tab:BP}. These are chosen among the allowed points in Fig.~\ref{Fig:Th_Exp_Allowed},  for BP1 $M_{H_5}$ is near the maximum value allowed while for BP2 features a much lighter fiveplet. Both points have a rather large branching ratio into two photons, $\textrm{Br}(H^0_5 \to \gamma \gamma) >  50\%$. Moreover, both points are consistent with the  dark matter constraints discussed in section~\ref{sec:DMpheno}. 

The production process of $H^0_5$ at the HL-LHC is dominated by $p p \to H^0_5 H^{\pm}_5 $.  The cross-section followed by $H^0_5 \to \gamma \gamma$,  is  $\sigma=0.0167$~pb and 0.0976~pb for BP1 and BP2, respectively. For the signal,  we  consider the two-photon final state  arising from the decay of  $H^0_5$.  For BP1 we also consider  cascade decays of $H^{\pm}_5$ state  that give rise to multi-photon final states, namely 
\begin{align}
   & {\rm BP1}: p p\to H_5^0+ H_5^\pm\to 2\gamma+ W^\pm H^{0}_3 (Z H^{\pm}_3) \to 2 \gamma + W^{\pm} + Z H  \to  3(4)\gamma+ X \\ \nonumber
   & {\rm BP2}: p p\to H_5^0+H_5^{\pm}\to 2\gamma+Z W \to 2\gamma+ X
\end{align}

In the above decay chain for BP1, the branching ratios for $H_5^\pm \to W^\pm H_3^0$ and $H_3^0\to ZH$ are nearly 100\%, moreover the branching ratios for  $H \to \gamma\gamma $ and $H \to \gamma Z$  reach together almost $100\%$. For BP2, the primary decay mode of $H^{\pm}_5$ is $Z W$ with $100\%$ branching ratio. 

In  the following, we implement a strong invariant mass cut around $M_{H_5}$ to select only those events originating from $H^0_5$.  Hence, an accurate reconstruction of the diphoton invariant mass can enhance the signal sensitivity. The photons are hard which facilitates event selection via a diphoton trigger~\cite{ATLAS:2021uiz}. 

\begin{figure}[t]
\centering
\includegraphics[width=0.45\textwidth]{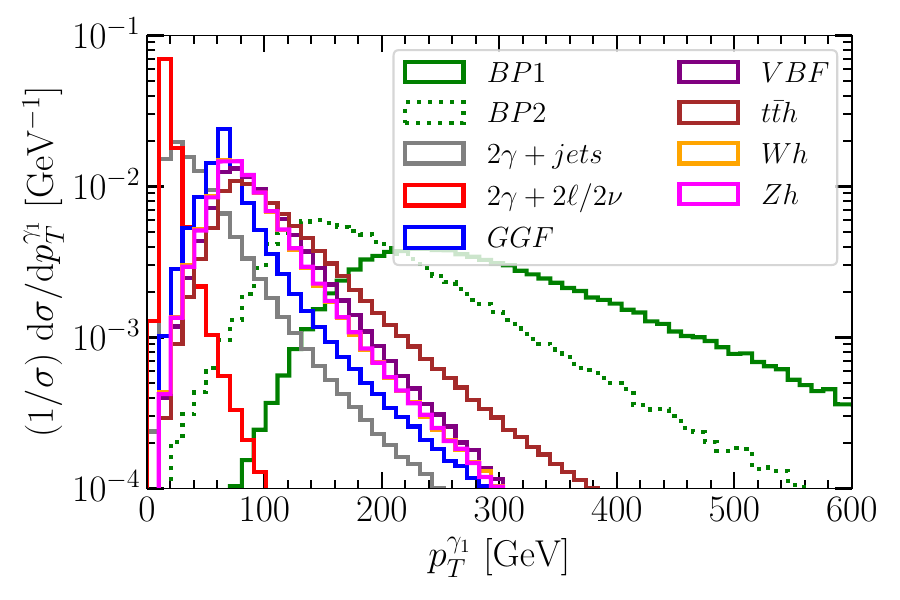}
\includegraphics[width=0.45\textwidth]{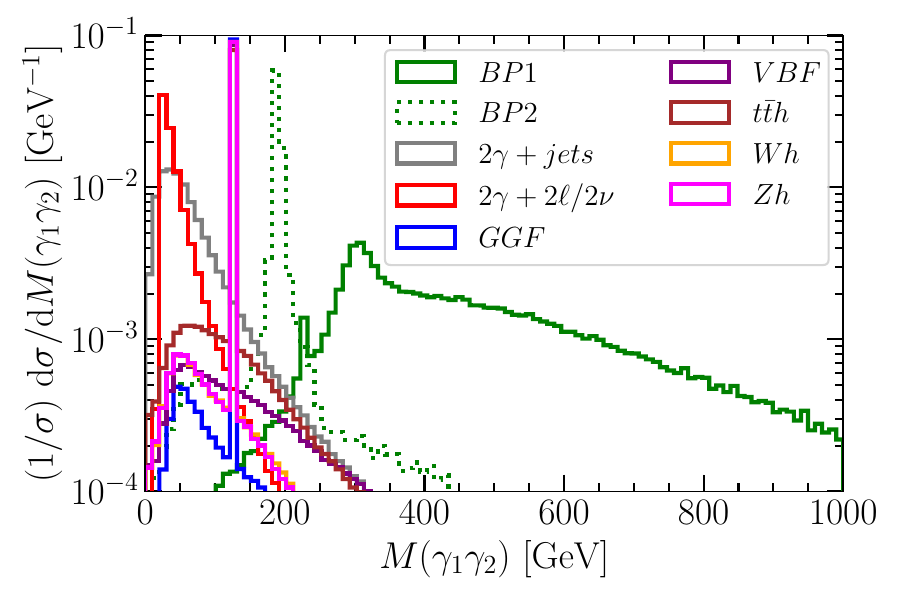}
\caption{Distribution of transverse momentum $p^{\gamma_1}_T$ of  the leading photon and  invariant mass of the  leading and sub-leading photon $M(\gamma_1 \gamma_2)$  for signal and background. }
\label{fig:dist}
\end{figure}

\begin{table}[!t]
\centering
	\begin{tabular}{|c|c|c|c|c|c|c|c|c|}\hline
 \multicolumn{2}{|c|}{$\sigma^s$ (pb) } &\multicolumn{7}{c|}{$\sigma^b$ (pb)} \\ \hline
  BP1  &BP2 &GGF   & VBF & $Zh$  &$ Wh $&$ t \bar{t}h$& $\gamma\gamma+2\ell(2\nu)$& $\gamma\gamma+jets$  \\ \hline
 $0.0167$& $0.0976$ &49.6&4.26 &0.9 &1.5 &0.6 &156.8&128.34\\ \hline
 		\end{tabular}
	\caption{Signal and background cross sections before applying selection cuts.}
	\label{tab:partonicxsec}
\end{table}
We simulate signal and background events with MadGraph5amc@NLOv3.5.1~\cite{Alwall:2014hca}. The dominant background for diphoton is $pp \to \gamma\gamma$. 
As our signal contains soft leptons and jets in association with two or more photons, we consider other background processes such as $pp\to \gamma\gamma + 2l/2 \nu/jets$ as well as SM Higgs production channels via gluon-fusion (GGF), vector boson fusion (VBF) and $Wh,Zh$,$t \bar{t}h$, and its decay to diphoton final state. Note that the background  
$pp \to \gamma \gamma +2\ell(2 \nu)$ does not involve the contribution from $pp\to Zh \to 2\ell + 2\gamma $, the latter is therefore generated separately. In Table~\ref{tab:partonicxsec}, we present the partonic cross-section for the signal $p p \to H^0_5 + all $, followed by $H_5^0$ decay to diphoton and for the SM background processes. The process $ \gamma\gamma + 2l(2 \nu)$ includes $pp\to \gamma\gamma$, $pp\to \gamma\gamma +2\ell$ and $pp\to \gamma\gamma +2\nu$, while  $ \gamma\gamma + jets $ includes $pp\to  \gamma\gamma +1jet $ and $pp\to  \gamma\gamma +2jets$. At the parton level we impose the following cuts, $p_T^{\gamma,\ell}\ge10$ GeV, $p_T^j\ge20$ GeV , $|\eta^{\gamma.\ell}|\leq2.5, |\eta^j|\leq5$.
Note that, for signal, the cross-section presented in Table.~\ref{tab:partonicxsec} includes $H^0_5 \to \gamma \gamma$ branching ratio, while for backgrounds involving SM Higgs boson, we only show the production cross-section for $g g \to h, p p \to Zh, Wh$ and $t \bar{t}h$. The background cross sections involving the SM Higgs state  are taken from \cite{CERN:twiki} and have been  multiplied with $\textrm{Br}(h \to \gamma \gamma)=0.25\%$.

The parton level events are passed to PYTHIA8.306~\cite{Sjdostrand:2007gs} to simulate parton shower and hadronization. In Fig.~\ref{fig:dist}, we present the distributions of transverse momentum $p^{\gamma_1}_T$ for the leading photon and of the invariant mass of the  leading and sub-leading photon $M(\gamma_1 \gamma_2)$. As can be seen, the majority of the signal events  populate  the region $p^{\gamma_1}_T > 100$  GeV whereas, the SM backgrounds peak in the region   $p^{\gamma_1}_T < 100$  GeV.  Thus, we expect $p_T$ cuts on the leading and sub-leading photon to reduce the  background as discussed below.   The $M(\gamma_1 \gamma_2)$ distribution  in Fig.~\ref{fig:dist} shows that the signal    peaks at $M(\gamma_1 \gamma_2)\sim M_{H_5}$. Therefore, a mass window around $M_{H_5}$  is required to select the signal events. Note that for BP1, where $H$ can arise from  the cascade decay of $H_5^\pm$, there is a small peak around $M_H$ which is due    to $H\to 2\gamma $, as can be seen from the solid green line in the right panel of Fig.~\ref{fig:dist}. 
After analysing the kinematics of the signal and background we consider the following sets of cuts to evaluate the signal sensitivity. 
\begin{itemize}
    \item $c_1:~N^\gamma\ge 2$
    \item $c_2:~c_1+ \{ p_T^{\gamma_{1,2}} \ge 80, 30~\text{GeV},~ |\eta^{\gamma_{1,2}}|<2.37$ \}
    \item $c_3:~c_2$+isolation \footnote{
    For isolation, we demand that the scalar sum of $p_T$ of all the stable particles (except neutrinos) found within a $\Delta R= 0.4$ cone around the photon direction is required to be less than $0.05~p_T + 6$ GeV.} for $ \gamma_{1,2} $ 
    \item $c_4:~c_3+  \{\text{ Reject event with } 1.37<|\eta^{\gamma_{1,2}}|<1.52$ \}
    \item $c_5:~c_4+ |M(\gamma_1 \gamma_2)-M_{H_5}| =\pm 20$ GeV
\end{itemize}

\begin{table}[t]
\centering
	\begin{tabular}{|c|c|c|c|c|c|c|c|c|}\hline
& \multicolumn{2}{c|}{	$\epsilon$  for signal } &\multicolumn{6}{c|}{$\epsilon$ for backgrounds} \\ \hline
&  BP1  &BP2 &GGF   & VBF & $Zh$  & $ Wh $ & $ t \bar{t}h$ & $2 \gamma+2\ell/2\nu~(+jets)$ \\ \hline
$c_1$ & 0.99 &0.99 &1.0     & 1.0    & 1.0     & 1.0    &  1.0    & 1.0    (0.99) \\ \hline
$c_2$ & 0.93 &0.72 &0.18 & 0.36 & 0.29& 0.29&  0.49 & 0.07 ($6.10^{-3}$) \\ \hline
$c_3$ & 0.69 &0.62 &0.17 & 0.33 & 0.26& 0.25&  0.30 & 0.05 ($6.10^{-3}$) \\ \hline
$c_4$ & 0.69 &0.61 &0.17 & 0.32 & 0.25& 0.25&  0.29 & 0.05 ($6.10^{-3}$) \\ \hline
$c_5$ & 0.11 &- & $1.1.10^{-5}$ & $10^{-4}$ & $6.5.10^{-5}$ & $8.8.10^{-5}$& $10^{-4}$ & 0.002 ($10^{-4}$) \\ \hline
$c_5$ & - &0.56 &$2.3.10^{-5}$ & $3.0.10^{-4}$ & $2.0.10^{-4}$& $3.0.10^{-4}$& $3.4.10^{-4}$ & $7.8.10^{-3}$ ($1.5.10^{-3}$) \\ \hline
\multicolumn{9}{|c|}{Events with $\mathcal{L}=3000/$fb, $\mathcal{S}=N^s/\sqrt{N^s+N^b+(N^b\delta b)^2}$} \\ \hline
 \multicolumn{5}{|c|}{BP1} &\multicolumn{4}{c|}{BP2}\\ \hline
 \multicolumn{5}{|c|}{$ N^s=5711, ~N^b=1032684$, $\mathcal{S}=0.55$} &\multicolumn{4}{c|}{$N^s=164434,N^b=3758548$, $\mathcal{S}=4.36$}\\ \hline
 		\end{tabular}
	\caption{Cut efficiency $\epsilon$ for signal and background. }
	\label{tab:cutflow}
\end{table}

We present a detailed cut-flow table for the two benchmark points and the SM backgrounds in Table~\ref{tab:cutflow}. As can be seen from this table, the signal cross-section is strongly suppressed for BP1, the global cut-efficiency is $11\%$,  while for BP2 the global cut-efficiency is much better, $56\%$. The reason is that the $M(\gamma_1 \gamma_2)$  distribution features a much narrower peak for BP2 than for BP1, see  Fig.~\ref{fig:dist}, right panel.
We find that after cuts $\sigma^s= 1.8\times10^{-3} (0.054)$ pb for BP1 (BP2).  After implementing the selection cuts, the largest background comes from $2 \gamma+2\ell/2\nu$ followed by $\gamma \gamma +jets$.  We calculate the signal significance as $\mathcal{S}=N^s/\sqrt{N^s+N^b+ (N^b\delta b)^2}$, where $N^s~(N^b)$ is the signal~(background) events with $\mathcal{L}=3000/$fb and $\delta b$ is the background uncertainty considered to be $1\%$. As shown in Table~\ref{tab:cutflow}, $\mathcal{S}=0.55$ and $\mathcal{S}=4.3$ can be obtained for BP1 and BP2, respectively.

We, therefore, conclude that  the heavier $H_5^0$ could not be probed at the HL-LHC via the two-photon channel while scenarios with lighter $H_5^0$ such as in BP2  offer better prospects. Note that optimized cuts and advanced signal identification techniques should improve the significance determined here. 


\section{Summary and Conclusion} \label{sec:summary}
In this work, we studied the collider and dark matter phenomenology of the GM-S model, a singlet extension of the Georgi-Machacek model, in its decoupling limit. In this limit, the common vev of the $Y=0$ and $Y=2$ triplets,  $v_\chi$,  is very small.  Based upon the transformation properties under the custodial symmetry, the physical scalars can be categorized into a 
fiveplet ($H^{\pm\pm}_5$, $H^{\pm}_5$ and $H^0_5$), a triplet ($H^{\pm}_3$ and $H^0_3$) and two CP-even singlets ($H,\,h$). In the decoupling limit,  the mixing between $h$ and $H$ is identically zero
as imposed by  perturbative unitarity of the quartic couplings. We performed a comprehensive study by taking into account theoretical constraints  (arising from, e.g., perturbative unitarity), measurement of oblique parameters, direct experimental searches/measurements  at the LHC, such as,  Higgs-to-diphoton rate measurements at the LHC, searches for a doubly charged Higgs and for heavy resonances decaying into diphotons  at the 13 TeV LHC. We find that:

\begin{itemize}
\item
Perturbative unitarity imposes an upper bound on the masses of the triplet and fiveplet, $M_{H_3}<280$ GeV, and $M_{H_5}<435$ GeV while oblique parameters measurements forbid a large mass splitting between these two states.
\item
The  charged scalars can give a significant contribution to the diphoton decay width 
of neutral scalars. 
Agreement with LHC measurements of the $h \to \gamma \gamma$ rate  imposes that $M_{H_5}-M \lsim 30$ GeV, while requiring that the total widths of the BSM scalars satisfy $\Gamma (H_i)/M_i < 0.5$  leads to $v_{\chi} > 4.8 \times 10^{-3}$ GeV. 

\item 
In the allowed scenarios,  the same-sign di-boson ATLAS search for doubly charged Higgs constrains the region where $M_{H_5} \sim 200 - 300$ GeV and $M_{H_3} \sim M_{H_5}$. 
\item 
The 
ATLAS search for heavy spin-0 resonances decaying into diphoton final states from Run~2 rules out $M_{H_3}\:(M_{H_5})<150\:(140)$~GeV. Moreover, it imposes a lower bound on $v_{\chi}$ of $v_{\chi} \ge 0.05$~GeV.
\end{itemize}

For the scan points that satisfy these theoretical and collider constraints,  we performed an additional scan on the dark matter parameters, imposing constraints from the observed DM relic density as well as direct and indirect DM detection experiments.
Our major findings are:  
  \begin{itemize}
    \item 
    The quartic coupling $\lambda_a$ is severely constrained from the direct detection 
    limit by LUX-ZEPLIN~\cite{LZ:2022lsv}. 
    As a consequence, the dominant contribution to the DM relic density comes from the annihilation channels $SS \to H H, H_5 H_5, H_3 H_3$, which depend on $\lambda_b$. 
    \item 
    The presence of new scalars that can be produced in DM pair annihilation, $S S \to H_5 H_5, H H,H_3 H_3$
    , improve the DM indirect detection prospects as the direct decay $H^0_5, H \to \gamma \gamma$ can induce a harder photon spectra as compared to SM final state. This significantly improves the prospect for CTA to probe the model. 
\end{itemize}

Last but not the least, we chose two benchmark points corresponding to $M_{H_5} =190$ and $300$~GeV and large BR$(H_5^0\to\gamma\gamma)$ and explored the detection prospects  at the High-Luminosity run of the LHC. 
For the lighter point, $M_{H_5} =190$~GeV, we found a significance of $>4\sigma$ in final states with at least two photons. 
 
In conclusion, the GM-S model in the decoupling limit  offers a rich phenomenology  that may be probed in near future direct and indirect detection experiments as well as at the high-luminosity run of the LHC.
Note finally, that we only considered  DM masses $\ge 100$~GeV, for which the observed SM-like Higgs boson cannot decay invisibly. The possibility of invisible Higgs decays in the GM-S model is left for future work.

\section*{Acknowledgements}

We thank Heather Logan for helpful discussions on the model and Christopher Eckner for discussions regarding the DM Indirect Detection. We would also like to thank Nicolas Berger, Christopher Robyn Hayes, Aruna K. Nayak and Tribeni Mishra for useful discussions regarding the ATLAS analysis. We thank Agnivo Sarkar for his contribution during the initial stage of this project and Jack Y. Araz for some useful discussions.
The authors also acknowledge the use of SAMKHYA: High-Performance Computing Facility provided by the Institute of Physics (IoP), Bhubaneswar. 
This work was funded in part by the Indo-French Centre for the Promotion of Advanced Research under grant no.~6304-2 (project title: Beyond Standard Model Physics with Neutrino and Dark Matter at Energy, Intensity and Cosmic Frontiers). RMG acknowledges furthermore support from the Indian National Science Academy (INSA) under the INSA senior scientist award. RP is supported in part by Basic Science Research Program through the National
Research Foundation of Korea (NRF) funded by the Ministry of Education, Science and
Technology (NRF-2022R1A2C2003567).
\appendix 
\section{Partial decay widths of BSM Scalars}
\label{appen1}

In the following we present the general formulae for the partial decay widths of scalar $H_I$~\cite{Hartling:2014aga,Hartling:2014zca}.

\begin{eqnarray}
\Gamma(H_I \to l_i^{\pm} l_j^{\pm})&=&\frac{M_{H_I}}{8 \pi}
\left\{ \left[ 1 - (x_i + x_j)^2 \right] |g^S|^2 + \left[ 1 - (x_i - x_j)^2 \right] |g^P|^2 \right\} \lambda^{1/2}(x_i^2, x_j^2) \nonumber\\
\Gamma(H_I \to V_1 V_2)&=&\mathcal{S} \frac{|g_{H_I V_1V_2}|^2 M_{H_I}^3}{64 \pi M_{V_1}^2 M_{V_2}^2} \left[ 1 - 2 k_1 - 2 k_2 + 10 k_1 k_2 + k_1^2 + k_2^2 \right] \lambda^{1/2}(k_1, k_2) \nonumber\\
\Gamma(H_I \to V H_2 )&=& 
 \frac{|g_{V H_I H_2}|^2 M_V^2}{16 \pi M_{H_I}}
\lambda\left( \frac{M_{H_I}^2}{M_V^2}, \frac{M_{H_2}^2}{M_V^2} \right)
\lambda^{1/2} \left( \frac{M_V^2}{M_{H_I}^2}, \frac{M_{H_2}^2}{M_{H_I}^2} \right) \nonumber\\
\Gamma(H_I \to H_2H_3)&=& \mathcal{S} \frac{|g_{H_I H_2H_3}|^2}{16 \pi M_{H_I}}
\lambda^{1/2}(X_2, X_3)
\end{eqnarray}

In the above equations, $V=W^\pm,Z$ and $H_I$ corresponds to BSM scalars. The couplings are $g^S=g^P=Y^\nu_{ij}$ while the kinematic quantities are given by  $x_i = m_{l_i}/M_{H_I}$, $\lambda(x,y) = (1 - x - y)^2 - 4xy$, $k_i = M_{V_i}^2/M_{H_I}^2$,  $X_i= M_{\Phi_i}^2/M_{H_I}^2$ . $\mathcal{S}$ is a symmetry factor which takes values $\mathcal{S} = 1~(1/2)$ if final state particles are distinct (identical). Partial decay widths for off-shell decays are as follows
\begin{eqnarray}
\Gamma(H_I \to V V^{*})&=&\mathcal{S} \delta_{V} \frac{3|g_{H_I VV}|^2 M_{H_I}}{64 \pi^{3}v^2}\Big[\frac{1-8k+20k^2}{(4k-1)^{1/2}}\arccos{\frac{3k-1}{2k^{3/2}}}-\frac{1-k}{6k}(2-13k+47k^2)\,\, \nonumber\\&& -\frac{1}{2}(1-6k+4k^2)\log(k)\Big], \nonumber \\
\Gamma(H_I \to V^* H_2)&=& \delta_V \frac{3 |g_{V H_I H_2} |^2 M_V^2 M_{H_I}}
{16 \pi^3 v^2} G_{H_2 V},
\end{eqnarray}
where the factor $\delta_{V}$ and $G_{ij}$ are given by,
\begin{eqnarray}
    \delta_{W}&=&3/2, \,\nonumber\\
    \delta_{Z}&=&3\left(\frac{7}{12}-\frac{10}{9}\sin^{2}\theta_{W}+\frac{40}{27}\sin^{4}\theta_{W}\right) ,\nonumber \\
G_{ij} &=& \frac{1}{4} \left\{ 2 (-1 + k_j - k_i) \sqrt{\lambda_{ij}} 
\left[ \frac{\pi}{2} + \arctan \left( \frac{k_j (1 - k_j + k_i) - \lambda_{ij}}{(1 - k_i) \sqrt{\lambda_{ij}}} \right) \right] \right. \nonumber \\
&& \qquad \left. + (\lambda_{ij} - 2 k_i) \log k_i 
+ \frac{1}{3} (1 - k_i) \left[ 5 (1 + k_i) - 4 k_j + \frac{2 \lambda_{ij}}{k_j} \right] \right\}.
\end{eqnarray}

In the above equation, $k_i = M_{i}^2/M_{H_I}^2$, and $\lambda_{ij} = -1 + 2 k_i + 2 k_j - (k_i - k_j)^2$.
The vertex factor involved in the above decay width formula in the limit $s_\alpha=0$ are given by
\begin{eqnarray}
&&g_{HW^+W^-} =c_W^2g_{HZZ}=\frac{4e^2 v_\chi}{\sqrt{3}s_W^2},  \nonumber \\
&&g_{H_5^0 W^+ W^{-}} = \sqrt{\frac{2}{3}} \frac{e^2}{s_W^2} v_{\chi}, ~~g_{H_5^0 Z Z} = -\sqrt{\frac{8}{3}} \frac{e^2}{s_W^2 c_W^2} v_{\chi}, \nonumber \\
&&g_{H_5^+ W^{-} Z} =-\frac{\sqrt{2} e^2 v_{\chi}}{c_W s_W^2}, ~~
	g_{H_5^{++} W^- W^-} =   \frac{2 e^2 v_{\chi}}{s_W^2}.
 \label{eq:HVVvertex}
\end{eqnarray}

\begin{eqnarray}
	&&g_{W^+ h H_3^{-}}= - \sqrt{2} \frac{e}{s_W} \left( c_\alpha \frac{v_\chi}{v} \right),~~ g_{W^+ H H_3^{-}}= - \sqrt{\frac{2}{3}} \frac{e}{s_W}\left( -  c_\alpha  \frac{v_\phi}{v} \right),~~ \nonumber  \\
	&&g_{W^+ H_3^0 H_5^{-}} = \frac{i}{2} \frac{e}{s_W} \frac{v_\phi}{v},~~	g_{W^+ H_3^+ H_5^{--} }= \frac{1}{\sqrt{2}} \frac{e}{s_W} \frac{v_\phi}{v}, \nonumber \\	
    &&g_{ZhH_3^0} = -i\sqrt{2}\frac{e}{s_Wc_W}\left(\frac{v_\chi}{v} c_\alpha  \right),~~ g_{ZHH_3^0} = i\sqrt{\frac{2}{3}}\frac{e}{s_Wc_W}\left(c_\alpha \frac{v_\phi}{v} \right),~~ \nonumber \\
	&&g_{ZH_3^0 H_5^0}   = i \sqrt{\frac{1}{3}} \frac{e}{s_Wc_W}\frac{v_\phi}{v},~~  g_{ZH_3^+H_5^{-}} = -\frac{e}{2 s_Wc_W}\frac{v_\phi}{v}, \nonumber \\
	&&g_{W^+ H_3^{-} H_5^0 }= - \frac{1}{2 \sqrt{3}} \frac{e}{s_W} \frac{v_\phi}{v}.
\end{eqnarray}

\begin{eqnarray}
    g_{H_3^0H_3^0H_5^0} &=& -\frac{2\sqrt{2}}{\sqrt{3} v^2} 
	\left(-8\lambda_5 v_\chi^3 + 4 M_1 v_\chi^2 
	+ (-4\lambda_5+2\lambda_3)v_\phi^2 v_\chi+3 M_2 v_\phi^2\right), \nonumber \\
	g_{H_3^+H_3^{-}H_5^0} &=& \frac{\sqrt{2}}{\sqrt{3} v^2}
	\left(-8\lambda_5 v_\chi^3+4 M_1 v_\chi^2
	+ (-4\lambda_5+2\lambda_3) v_\phi^2 v_\chi+3 M_2 v_\phi^2\right), \nonumber \\
	g_{H_3^0H_3^+H_5^{-}} &=& -i\frac{\sqrt{2}}{v^2}
	\left(-8\lambda_5 v_\chi^3+4 M_1 v_\chi^2
	+ (-4\lambda_5+2\lambda_3) v_\phi^2 v_\chi+3 M_2 v_\phi^2\right), \nonumber \\
	g_{H_3^+H_3^+H_5^{--}} &=& -\frac{2}{v^2} 
	\left(-8\lambda_5 v_\chi^3 + 4 M_1 v_\chi^2  
	+ (-4\lambda_5+2\lambda_3) v_\phi^2 v_{\chi} + 3 M_2 v_\phi^2  \right) ,
 \nonumber \\
	g_{HH_5^0H_5^0} &=& g_{HH_5^+H_5^{-}} = g_{HH_5^{++}H_5^{--}} 
	= 8\sqrt{3}\left(\lambda_3+\lambda_4\right) v_\chi  + 2\sqrt{3}\,M_2 ,\nonumber \\
 g_{HH_3^0H_3^0} &=& g_{HH_3^+H_3^{-}} 
	=  \frac{8}{\sqrt{3}} \frac{v_\phi^2 v_\chi}{v^2}  \left(\lambda_3 + 3\lambda_4\right)
	+ \frac{4}{\sqrt{3}} \frac{v_\chi^2 M_1}{v^2}
	 \nonumber\\
	&&  +\frac{16}{\sqrt{3}} \frac{v_\chi^3}{v^2}
	\left(6\lambda_2 +\lambda_5\right)   
	- \frac{2\sqrt{3}\,M_2 v_\phi^2}{v^2}  +\frac{8}{\sqrt{3}} \lambda_5 \frac{v_\chi v_\phi^2}{v^2}.
\end{eqnarray}

\medskip

\clearpage
\bibliographystyle{apsrev4-1}
\bibliography{references}
\end{document}